\RequirePackage{fix-cm}
\documentclass[smallextended]{svjour3}       
\smartqed  
\usepackage{graphicx}
\usepackage{amsmath}
\usepackage{algpseudocode}
\usepackage{url}
\usepackage{caption}
\usepackage{acronym}
\usepackage{multirow}
\usepackage{hhline}
\usepackage{framed}
\usepackage{enumitem}
\usepackage{subfigure}
\usepackage{natbib}
\usepackage{longtable}
\usepackage{calc}

\usepackage{amssymb}
\usepackage{amsthm}

\usepackage{float}
\floatstyle{ruled}
\newfloat{algorithm}{thp}{lop}
\floatname{algorithm}{Algorithm}

\journalname{Empirical Software Engineering}

\begin{document}

\newtheorem{thm}{Theorem}
\newtheorem{defi}[thm]{Definition}

\newcommand{\etal}{~\emph{et al}}

\title{A systematic mapping study on cross-project defect prediction}


\author{Steffen Herbold}

\institute{Steffen Herbold \\
Institute of Computer Science\\
University of G\"{o}ttingen, Germany\\
\email{\{herbold\}@cs.uni-goettingen.de}}

\date{Received: date / Accepted: date}

\maketitle

\acresetall
\section{Introduction}
%
%

In order to utilize the often limited resources available
for the quality assurance of a software efficiently, test managers require tools
that support decision making regarding the focus of the quality assurance.
One such tool is defect prediction, i.e., the prediction of the location of
defects\footnote{The term defect is used throughout this paper. In the
literature, the terms fault and bug are also used interchangeably.}.
In recent years, the prediction of defects in a target product based on data
from other products, the so called \ac{CPDP} became a popular research topic.
Within this article, we provide a systematic mapping study of the literature onon \ac{CPDP}.
Three previous literature reviews were performed in the area of defect
prediction. Our mapping study differs from the others in the following aspects:
\begin{itemize}
  \item \emph{Timeframe:} the previous reviews cover the time up to
  1999 \citep{Fenton1999}, for the timeframe
  2000--2007 \citep{Catal2009a}, and the timeframe
  2000--2010 \citep{Hall2012}. Our study covers the timeframe
  2006--2015, whereas 2006 is the year of the first publication we
  identified that specifically addressed \ac{CPDP}. 
  \item \emph{Focus:} The previous reviews were focused on defect
  prediction in general. Moreover, the topic \ac{CPDP} was very new in the
  previously considered timeframes. The focus of our review is strictly on
  \ac{CPDP}, which was only a sidenote or too new to be covered by the previous
  reviews. 
  \item \emph{Analysis:} Following the guidelines by \cite{Kitchenham2007} we performed a detailed analysis of the
  state-of-the-art of \ac{CPDP} to provide an in-depth look into the
  approaches proposed, state of practice of case studies, and the results
  achieved. 
\end{itemize}

We make five significant contributions by presenting:
\begin{itemize}
  \item A set of 50 studies addressing \ac{CPDP} published from July 2002 to
  December 2015. 
  \item A summary of the approaches proposed for \ac{CPDP} which can be
  used by researchers as foundation for future investigations of \ac{CPDP}.
  \item A review of the state of practice of case study configurations and
  reporting. The review brings to light a very heterogeneous way of
  conducting and reporting case studies that precludes in-depth comparisons of the
  performance of approaches.
  \item An extension of the taxonomy proposed by \cite{Turhan2012} on
  ways that researchers can address cross-project problems based on the findings
  of our review.
\end{itemize}

The remainder of this article is structured as follows. In the next section, we
present our systematic mapping study methodology. In Section
\ref{sec:foundations}, we give the foundations required for our mapping study.
Afterwards, the results of our mapping study are presented in Section
\ref{sec:related-work}. and discussed in Section~\ref{sec:metastudy}. Finally,
we summarize our results and present our conclusions in Section~\ref{sec:conclusion}.

\section{Methodology}
\label{sec:methodology}
Our review follows the guidelines for
systematic literature reviews proposed by \cite{Kitchenham2007}. In
the following, we define our underlying research questions, inclusion
and exclusion criteria, how we identified papers, and which data was collected
for our study. We do not define our study as systematic literature review but as
a mapping study, as no synthesis of the results was possibly due to the
heterotenity of our findings. 

\subsection{Research Questions}

The motiviation of this mapping study was answering two questions: 1) which
techniques were already considered in the state-of-the-art of cross-project
defect prediction; and 2) how well do the approaches perform? Ideally, we wanted
to synthesize the findings to determine the best approaches from the literature.
However, during our initial considerations of the literature, we realized that
the case study settings are very diverse in terms of data, classifiers, and
performance measures. This precludes concrete comparisons of performances
between studies on a large scale. We would like to stress that this does not
imply that the case studies were carried out in an improper way or results were
not reported adequately. This was not the case for the publications we
considered, the case studies on their own were well done and not problematic.

However, due to the diversity of the considered aspects,
comparisons were almost impossible, meaning that considering all case studies
as a whole is problematic. For example, if approach A is evaluated only on data
set X and approach B is evaluated only data set Y, it is unclear how exactly A
performs on Y and B performs on X. Due to this, only subsets of studies would be
comparable to each other. Thus, we dropped the comparison of performance as
aspect from this mapping study. 

To study which techniques were already considered in the state of the art, we
defined the following five research questions. 
\begin{itemize}
  \item \textbf{RQ1.} Which approaches were already considered for \ac{CPDP}?
  \item \textbf{RQ2.} Which classifiers were the most popular for \ac{CPDP}
  studies?
  \item \textbf{RQ3.} Which data sets were used within \ac{CPDP} studies?
  \item \textbf{RQ4.} Which performance metrics were used to assess \ac{CPDP}?
  \item \textbf{RQ5.} Which baselines were proposed approaches compared to?
\end{itemize}

\subsection{Inclusion and Exclusion Criteria}

To identify which papers should be part of our review, we defined the following
criteria for inclusion:
\begin{itemize}
  \item publications with a case study that includes \ac{CPDP} experiments;
  or
  \item publications that discuss theoretical aspects of \ac{CPDP}; or 
  \item publications on tooling that specifically addresses \ac{CPDP}; or
  \item fully automated defect prediction approaches that do not require any
  labelled data from the target product.
\end{itemize}

Additionally, we used the following exclusion criteria:
\begin{itemize}
  \item publications that only address \ac{WPDP} or only mention \ac{CPDP}
  as aspect for future consideration; and
  \item publications that were not peer-reviewed; and
  \item publications that are not fully published in English. 
\end{itemize}

\subsection{Identification of Papers}

We used Google Scholar for the identification of papers. The advantage of using
Google Scholar in comparison to directly accessing the databases provided by,
e.g., IEEE Xplore or ACM is that these databases are indexed by Google Scholar
anyway.
Moreover, Google Scholar searches can include results that are not contained by
any of the standard indexes in case of publications at smaller conferences and
workshops, especially if they are restricted to a certain region. Table
\ref{tbl:searchterms} summarizes the terms we used for our search and the number
of hits. 

\begin{table}
\centering
\begin{tabular}{|r|c|}
\hline
Search Terms & No. Results \\
\hline
cross project defect prediction & ca. 122,000 \\
cross project fault prediction & ca. 166,000 \\
cross project bug prediction & ca. 37,100 \\
cross company defect prediction & ca. 85,300 \\
cross company fault prediction & ca. 83,000 \\
cross company bug prediction & ca. 27,600 \\
\hline
\end{tabular}
\caption{Terms used for searching Google Scholar}
\label{tbl:searchterms}
\end{table}

Due to the power of the search engine, we could not consider all results.
Instead, we focused on the first 1,000 hits of the search for each query. Hence,
we considered a total of 6,000 hits achieved. We have to note that huge parts of
the query results were overlapping, due to the semantic recognition by the
search engine of the similar terminology. The titles of all hits were
considered. In case the title somehow indicated a relation to defect prediction,
the abstract and the actual publication were scanned for cross-project aspects.
For this, we mainly considered the case study setup, but also the general
structure of the approach. After this filtering, we selected the 49 publications
that were included in our review.

Moreover, we note that we found no results that were included after
about hits 850 for any query, which gives us confidence that we did not miss
any papers. Additionally, we checked the related work cited in each of the
publications we included to provide a cross-check if we missed something.
All identified papers were already covered by our Google Scholar search results.

\subsection{Data Collection}

Since we decided that we could not directly assess the quality of approaches
realistically, due to the diversity in the case studies, we do not define
quality check criteria as proposed by \cite{Kitchenham2007}. Instead, we define aspects that are important to
describe the techniques that were used and how they were evaluated. Therefore,
we collected the following data for each publication:  
\begin{itemize}
  \item \textbf{Approach.} The approach towards \ac{CPDP} that was considered in
  the publication. We summarize the approach of all publications in order to
  provide a good overview of the work that was done and, thereby, enable
  future research to utilize this collected body of knowledge. Moreover, to determine reporting guidelines, the work carried
  out so far must be taken into account to ensure that it can be addressed by
  the reporting guidelines. During our summary of the approaches, we did not
  always use the same mathematical formulations or wording as the authors.
  Instead, we harmonized the different representations of approaches into a
  common language and mathematical terminology. 
  \item \textbf{Approach Type.} \cite{Turhan2012} (discussed in
  Section~\ref{sec:turhan2012}) provided a categorization of six ways to treat
  differences between data sets to improve the shifts between different project
  contexts: outlier detection; relevancy filtering; instance weighting;
  stratification; cost curves; and mixture models. As part of our review, we
  collect for each publication which of these type of techniques was proposed.
  In case a method does not fit the taxonomy proposed by Turhan, we use the
  category ``other'' and state in parenthesis what kind of approach it is.
  \item \textbf{Classifier.} The machine learning or statistical classifier that
  was used in the work. We collect this data because many approaches considered
  in the literature do not define new classification models, but instead data
  treatments. Since the classification models can impact the
  results, it is important for consistent reporting and comparability to
  identify commonly used classification models.
  \item \textbf{Data.} The data used in case studies has a major impact on
  the external validity of results and the comparability. If two approaches are
  evaluated on different data sets, it is unclear if similarities or differences
  in the results are due to the approaches or due to the data. 
  \item \textbf{Case Study Setup.} Important factors regarding the comparability
  of studies due to the case study setup are, e.g., how the data in a case study
  is used and to which baselines results are compared.
  \item \textbf{Performance Measures.} Since different performance measures
  consider different aspects, comparisons between studies with different
  performance measures are very difficult. We collect information about the
  performance measures with the goal to identify a popular set of measures that
  covers all important aspects of prediction models which can be used for
  comparable studies. 
  \item \textbf{Results.} We summarize the results achieved as well as
  which numbers were reported as part of a publication. The summary of the
  results themselves provide good hints to researchers as to how techniques may
  perform when compared to each other, even though we were not able to create a
  general overall performance comparison through this review. We report the mean
  performance of the results if possible. Otherwise, we try to report other
  summarative statistics like median performance, if those are available. 
\end{itemize}

\section{Foundations}
\label{sec:foundations}

In this chapter, we introduce the foundations required for our mapping study.
We start with a brief discussion of the terms \ac{WPDP} and \ac{CPDP}.
Then, we formalize the problem of \ac{CPDP} in order to be able to better
discuss proposed approaches. Afterwards, we list the performance measures used
in the literature. Finally, we give an overview about the publicly available defect
prediction data sets, that were used in publications on \ac{CPDP}. 

\subsection{Defect Prediction Types}

Within this section, we want to clarify the terms \ac{WPDP} and \ac{CPDP}.
\ac{WPDP} is based only on data from the same project. Two kinds of \ac{WPDP}
are considered in the literature.
\begin{itemize}
  \item Cross-validation studies, in which data from one revision of a product is
  split into partitions and one partition is used as test data and the other
  partitions are used as training data.
  \item Studies in which previous revisions of the same software product are
  used to predict the defects in a newer revision, e.g., the usage of version
  1.0 to predict defects in version 2.0. 
\end{itemize}
In the \ac{WPDP} setting, no data from other products is used to train the
classifer.
For \ac{CPDP} we differentiate between \ac{MPDP}, mixed \ac{CPDP} and strict
\ac{CPDP}. For \ac{MPDP}, data from the target product is used together with
data from other products. For mixed \ac{CPDP}, no data from the target product
itself is allowed, but data from older revisions. This data is used together
with data from other products. For strict \ac{CPDP}, only data from other
products is allowed, old revisions from the target product may not be part of
the training data.


\subsection{Formalization of \ac{CPDP}}
\label{sec:cpdp-formalization}
Let $S$ be the set of software entities $s$ of which a software
product\footnote{We use the term software product within this paper to denote a
specific release of a software project. Hence, each product is
unique, but a project may have developed multiple products.}
is comprised.
Let $M := \{m_1, \ldots, m_p\}$ be a set of software metrics \citep{Fenton1997}.
For a software product $S$, the metrics assign a numerical value to all of its
entities, i.e., $m_i: S \to \mathbb{R}, i=1, \ldots, p$. Hence, with $M$, each
entity $s$ can be transformed into a $p$-dimensional vector $M(s) := (m_1(s),
\ldots, m_p(s))$ and we denote $M(S) := \{(m_1(s), \ldots, m_p(s)): s \in S)\}
\subset \mathbb{R}^p$.

We say an entity is defect-prone if it contains at least one defect. More
formally, we have a concept $c_S: S \to \{0,1\}$, such that
\begin{equation}
c_S(s) = \begin{cases}1 & \mbox{if}~s~\mbox{is defect-prone} \\ 0 &
\mbox{otherwise}\end{cases}
\end{equation}
for all entities $s \in S$. In \ac{CPDP}, we have a set of software products
$S^{cand} = \{S^1, \ldots, S^n\}$ for which we know both the metric data as well as the
concept, i.e., we have $M(S),c_S$ fully specified for each $S\in
S^{cand}$.
We want to predict the defects of a software product $S^*, S^* \not\in
S^{cand}$, for which we know the metric data $M(S^*)$, but not the
concept $c^* := c_{S^{*}}$. We call $S^*$ the target
product and $c^*$ the target concept. In order to estimate the target concept
$c^*$, a hypothesis $h: \mathbb{R}^p \to \{0,1\}$ is calculated.

\begin{defi}[\ac{CPDP} Learner]
A \ac{CPDP} learner takes as input the metric data
$M(S)$ and concepts $c_S$ for all $S \in S^{cand}$ and the metric data
$M(S^*)$ of the target product $S^*$. The output of the training procedure is a
concept $h$ that estimates the target concept $c^*$.
\end{defi}

Please note that while $S^{cand}$ is the input for the \ac{CPDP} learner,
it should not be confused with the training data, which we denote with
$S^{train}$. The difference is that one popular approach for \ac{CPDP} is
relevancy filtering (see Section~\ref{sec:turhan2011}), which selects a subset
of the candidate data as training data. Hence, $S^{train} \subseteq
\bigcup_{i=1,\ldots,n} S^{cand}$. 

In addition to the above definitions, we use the following notations in the
remainder of this paper.
\begin{itemize}
  \item For convenience and better readability, we often simply write 
  $S^{cand}$ instead of $\bigcup_{i=1,\ldots,n} S^{cand}$ when we refer to
  relevancy filters.
  \item Similarly, we write $\bigcup S^{cand}$ instead of
  $\bigcup_{i=1,\ldots,n} S^{cand}$ in the following. 
  \item $char(m(S))$ to denote the distributional characteristic $c$ of the
  metric $m$ for the software entities $S$, e.g., $mean(m(S))$ and $max(m(S))$.
\end{itemize}

\subsection{Performance Measures}
\label{sec:performance-measures}

Most of the performance measures are based on the confusion matrix, i.e., the
number of \emph{\ac{tp}} predictions, \emph{\ac{fp}}
predictions, \emph{\ac{tn}} predictions, and \emph{\ac{fn}} predictions. The
used measures based on these values are the following. 
\begin{itemize}
  \item $recall = pd = tpr = completeness = \frac{tp}{tp+fn}$
  \item $precision = correctness = \frac{tp}{tp+fp}$
  \item $pf = fpr = \frac{fp}{tn+fp}$
  \item $F\text{\emph{-}}measure = 2 \cdot \frac{recall \cdot
precision}{recall+precision}$
  \item $G\text{\emph{-}}measure = 2 \cdot \frac{recall \cdot
  (1-pf)}{recall+(1-pf)}$
  \item $balance = 1-\frac{\sqrt{(1-recall)^2+pf^2}}{\sqrt{2}}$
  \item $accuracy = \frac{tp+tn}{tp+fp+tn+fn}$
  \item $error = \frac{fp+fn}{tp+fp+tn+fn}$
  \item $error_{Type I} = \frac{fp}{tp+fn}$  
  \item $error_{Type II} = \frac{fn}{tn+fp}$
  \item $MCC = \frac{tp \cdot tn - fp \cdot
  fn}{\sqrt{(tp+fp)(tp+fn)(tn+fp)(tn+fn))}}$
\end{itemize}
Moreover, two measures for successful predictions experiments based on the above
are used.
\begin{itemize}
  \item $succ_{0.75}$ defined as the percentage of predictions that achieve
  $recall>0.75$, $precision>0.75$, and $accuracy>0.75$.
  \item $succ_{0.7, 0.5}$ defined as the precentage of predictions that achieve
  $recall>0.7$ and $precision>0.5$.
\end{itemize}

Moreover, two variantes of the \emph{\acf{AUC}} are used. \emph{\ac{AUC}} is
distributed between zero and one. The variants of \emph{AUC} are defined using
the \ac{ROC}.
\begin{itemize}
  \item \emph{AUC} uses the \emph{pf} versus \emph{recall}
  as ROC.
  \item $AUC_{Alberg}$ uses the \emph{recall} and the percentage of   modules
  considered to define the ROC \citep{Ohlsson1996}.
\end{itemize}

As an alternative to \emph{AUC}, the \emph{H-measure} was
proposed by \cite{Hand2009}. The key difference between \emph{AUC} and
\emph{H-measure} is that \emph{AUC} implicitly uses classifier-dependent
misclassification costs, whereas the \emph{H-measure} uses a prior distribution
for costs.

Another measure that is used is the goodness of fit using the \ac{HL} test
\citep{Hosmer1980} and Huberts statistical procedure \citep{Sharma1996}. This
way, two classification models are compared directly with each other and it is
evaluated if the result is significantly different and how large the effect
size, i.e., the difference between the results is. 

Furthermore, measures related to the cost required for reviewing effort are
used.
\begin{itemize}
  \item $NECM = \frac{C_1\cdot fp + C_2\cdot fn}{tp+fp+tn+fn}$ is the normalized
  expected cost of misclassification with $C_1$ the cost of a false positive
  (Type I error) and $C_2$ the cost of a false negative (Type II error).
  \item $NECM_{C_{ratio}} = \frac{fp + C_{ratio}\cdot fn}{tp+fp+tn+fn}$ is
  another way to express the above, with $C_{ratio} = \frac{C_2}{C_1}$. In case
  this variant is used with a concrete ratio, e.g., $C_{ratio}=10$ we
  simply write $NECM_{10}$.
  \item $cost = \sum_{s^* \in S^*} h(s^*)\cdot LOC(s^*)$ with $LOC(s^*)$
  being the \ac{LOC} of the inspected entities.
  \item $NofB_{20\%}$, i.e., the number of bugs found when inspecting 20\% of
  the code.
  \item $NofC_{80\%}$, i.e., number of classes visited until 80\% of the bugs
  are found.   
\end{itemize}

Finally, one performance measure related to the consistency of results between
experiments is used. Here, the aim is to evaluate how big the difference between
classification results are. Using this intution, the following performance
metric is defined~\citep{He2015}. 
\begin{equation}
consistency = \frac{tp\cdot(tp+fp+tn+fn)-(tp+fn)^2}{(tp+fn)\cdot(tn+fp)}
\end{equation}

\subsection{Data Sets}
Nine public data sets were used in the investigated publications on \ac{CPDP}.
We only give a general overview of the data, including the number of products, as well as the number
and general types of metrics after the taxonomy by \cite{Fenton1997}. Complete lists of metrics, including descriptions
are provided in the publication or at the download location of the respective
data sets. 

\subsubsection{NASA}
\label{sec:nasa-data}
This dataset was donated by NASA through the NASA \ac{MDP}. The data contains
information about eleven products implemented in C++ and one product implemented
in Java. The metrics measure static product metrics. 21 metrics are available
for all twelve products, and depending on the product there are up to 39 metrics
total. Additionally, the data contains for each module the information whether
it was defective or not. The data is publicly available
online\footnote{http://openscience.us/repo/defect/mccabehalsted/}.

\subsubsection{SOFTLAB}
\label{sec:softlab-data}
This dataset was donated by SOFTLAB, a Turkish software company. The data
contains information about three software products in C. 30 static
product metrics are available for each module of the products together with the
information if the module was defective or not. The data is publicly
available online\footnote{http://openscience.us/repo/defect/mccabehalsted/}.

\subsubsection{JURECZKO}
\label{sec:jureczko-data}
This dataset was donated by \cite{Jureczko2010}. It
consists of data about 48 product releases of 15 open source projects, 27
product releases of six proprietary projects and 17 academic products
implemented by students, i.e., 92 released products in total. As metrics, they
collected 20 static product metrics for Java classes, as well as the number of
defects found in each class. The data is publicly available
online\footnote{http://openscience.us/repo/defect/ck/}.

\subsubsection{RELINK}
\label{sec:relink-data}
This dataset was published by cite{Wu2011}. The data contains defect
information about three products. The defect labels were manually verified and
not just automatically generated from \ac{SCM} commit comments. 20 static
product metrics are available for each module together with the information
if an entity was defective or not. The data is publicly available
online\footnote{\url{http://www.cse.ust.hk/~scc/ReLink.htm}}.

\subsubsection{AEEEM}
\label{sec:aeeem-data}
This dataset was published by \cite{DAmbros2010}.
The data set contains information about five Java products. 61 software metrics
are available, including static product metrics, process
metrics like the number of previous defects, the entropy of code changes, and
source code churn. The data is publicly available online\footnote{\url{http://bug.inf.usi.ch/}}. An extension of this data with
project factors was performed by \cite{Zhang2014, Zhang2015}, which
is also publicly available
online\footnote{\url{http://fengzhang.bitbucket.org/replications/unimodel.html}}.

\subsubsection{MOCKUS}
This data set is based on data collected by \cite{Mockus2009}. The
data contains information about roughly 235,000 projects hosted on SourceForge
and GoogleCode.
From this huge body of projects, \cite{Zhang2014, Zhang2015} 
extracted 1,385 projects by filtering based on the programming language,
projects with a small number of commits or short lifespan, and limited defect data or
fix-inducing commits. The metrics obtained in the data set are 21 product
metrics and 5 process metrics. The data is publicly available
online\footnote{\url{http://fengzhang.bitbucket.org/replications/unimodel.html}}.

\subsubsection{ECLIPSE}
This data set was published by \cite{Zimmermann2007}. The data
contains defect information about three Eclipse releases, collected on file and
package level. 31 static product metrics are available on
the file level, 40 metrics on the product level. The data contains information
about the number of pre- and post-release defects for all measured entities.
Additionally, metrics about the size of the \ac{AST} and the frequency of node
types in the \ac{AST} are available. The data is publicly available
online\footnote{\url{https://www.st.cs.uni-saarland.de/softevo/bug-data/eclipse/}}.

\subsubsection{NETGENE}
This data set was published by \cite{Herzig2013}. The data contains
defect information about four open source projects that follow strict and
industry like development processes. The data contains a total of 465 metrics,
including static product metrics, network metrics, as well as genealogy
metrics, i.e., metrics related to the history of a file, e.g., the number of
authors or the average time between changes. The data is publicly available
online\footnote{\url{https://hg.st.cs.uni-saarland.de/projects/cg_data_sets/repository}}.

\subsubsection{AUDI}
This data set was published by \cite{Altinger2015}. The data
contains defect information about three automotive projects developed by Audi
Electronics Venture GmbH. The data contains a total of 17 static source code
metrics. A special characteristic of this
data set is, that the source code for which the metrics were measured was not
written directly, but instead generated automatically from Matlab/Simulink
models. Moreover, the generated source code follows the very strict MISRA coding
guidelines~\citep{MISRA}. The data is publicly available
online\footnote{\url{http://www.ist.tugraz.at/_attach/Publish/AltingerHarald/MSR_2015_dataset_automotive.zip}}.

\section{Literature Review}
\label{sec:related-work}

In this section, we provide the review of the the state of the art of \ac{CPDP},
ordered by the time of the publications. The publications and the information in
this overview were collected according to the methodology discussed in
Section~\ref{sec:methodology}. Some of the publications were first published at
a conference and then invited to a journal to submit an extended version of
their findings. We summarize both these publications together, due to the fact
that the general approach, data, and case study setup are similar in both and
that the conference publication is a subset of the journal extension.

\subsection{Briand\etal, 2002}
\label{sec:briand2002}
\noindent\textbf{Approach.} \cite{Briand2002} if a model trained for
one project (Xpose), is suitable for the prediction of another project (Jwriter). To
this aim, they evaluate the ranking of defect prone-entities produced by the
MARS model for Xpose for the Jwriter project. 

\noindent\textbf{Approach type.} None.

\noindent\textbf{Classifier.} Linear regression, \ac{MARS}.

\noindent\textbf{Data.} The authors consider 2 java products for which 15 static
product metrics are measured. The data is not publicly available. 

\noindent\textbf{Case study setup.} The authors trained the MARS based on the
Xpose project and evaluated it based on the Jwriter project. The paper also
performs other experiments, however, they are not in the cross-project context. 

\noindent\textbf{Performance measures.} $benefit$, $precision$, and $recall$.

\noindent\textbf{Results.} The authors do not report mean results for the
performance measures, but report curves based on different thresholds that are
used for the classification with the linear regression and \ac{MARS}. The
results show that both the linear regression and \ac{MARS} are better than
random guessing during the review and the the $benefit$ 17.6 times the average
costs of not finding and fixing a defect during the inspection of the source
code. 

\subsection{Nagappan\etal, 2006}
\label{sec:nagappan2006}
\noindent\textbf{Approach.} \cite{Nagappan2006} investigate which
metrics are suitable to predict post-release defects. As part of this study, they also investigate
the similarity between prediction models trained for different products.
Concretely, they calculated whether the trained predicton models are
correlated.
They assume that if two predictors are correlated, they work for both products. 

\noindent\textbf{Approach type.} Other (metric type influence).

\noindent\textbf{Classifier.} Logistic Regression.

\noindent\textbf{Data.} The authors consider 5 large-scale products from
Microsoft for which 18 static product metrics are measured, mainly
concerning complexity and size.
The data is not publicly available.

\noindent\textbf{Case study setup.} The authors first performed
\ac{PCA}~\citep{Pearson1901} to reduce the metrics to their principal
components.
Then, a logistic regression model was trained for each product. Using Spearman
and Pearson correlation the authors tested if the coefficients of the models
are correlated. 20 pair-wise comparisons between products were performed with
two correlation measures, i.e., 40 model correlations were analyzed. 

\noindent\textbf{Performance measures.} Not applicable. 

\noindent\textbf{Results.} For two products, both Spearman and Pearson
correlation determined that the models are correlated. For one product only
Spearman correlation was significant, for another only Pearson correlation. In
total, only 6/40 correlations were significant, i.e., 15\%. The authors conclude
that \ac{CPDP} is possible if the data comes from similar products. 

\subsection{Khoshgoftaar\etal, 2008}
\label{sec:koshgoftaar2008}
\noindent\textbf{Approach.} \cite{Khoshgoftaar2008}
propose to use a combination of multiple classifers as well as data from
multiple products for \ac{CPDP}. They consider four different scenarios: (1) one
classifier trained for one product; (2) multiple classifiers trained on a single product;
(3) one classifier trained for multiple products; and (4) multiple classifiers
trained on multiple products. In case multiple classifiers and/or products are
used, majority voting is used to determine the classification. 

\noindent\textbf{Approach type.} Other (classification model).

\noindent\textbf{Classifier.} 1-\ac{NN}, Alternating Decision Tree, Bagging, C4.5 Decision
Tree, Decision Table, Logistic Regression, $k$-NN,
Lines-of-Code, Locally Weighted Learning with Decision Stumps,
Logistic Regression, Na\"{i}ve Bayes, One Rule, Partial Decision Tree,
Random Forest, Repeated Incremented Reduced Error Pruning, Ripple Down Rules, 
\ac{SVM}, and Tree-Disc Classification Tree.

\noindent\textbf{Data.} Seven products from the NASA data. 

\noindent\textbf{Case study setup.} 
The authors use five experiment configurations: (1)--(4) as described above;
and (5) \ac{WPDP} with 10x10 cross validation. For (1) and (3) each of the
classifier is tested and (2) and (4) with all classifiers are used at once. For
(3) and (4) one product is used as target product and the other six for training. 
The authors utilize \ac{ANOVA}~\citep{Fisher1918} to determine the statistical
significance of their results.

\noindent\textbf{Performance measures.} $error_{Type I}$, $error_{Type II}$,
$NECM_{15}$, $NECM_{20}$, and $NECM_{25}$. 

\noindent\textbf{Results.} The authors report the mean performance of all
classifiers together on each product, as well as the overall mean performance
across all products.
The performances of the individual classifiers are only reported for \ac{WPDP}.
The single classifiers trained on a single product (1) achieved a mean
performance of 0.685 $NECM_{15}$, 1.221 $NECM_{20}$, 1.459 $NECM_{25}$, 0.305
$error_{Type I}$, and 0.371 $error_{Type II}$. The multiple classifiers trained
on a single product (2) achieved a mean performance of 0.906 $NECM_{15}$, 1.132
$NECM_{20}$, 1.357 $NECM_{25}$, 0.260 $error_{Type I}$, and 0.353 $error_{Type
II}$.
The single classifiers trained on multiple products (3) achieved a mean
performance of 0.821 $NECM_{15}$, 1.002 $NECM_{20}$, 1.183 $NECM_{25}$, 0.317
$error_{Type I}$, and 0.283 $error_{Type II}$. The multiple classifiers trained
on multiple products (4) achieved a mean performance of 0.836 $NECM_{15}$, 1.038
$NECM_{20}$, 1.239 $NECM_{25}$, 0.262 $error_{Type I}$, and 0.318 $error_{Type
II}$.
The authors conclude that using the majority vote of a single classifier on
multiple products (3) yields the best overall results. Using multiple
classifiers does not make a big difference. Using \ac{ANOVA}, the
authors determine that their findings are statistically significant. 

\subsection{Watanabe\etal, 2008} 
\label{sec:watanabe2008}
\noindent\textbf{Approach.} \cite{Watanabe2008} propose to compensate differences
between products through a standardization technique that rescales the data. In
a scenario with only one candidate product as training data, they propose to use
this product as reference for the standardization of the target data. This shall
increase the homogeneity between the target product and the candidate product.
As formula for standardization, the authors propose to multiply each metric
value of the target product with the mean value of the candidate product and
divide this by the mean of the target product itself, i.e., 
\begin{equation}
\hat{m}_i(s^*) = \frac{m_i(s^*) \cdot mean(m_i(S))}{mean(m_i(S^*))}
\end{equation}
for all $s^* \in S^*$.

\noindent\textbf{Approach type.} Other (instance transformation).

\noindent\textbf{Classifier.} C4.5 Decision Tree.

\noindent\textbf{Data.} For both projects, data about
three versions was mined. The mined data contained seven metrics and bug labels
based on the comments of the \ac{SCM} logs. The data is not publicly available.

\noindent\textbf{Case study setup.} The authors use two experiment
configurations: (1) \ac{CPDP} with their standardization; and (2) \ac{CPDP}
without their standardization. All versions of each project were used
once as training data to predict the defects of all versions of the other
projects. 

\noindent\textbf{Performance measures.} \emph{recall} and \emph{precision}.

\noindent\textbf{Results.} With the standardization (1), the
authors report a mean performance 0.65 \emph{recall} and 0.73 \emph{precision}.
Without standardization (2), the mean performance is 0.50 \emph{recall} and 0.75
\emph{precision}. Hence, they observed a mean gain of 0.15 in recall and 0.02 in
precision in their experiments. 

\subsection{Turhan\etal, 2009}
\label{sec:turhan2009}
\noindent\textbf{Approach.} 
\cite{Turhan2009} propose to use training data which is similar
to the target product. To this aim, they devised a method based on the $k$-NN
algorithm for relevancy filtering of the training data from the union of
all available candidate products $S^{cand}$. Turhan\etal~measure the
similarity between entities $s, s'$ with the Euclidean distance
between the metric vectors of the entities, i.e., $dist(M(s), M(s'))$. Based on
the distance, they select for each entity $s \in S^*$ the $k$ closest entities
from the candidate data. Before the relevancy filter is applied, the
logarithm is applied to all metric data taking pattern from
\cite{Menzies2007}.

\noindent\textbf{Approach type.} Relevancy Filtering.

\noindent\textbf{Classifier.} Na\"{i}ve Bayes.

\noindent\textbf{Data.} Seven products from the NASA data and all three
products from the SOFTLAB data.

\noindent\textbf{Case study setup.} The authors used three experiment
configurations: (1) \ac{WPDP} with 90\% of the target data for training and
10\% as test data repeated 20 times; (2) \ac{CPDP} with all data not from the
target product as training data; and (3) \ac{CPDP} with the $k$-NN relevancy
filter on 10\% of the target data as test data repeated 100 times. The NASA and
the SOFTLAB data were not considered as a single data set but used in different
experiments, i.e., no data from NASA was used to predict defects in the SOFTLAB
data and vice versa. Hence, all experiments we report on here are performed once
on the NASA data and once on the SOFTLAB data.

\noindent\textbf{Performance measures.} \emph{recall} and \emph{pf}\footnote{They
also use \emph{balance} in the paper. However, we omit it from our comparison
since no performance results related to the \ac{CPDP} experiments
were reported for the \emph{balance}.}.

\noindent\textbf{Results.} On the NASA data, the \ac{WPDP}
(1) achieves a median performance of 0.75 \emph{recall} and
0.29 \emph{pf}. The \ac{CPDP} with all data (2) achieves a median 0.97
\emph{recall} and with 0.64 \emph{pf}. No summary statistics are reported for the complete data
set for the \ac{CPDP} with $k$-NN relevancy filter (3). Here, only results on
a product level are reported. However, it can be seen in the results that the
\emph{recall} for \ac{CPDP} with $k$-NN relevancy filter is consistently
lower than that of \ac{CPDP} with all other data, whereas for \emph{pf} the
opposite is the case. For the SOFTLAB data, the \ac{WPDP} (1) achieves
a median performance of 0.88 \emph{recall} and 0.29 \emph{pf}. The
\ac{CPDP} with all data (2) achieves a median performance of 0.95 \emph{recall}
and 0.65 \emph{pf}. For the \ac{CPDP} with $k$-NN relevancy filter (3),
no overall results for the complete SOFTLAB data are reported. However, based on
the individual results for each product, the same conclusions as for the
NASA data can be drawn.

\subsection{Zimmermann\etal, 2009}
\label{sec:zimmermann2009}
\noindent\textbf{Approach.} 
\cite{Zimmermann2009} consider pair-wise \ac{CPDP}
without any data processing, i.e., one product is used to predict the defects of
another product. Based on the performance of pair-wise predictions,
Zimmermann\etal~propose to train a decision tree for the relevancy filter to
determine the product best suitable for training. This decision tree is
based on project factors (e.g., if a database is used) as well as the metric data. Project
factors considered are, e.g., the organization that developed the code, whether
internationalization was applied, or if the project uses a database. Factors
from the training data itself are, e.g., the number of observations and the code
churn.

\noindent\textbf{Approach type.} Relevancy filtering.

\noindent\textbf{Classifier.} Logistic Regression.

\noindent\textbf{Data.} The authors consider twelve large-scale
and well-known software products, e.g., Mozilla Firefox, Internet Explorer, the
Windows kernel, and Apache Tomcat. For some of the projects, multiple versions
are available, which means a total of 28 data sets are part of the case study.
Six software metrics were measured for each data set. All of these metrics are
designed in a way that they are relative with respect to the project size or
number of commits, e.g., $\frac{\text{added \ac{LOC}}}{\text{total
\ac{LOC}}}$.
The metrics cover code churn, pre-release defects, and code complexity. The data is
not publicly available.

\noindent\textbf{Case study setup.} The authors used two experiment
configurations: (1) pair-wise \ac{CPDP} with all version except old versions of
the same product (strict \ac{CPDP}) and (2) a decision tree trained as described
above to select training data. For both, the authors remove all trivial
prediction models, i.e., models that only predict defect or no defect for
instances.  

\noindent\textbf{Performance measures.} \emph{recall}, \emph{precision},
\emph{accuracy}, and $succ_{0.75}$. 

\noindent\textbf{Results.} Due to the removal of trivial models and old
versions for predictions, the authors only consider 622 combinations of
the pair-wise predictions, and not $28\cdot27=756$.
The success rate of the pair-wise \ac{CPDP} (1)
was 0.034, i.e., only 21 of the 622 predictions achieved were successful
according to $succ_{0.75}$. The mean \emph{precision} of the predictions is
\emph{0.374}, \emph{recall} and \emph{accuracy} are not reported. With their
decision tree procedure (2), they can raise the success rate for small areas of
products. For example, they achieve a success rate of 0.324 in case the test
product has more pre-release bugs than the training product, the test and
training products are either implemented for different operating systems or both
for Windows, and the standard deviation of the relative complexity is higher in
the target product.
Moreover, they analyzed the effect of project factors on the metrics in general.
They observed that on the one hand, differences between project factors are
often good for the \emph{recall}, but bad for the \emph{accuracy} of the models.
On the other hand, having the same project factors is often good for the \emph{precision}.
Finally, they also analyzed the effect of difference in distributional
characteristics of the numerical characterics used for the predictions on the
results. Here, they observed that higher median values in the test data increase
the \emph{precision} and \emph{recall}, but decrease the \emph{accuracy}.

\subsection{Camargo Cruz and Ochimizu, 2009}
\label{sec:camargocruz2009}
\noindent\textbf{Approach.} \cite{CamargoCruz2009} propose to apply a power
transformation~\cite{Hoaglin1983} to the metric data and then standardize it.
The power transformation is based on the logarithm and the observation that software metrics, especially the size and
complexity, often follow exponential distributions~\citep{Kan2003}, which is
the same as what \cite{Turhan2009} do for the treatment of the data.
The standardization of the data is based on the median. Concretely, the
difference between the median value of a reference product $S^{ref} \in
S^{cand} \cup S^*$ and other products $S$. Together with the power
transformation, the standardization formula is
\begin{equation}
\begin{split}
\hat{m}_i(s) &= \log (1+m_i(s))\\
&+ median(\log(1+m_i(S))) \\
&-median(\log(1+m_i(S^{ref})))
\end{split}
\end{equation}
for all $s \in S, S^{ref}$. The authors consider only pair-wise
\ac{CPDP} and propose to use the training product as reference.  

\noindent\textbf{Approach type.} Other (instance transformation).

\noindent\textbf{Classifier.} Logistic Regression.

\noindent\textbf{Data.} The authors use defect data about
classes from seven different Java products. For all products, three
static product metrics are available. The defect data was obtained from both the
\ac{SCM} logs as well as the \ac{ITS}. Moreover, the authors did not use the defect label
directly, but rather consider the most/least defective classes. For this, they
used the median of the defects contained in the entire data set of a product as
reference point. If a class contains less defects than the median, it belongs to
the least defective classes, if it is greater or equal to the median, it belongs
to the most defective classes. The data is not publicly available.

\noindent\textbf{Case study setup.} The authors used two experiment
configurations: (1) \ac{CPDP} with their standardization; and (2) \ac{CPDP}
without their standardization. As training data, the authors only use one
product of the data. As target products, two others are used. The other four
products are not used for the evaluation of the defect prediction. 

\noindent\textbf{Performance measures.} \ac{HL} test and Huberts statistical
procedure. 

\noindent\textbf{Results.} Through the \ac{HL} test the authors conclude
that the \ac{CPDP} model with standardization (1) has a better goodness of fit
than the \ac{CPDP} model without standardization (2). Through the Hubert
statistic, they observe a small advantage due to the standardization for one of
the target products and a larger advantage on the other product. 

\subsection{Jureczko and Madeyski, 2010}
\label{sec:jureczko2010}
\noindent\textbf{Approach.} \cite{Jureczko2010}
proposed relevancy filtering through the clustering of products based on
self-organizing maps\footnote{Self-organizing maps are also frequently referred to as Kohonen's neural network, Kohonen network, or
Kohonen map.}~\citep{Kohonen1982}, as well as with $k$-means clustering
with a value of $k=2$, i.e., a separation of the data into two groups. For the
training of the defect prediction model, the authors propose to use only data
from the same cluster.

\noindent\textbf{Approach type.} Relevancy filtering.

\noindent\textbf{Classifier.} Linear Regression.

\noindent\textbf{Data.} All 92 products from the JURECZKO data.

\noindent\textbf{Case study setup.} The authors used three experiment
configurations: (1) \ac{CPDP} with data from all other products; (2)
\ac{CPDP} with $k$-means clustering as relevancy filter; and (3)
\ac{CPDP} with self-organizing maps as relevancy filter. To test whether
the cluster model is better than using all data, the authors use the
Shapiro-Wilk test~\citep{Shapiro1965}.

\noindent\textbf{Performance measures.} $NofC_{80\%}$. 

\noindent\textbf{Results.} The authors report only the mean
performance per cluster, the overall mean values are not reported. With the
separation of the data into two clusters with the $k$-means clustering (2), the
authors did not note any increase in prediction performance over using all data
(1). Regardless of the aproach, about 47\%-49\% of the classes needed to be
visited with the clusters and with using all data. With the self-organizing map
(3), four clusters were created. However, the authors note that ``there are
releases that were classified into none of those
clusters''~\citep{Jureczko2010}.
No information on how many of the products in the data could not be
clustered is given.
However, from the other data we believe that 58 products could be clustered and
the remaining 34 products could not be clustered. For two of the four clusters
that contain 26 of the products the performance of using the cluster is worse
than using all data. For the other two clusters, containing 32 products, the
performance is increased by 11\% and 5\% respectively.

\subsection{Liu\etal, 2010}
\label{sec:liu2010}
\noindent\textbf{Approach.} 
\cite{Liu2010} propose a \ac{CPDP} classifier created through a genetic
program. The genetic program generates an S-expression-tree
\citep{Rivest1997} to create an equation that models the defects using metric
data. The authors propose three different ways to use the genetic program. The
first is to use all candidate products $S^{train}=S^{cand}$ to train the
genetic program. The second is to use one of the candidate products $S^{val} :=
S^i \in S^{cand}$ for model selection and $S^{train}=S^{cand} \setminus
S^{val}$ for training. This means, that multiple genetic programs are trained
with $S^{train}$ and their performance is evalutated on $S^{val}$. This is
repeated for all possible candidate products once as $S^{val}$. Then, the best of these
models is used, i.e., the model that achieved the best
performance on $S^{val}$. The third approach is called validation-and-voting. 
The general strategy is the same as above, except that not only the best model
on $S^{val}$ is used, but instead all models of these models are used, i.e.,
multiple genetic programs are used, one for each $S^{val}$. 
The majority vote of these genetic programs defines the classification. 
However, all genetic programs that do not achieve a minimal performance
criterion\footnote{The authors propose to use type I and type II error below
0.5 as minimal criterion.} are dropped and not allowed to vote. 

\noindent\textbf{Approach type.} Other (classification model).

\noindent\textbf{Classifier.} Genetic program (see above) as well as 
1-NN, Alternating Decision Tree, Bagging, C4.5 Decision
Tree, Decision Table Logistic Regression, $k$-NN
Learning, Lines-of-Code, Locally Weighted Learning with Decision Stumps,
Logistic Regression, Na\"{i}ve Bayes, One Rule, Partial Decision Tree,
Random Forest, Repeated Incremented Reduced Error Pruning, Ripple Down Rules, 
\ac{SVM}, and Tree-Disc Classification Tree.

\noindent\textbf{Data.} Seven products from the NASA data.

\noindent\textbf{Case study setup.} The authors used four experiment
configurations: (1) \ac{CPDP} with the genetic program with all products for
training; (2) \ac{CPDP} with the genetic program and one product for model
selection; (3) \ac{CPDP} with genetic program and validation-and-voting; and
(4) \ac{CPDP} with each of the classifiers listed above other than the
genetic program.  To evaluate the statistical significance of the differences in
performance, the authors use a standard t-test~\citep{Student1908}.

\noindent\textbf{Performance measures.} $error$, $error_{Type I}$,
$error_{Type II}$, $NECM_{15}$, $NECM_{20}$, and $NECM_{25}$.

\noindent\textbf{Results.} The authors report the values achieved on
each product for the $error$, $error_{Type I}$, $error_{Type II}$ and $NECM$ for
configurations (1)--(3). For experiment configuration (4), only the mean values
over all classifiers are reported for all performance metrics as well as the
mean performance on all products in terms of $error$, $error_{Type I}$, and
$error_{Type II}$ each classifier individually. 
The genetic program trained with all products (1) achieves a mean performance of 
0.337 $error$, 0.339 $error_{Type I}$, 0.279 $error_{Type II}$, 0.869
$NECM_{15}$, 1.029 $NECM_{20}$, and 1.247 $NECM_{25}$. The genetic program with
one product for model selection (2) achieves a mean performance of 0.328
$error$, 0.325 $error_{Type I}$, 0.327 $error_{Type II}$, 0.926 $NECM_{15}$,
1.137 $NECM_{20}$, and 1.350 $NECM_{25}$. The genetic
program with validation-and-voting (3) achieves a mean performance of 0.349
$error$, 0.367 $error_{Type I}$, 0.211 $error_{Type II}$, 0.733 $NECM_{15}$,
0.869 $NECM_{20}$, and 1.006 $NECM_{25}$. Among the genetic programs (1)--(3),
the validation-and-voting (3) performs consistently best and outperforms or is
equal to the other genetic algorithms on all data sets for all cost models. This
result is confirmed by the t-test. The best of the other
classifiers (4) is the Tree-Disc Classification Tree with a mean performance of
0.210 $error$, 0.212 $error_{Type I}$, 0.205 $error_{Type II}$. The $NECM$
values are not available. 

\subsection{Menzies\etal, 2011, 2013}
\label{sec:menzies2011}
\noindent\textbf{Approach.} An approach not only for \ac{CPDP}, but
rather for defect prediction in general and also effort prediction was first
proposed by \cite{Menzies2011} and later extended \cite{Menzies2013}. In their work, Menzies\etal~propose to first determine
local regions of the data of high homogeneity before building a prediction
model. Concretely, they use the WHERE algorithm to determine the homogeneous
regions within the data. WHERE first applies
the Fastmap algorithm~\citep{Faloutsos1995} to map the input data to two
dimensions. The idea behind this is to have a fast approximation of a
dimensionality reduction as achieved with \ac{PCA}. Afterwards, the
QuadTree clustering algorithm~\citep{Schikuta1993} is used to determine clusters
within the data. After the QuadTree algorithm is terminated, a post-procedure to
join small clusters is executed that determines the final clustering and,
thereby, the homogeneous regions in the data. The clusters determined through
the WHERE algorithm are then analyzed with the WHICH algorithm~\citep{Huang2010}
to create rules for predictions. The rules learned with WHICH describe the
non-defective data and, thereby, foster
insights on attributes of non-defective data. 

\noindent\textbf{Approach type.} Mixture model.

\noindent\textbf{Classifier.} WHICH algorithm, see above. 

\noindent\textbf{Data.} Seven products from the JURECZKO data; the effort
prediction part of the case study is out of scope of this review. 

\noindent\textbf{Case study setup.} The authors consider three experiment
configurations: (1) rules created on all data; (2) \ac{MPDP} data from the
target product is allowed and (3) \ac{CPDP} where no data from the target
product is allowed. For configurations (2)--(3) all seven products in the data
are merged into a single data set. Then, the data is clustered with WHERE. To
learn rules with WHICH for a cluster $C$, the clostest cluster
$C'$\footnote{based on the median score of the dependent variable, for details,
please consult the original article by \cite{Menzies2013} directly.} is used for
training. In experiment configuration (2) all data from the cluster is used for
training, for configuration, i.e., (3) data from the target product is removed
from the training cluster.

\noindent\textbf{Performance measures.} None of the performance measures introduced in
Section \ref{sec:performance-measures} ist used. To measure the performance of
the rules the authors measured the median, the interquartile
range\footnote{called stabiliy in the article} and the worst case, i.e., the
100th percentile of the maximum of the defect values seen in the data set
without applying the learned rule. Since the learned rules describe data that
should not contain defects, this value should be low.

\noindent\textbf{Results.} For the \ac{MPDP} (2), the overall gain of
using the local rules obtained after applying WHERE in comparison to global
rules without clustering (1) in terms of the median of the defect values seen is
about 13\%, the interquartile range is decreased by about 71\% and the worst-case performance is reduced by
66\%\footnote{In the paper, the authors report
that the fraction of the sum of the medians of the cluster data by the sum of
the medians of the global data is 0.64, i.e., a gain of not 13\% but 36\%.
However, this takes the effort prediction rules into account. With only the
defect prediction products, this value changes to 0.87, i.e., a gain of 13\%.
The same reporting differences hold true for the interquartile range and the
worst case, where we also only report the results for the defect prediction
data without the effort prediction results.}. For the \ac{CPDP} (3), they
observe a further gain in terms of the median of 60\% over using all data (1).
However, as the authors also note that this high percentage may be misleading
due to the very small numbers involved, i.e., the absolute gain is only a
reduction from eleven to four. This further decrease leads to an 
increasing interquartile range, which the authors interpret as a loss of
stability of the results. The worst-case scenario is also reduced by an
additional 70\% for \ac{CPDP}.

\subsection{Turhan\etal, 2011}
\label{sec:turhan2011}
\noindent\textbf{Approach.} \cite{Turhan2011} suggest to create a
\ac{MPDP} model by augmenting \ac{WPDP} training data with cross-project data,
before training a classifer. In order to ensure that the cross-project data does not overwhelm the within-project data,
the authors suggest to create a family of classifiers trained with different
amounts of cross-project data. Concretely, they suggest to first select
candidate entities from other products using the $k$-NN relevancy filter
suggested by \cite{Turhan2009} (see
Section~\ref{sec:turhan2009}). Then, $k=10, 20, 30,\ldots$ instances are
randomly drawn from the selected cross-project candidate data to augment the
within-project data. For each of these augmented data sets, a \ac{WPDP}
classifier is trained and evaluated on data from the target product. The one
with the best performance is then selected as classifier. Before training or
relevancy filtering, the authors suggest to perform log-transformations for all
metric data.

\noindent\textbf{Approach type.} Relevancy filtering.

\noindent\textbf{Classifier.} Na\"{i}ve Bayes.

\noindent\textbf{Data.} Seven products from the NASA data and all three
products from the SOFTLAB data.
 
\noindent\textbf{Case study setup.} The authors use two experiment
configurations: (1) \ac{WPDP} with 10x10 cross validation and (2)
\ac{MPDP} with 10x10 cross validation, where the training data is
extended with cross-project data.

\noindent\textbf{Performance measures.} \emph{recall}, \emph{pf}, and
\emph{balance}.

\noindent\textbf{Results.} Using the Mann-Whitney-U
test~\citep{Mann1947}\footnote{This test also known as Wilcoxon test, Wilcox Rank Sum test, and
Mann-Whitney-Wilcoxon test, due to the parellel work by \cite{Mann1947} and \cite{Wilcoxon1945}.}, the authors
determine that \emph{recall}, \emph{pf}, and \emph{balance} are not
statistically significantly different in both configurations for seven of the ten products
used in the study. For one product, they noted an increase through the
cross-project data augmentation in terms of \emph{balance}, but a decrease in
terms of \emph{pf}, whereas \emph{recall} was unchanged. For one product, they
noted an increase in \emph{recall} and \emph{balance}, whereas \emph{pf} was
unchanged. For one product, all three performance measures were improved through
the cross-project data.  The authors conclude that mixing cross-project with
within-project data can improve the results of predictions, but only to a limited degree and not reliably. 

\subsection{Premraj and Herzig, 2011}
\label{sec:premraj2011}
\noindent\textbf{Approach.} \cite{Premraj2011} conducted a
study on the effect of network metrics on defect prediction models. Network
metrics consider the underlying graph structure that connects software entities.
In comparison to other product metrics, the network metrics take the
interactions and information flow between software entities into account.

\noindent\textbf{Approach type.} Other (metric type influence).

\noindent\textbf{Classifier.} $k$-NN, Logistic Regression, Na\"{i}ve Bayes, Recursive
Partitioning, \ac{SVM} with a \ac{RBF} kernel, and Tree Bagging.

\noindent\textbf{Data.} In their study, the authors collect two versions of 
three open source Java projects, i.e., six products in total. The data contains
nine product metrics, 25 network metrics, as well as information about the
post-release defects. The authors published the data online, however, at this
time we could not find a working internet address for the data.

\noindent\textbf{Case study setup.} The authors used three experiment
configurations\footnote{Additional experiments related to the impact of
network metrics on \ac{WPDP} were conducted in the publication, but are not
reported on here.}: (1) \ac{CPDP} with only the product metrics; (2) \ac{CPDP}
with only the network metrics; and (3) \ac{CPDP} with product and network
metrics. Only the newest version of the three products was used for the
\ac{CPDP} context. The predictions were made pair-wise with all possible
combinations of the three remaining products.

\noindent\textbf{Performance measures.} \emph{recall}, \emph{precision}, and
\emph{F-measure}.

\noindent\textbf{Results.} Premraj and Herzig determined that there is
no significant difference between using product metrics (1) and network metrics
(2) for \ac{CPDP}. No numerical values are reported, the evaluation is performed visually based on plots. Only for one combination of
products they saw a difference for the \emph{precision}, for the rest the results were almost the
same. They did observe that using both product and network metrics together (3)
decreases the performance of the prediction and, therefore, argue that the
metrics used for predictions should be carefully selected and that more is not
always better. 

\subsection{Turhan, 2012}
\label{sec:turhan2012}
\cite{Turhan2012} created an overview of the problems due to
\emph{data set shift} for prediction models in software engineering, not only for
\ac{CPDP}, but also for cross-project effort prediction. The article does not
contain \ac{CPDP} techniques or a case study, which is why we break with the
reporting pattern for this work. 

Turhan identified six types of data set shift: (1) \emph{covariate shift}, i.e.,
different distributions in the training and target data; (2) \emph{prior
probability shift}, i.e., the probabilities of the dependent variable differ (in
case of \ac{CPDP}, this means of the probability of defect is different in
training and target data); (3) \emph{sample selection bias}, i.e., problems due
to a different nature in the training data and the target data, e.g., due to
different maturity of the development process; (4) \emph{imbalanced data}, i.e.,
an unequal number of defect-prone and non-defect-prone entities; (5) \emph{domain
shift}, i.e., problems that can arise if the same thing is done in a different
manner, which can lead to inconsistencies; and (6) \emph{source component
shift}, i.e., when data from different contexts is merged, which can lead to too general
data from which specific conclusions cannot be drawn. 

Moreover, Turhan identified representative groups of techniques that can be
applied to treat the above discussed problems. As broad groups, Turhan
identified instance-based and distribution-based techniques. For each of these
broad groups, Turhan identified three subgroups. For the instance-based
techniques, the following subgroups were identified. (1) \emph{Outlier
detection}, i.e., the cleaning of the data through the removal of points that do
not model the normal behavior but special conditions. Turhan notes that outliers
may lead to covariate shifts in the data and, thereby, false generalizations.
(2) \emph{Relevancy filtering}, i.e., filtering the available training data based on
the target data. The idea is to only keep the data for training, that is
actually relevant for the target data. (3) \emph{Instance weighting}, i.e.,  to
use different weights for each instance in the training data, depending on the
relevance of the instance for the result. In case this is based on the target
data, instance weighting is a milder form of a relevancy filter, where instances
are completely removed instead of only receiving a low weight. For the
distribution-based techniques, the following subgroups were identified. (1)
\emph{Stratification}, i.e., avoiding probablitity shifts by sampling through
certain guarantees that the underlying distribution did not change. (2)
\emph{Cost curves}, i.e., decision support through a curve that describes
predictor performances ``over the full range of possible class distributions and
misclassification costs''~\citep{Drummond2006}. They can provide visual support
for decision making and the analysis of predictor models. (3) \emph{Mixture
Models}, i.e., using not a single model, but different models for different
parts of the data, and thereby addressing the problem of source component shift. 

\subsection{Ma\etal, 2012}
\label{sec:ma2012}
\noindent\textbf{Approach.} 
\cite{Ma2012} propose an instance weighting approach based
on the idea of \emph{data gravitation}. The general idea is to
apply a weighing scheme to the training data based on the similarity with the
target data. The authors take the term \emph{weight} literally by adapting
Newton's Universal Gravitation law \citep{Newton1687} to calculate the force
between software entities from the training data and the target data. The
authors calculate the number of similar attributes $simatts_s$ of an entity $s
\in S$ as the number of metrics where the value of $s$ is within the bound of
the minimal and maximal value of the attribute in the target product:
\begin{equation}
simatts_s = \sum_{j=1}^p sim_{sj}
\end{equation}
where
\begin{equation}
sim_{sj} = \begin{cases}1 & \mbox{if}~min_j \leq m_j(s) \leq max_j \\ 0 &
\mbox{otherwise}\end{cases}
\end{equation}
and
\begin{equation}
\begin{split}
min_j &= \min_{s' \in S^*} m_j(s') \\
max_j &= \max_{s' \in S^*} m_j(s').
\end{split}
\end{equation}
They then calculate the weight of each training instance as 
\begin{equation}
w_s = \frac{simatts_s}{(p-simatts_s+1)^2}.
\end{equation}

\noindent\textbf{Approach type.} Instance weighting.

\noindent\textbf{Classifier.} Na\"{i}ve Bayes. 

\noindent\textbf{Data.} Seven products from the NASA data and all three
products from the SOFTLAB data. 

\noindent\textbf{Case study setup.} The authors used three experiment
configurations: (1) \ac{CPDP} with 90\% of the available training
data; (2) \ac{CPDP} with the $k$-NN relevancy
filter proposed by \cite{Turhan2009} (see
Section~\ref{sec:turhan2009}); and (3) \ac{CPDP} with the proposed 
weighting approach with 90\% of the available training data. The products from
the NASA data are used together as a training data. Using this data, the authors perform defect predictions on each
of the product from the SOFTLAB data. Moreover, the authors once use each
product from the NASA data as target product and the remaining products from the
NASA data for training. 

\noindent\textbf{Performance measures.} \emph{recall}, \emph{pf},
\emph{F-measure}, and \emph{AUC}.
 
\noindent\textbf{Results.} The authors report the values for all
performance metrics for the predictions on the SOFTLAB data and the
\emph{F-measure} and \emph{AUC} for the predictions on the NASA data. 
On the SOFTLAB data, the \ac{CPDP} without further treatment (1) achieves a mean
performance of 1.000 \emph{recall}, 0.979 \emph{pf}, 0.306 \emph{F-measure} and
0.510 \emph{AUC}. The $k$-NN relevancy filter (2) achieves a mean
performance of 0.898 \emph{recall},0.497 \emph{pf}, 0.415 \emph{F-measure} and
0.684 \emph{AUC}. The proposed weighting (3) achieves a mean performance
of 0.902 \emph{recall}, 0.413 \emph{pf}, 0.473 \emph{F-measure} and
0.744 \emph{AUC}.
On the NASA data, the \ac{CPDP} without further treatment (1) achieves a mean
performance of 0.335 \emph{F-measure} and 0.651 \emph{AUC}. The $k$-NN
relevancy filter (2) achieves a mean performance of 0.324 \emph{F-measure} and
0.639 \emph{AUC}. The proposed weighting (3) achieves a mean
performance of 0.348 \emph{F-measure} and 0.669 \emph{AUC}.  Only in terms
of \emph{recall}, the proposed weighting is outperfomed by the other two models.
However, \ac{CPDP} without further treatment (1) has mean \emph{recall} of 1.0
and a mean \emph{AUC} of 0.504, which indicates that the trained models
degenerate to trivial predictions that always predict defects. Moreover, the
\emph{recall} of the weighting and the $k$-NN approach are very close to each
other.

\subsection{Peters and Menzies, 2012}
\label{sec:peters2012}

\noindent\textbf{Approach.} A general problem in the way of \ac{CPDP} is
sharing data about proprietary products. Due to the potentially
sensitive information that can be glanced from defect data, companies are
unwilling to share such information. \cite{Peters2013a} developed
the approach MORPH to facilitate data sharing while ensuring privacy.
The task of MORPH is to mutate the metric values of the entities $s \in S$
\begin{equation}
m_i(s) = m_i(s) + r \cdot (m_i(s)-m_i(s^{NUN}))
\end{equation}
with $r \in [\alpha, \beta] \cup [-\beta,-\alpha]$ and $s^{NUN}$ the nearest
unlike neighbor of $s$, i.e., the NN of that has a different
class label defined as 
\begin{equation}
s^{NUN} = \arg\min_{s \in S: s'\neq s \wedge c_S(s')\neq c_S(s)} dist(s, s').
\end{equation}
The authors use the values $\alpha=0.15$ and $\beta=0.35$.

\noindent\textbf{Approach type.} Other (data privacy).

\noindent\textbf{Classifier.} Logistic Regression, Na\"{i}ve Bayes, and Random Forests.

\noindent\textbf{Data.} Ten products of the JURECZKO data. 

\noindent\textbf{Case study setup.} 
The case study has two goals. The first goal
is to evaluate if the performance of the defect prediction suffers from the
privatization of the data through MORPH. To this aim, the authors used two
experiment configurations: (1) \ac{CPDP} with each of the products in the data
set once as target product and the other nine as training data; and (2)
\ac{WPDP} with 10x10 cross-validation. The defect prediction is performed with the original data as
well as with privatized data. The second goal is to evaluate how well MORPH
privatizes the data. To this aim, the authors used one more experiment
configuration: (3) data swapping with $p \in \{10\%, 20\%, 40\%\}$ of the data
swapped.

\noindent\textbf{Performance measures.} \emph{F-measure} for prediction
performance. The privacy is evaluated by measuring how many queries to the privatized data,
e.g., for certain subranges of attributes, return the same instances as to the
original data.

\noindent\textbf{Results.} The authors do not report numerical results.
Line graphs of the privacy and the \emph{F-measure} for the different
configurations and privacy methods are given instead. Regarding prediction
performance, the graphs show only very small differences between the original
(1) and the privatized data (2,3). Therefore, the authors conclude that MORPH
and data swapping both do not degrade the prediction performance. Moreover, the
privacy evaluation shows that MORPH (2) is comparable with swapping $p=40\%$ of the
data and better than swapping less data (3). 

\subsection{Z. He\etal, 2012}
\label{sec:he2012}
\noindent\textbf{Approach.} \cite{He2012} studied the general
feasibility of \ac{CPDP}, based on the findings of very low success by
\cite{Zimmermann2009}. While Zimmermann\etal~only considered
pair-wise predictions in their study, He\etal~consider combinations of one, two,
or three products as training data. Combinations with more products are not
taken into account due to the exponentially rising number of combinations. They
disallow combinations with training data from the target product, i.e., other
versions. Then, taking further pattern from the study by Zimmermann\etal,
the authors train a decision tree for the identification of training data that
has a high chance for success. In comparison to Zimmermann\etal, they base their
decision tree solely on distributional characteristics. They use 16
distributional characteristics for each of the 20 metrics. Thus, they
create a 320-dimensional vector for each combination of products they
evaluated. As label for the learning, they use whether the combination of data
was successful in terms of achieving high recall and precision or not. 

\noindent\textbf{Approach type.} Relevancy filtering.

\noindent\textbf{Classifier.} C4.5 Decision Tree, Decision Table,
Logistic Regression, Na\"{i}ve Bayes, and \ac{SVM}.

\noindent\textbf{Data.} 34 products of  JURECZKO data. 

\noindent\textbf{Case study setup.} The authors used three experiment
configurations: (1) each product once as target product with all combinations of
one, two, and three products as training data to determine the best possible
combination for each product and the best possible result that can be achieved;
(2) \ac{WPDP} with 5x5 cross validation; (3) \ac{WPDP} with data from old
versions of the same target product; (4) the decision tree approach described
above evaluated using 5x5 cross-validation with products as hold-out data during
of the cross-validation. 

\noindent\textbf{Performance measures.} \emph{recall}, \emph{precision},
\emph{F-measure} and $succ_{0.7,0.5}$.

\noindent\textbf{Results.} The best mean best-case \ac{CPDP}
performance in the experiment using all configurations (1) was achieved with
the Decision Table. The mean performance over all target products is 0.735 
\emph{recall}, 0.560 \emph{precision}, and 0.627 \emph{F-measure}. For 18 of
the 34 products at least one combination of training data that fulfills the
criteria for a successful defect prediction was found, i.e., only $succ_{0.7,
0.5} = 2.47\%$ of all combinations led to a successful prediction. The best \ac{WPDP} performance with 5x5 cross
validation (2) is achieved with the C4.5 Decision Tree, with a mean performance
of 0.486 \emph{recall}, 0.558 \emph{precision}, and 0.512 
\emph{F-measure}. For the \ac{WPDP} based on old versions of the same
projects (3), the best performance is also achieved by the C4.5 decision tree,
with mean values of 0.585 for \emph{recall}, 0.512 for \emph{precision}, and
\emph{0.496} for \emph{F-measure}.
In a product-level comparison, they determined that for 18
products the best-case \ac{CPDP} achieved with configuration (1)
performed significantly better in terms of the \emph{recall} and
\emph{F-measure} than the \ac{WPDP} with cross validation (2).
For \emph{precision}, they observe a different effect, here the \ac{CPDP} yields
better results for four products, but is worse for ten products. The product-level comparison between
\ac{CPDP} (1) and \ac{WPDP} with old versions (3) are similar for
\emph{recall} (14 times better \ac{CPDP} better, one time worse) and
\emph{F-measure} (19 times \ac{CPDP} better). For \emph{precision}, 
\ac{CPDP} is better for 11 products and lower for three products.

For their decision tree approach (4) the authors report a mean performance of
0.683 \emph{recall} and 0.739 \emph{precision}. The
created decision tree is huge with 642 leaf nodes of which 270 lead to a
successful prediction. Hence, the tree effectively contained 270 rules for good
predictions. Through these results, the authors concluded that distributional
characteristics can be used to determine good training data for \ac{CPDP}.

\subsection{Rahman\etal, 2012}
\label{sec:rahman2012}
\noindent\textbf{Approach.} \cite{Rahman2012} propose performance
measures for the evaluation of \ac{CPDP}. The proposed performance measures
\emph{AUCEC} and \emph{AUCECF} are variants of \emph{AUC} that take the cost
effectiveness into account. The motivation for \emph{AUCEC}/\emph{AUCECF} is hence the
same as for defect prediction in general. In practice, only parts of a
software product are investigated in detail for quality assurance instead of
the whole product.
Assuming that a model ranks files by their defect-proneness, it becomes possible
to inspect only few files and still find most of the defects. In case of ties in the ranking by a predictor, the authors propose to rank the smaller
file higher, because less data needs to be inspected to find a defect with the
same probability. \emph{AUCEC} and \emph{AUCECF} formalize this idea as a
performance measure. \emph{AUCEC} is measured using an \ac{ROC} curve where the
percentage of defects found is measured against the percentage of the \ac{LOC}
inspected. Similarly, the authors define \emph{AUCECF} using the percentage of
the number of files inspected instead of \ac{LOC}. This is identical to the
definition of $AUC_{Alberg}$. 

\noindent\textbf{Approach type.} Cost Curves.

\noindent\textbf{Classifier.} Logistic Regression.

\noindent\textbf{Data.} The authors collected defect data for 38 versions
of nine projects. They collected process metrics (e.g., number of commits;
number of developers; added, deleted and changed \ac{LOC}) and the \ac{LOC} of
files and combined the data with information about added features, improvements
and defects from  the \ac{ITS} of the projects. The data is not publicly
available.

\noindent\textbf{Case study setup.} 
The authors use five experiment configurations: (1) with a threshold of
0.5 for the cutoff probability, i.e., all files predicted with a likelihood of
at least 0.5 as defective are considered to be defective\footnote{This is the
usual way of how logistic regression models are used for prediction, in case
it is not specifically stated otherwise.}; (2) with the best possible threshold
that maximizes the \emph{F-measure} on the training data; (3) with the best possible threshold
that maximes the \emph{F-measure} on the target data; (4) \ac{WPDP} based on
old versions of the project; and  (5) \ac{CPDP} with a single metric as
training data.

\noindent\textbf{Performance measures.} \emph{F-measure}, \emph{AUC},
\emph{AUCEC}, and \emph{AUCECF}.

\noindent\textbf{Results.} No numerical values are reported, instead
visual analysis is performed supported with Mann-Whitney-U tests for statistical
significance of the results. The case study shows that 
\ac{CPDP} (1)--(3) performs worse in terms of \emph{F-measure}, \emph{AUC}, and
$AUC_{Alberg}$ than \ac{WPDP} (4). In terms of \emph{AUCEC}, the authors do not
see any significant differ between \ac{CPDP} and \ac{WPDP}. The single metric
models show (5) that while
the size metrics are effective if the performance is measured with \emph{AUC},
they do not perform well using the cost sensitive \emph{AUCEC} metric.
From this, the authors infer that product metrics related to the size might be bad choices for
prediction models in cost-sensitive settings.

\subsection{Uchigaki\etal, 2012}
\label{sec:uchigaki2012}
\noindent\textbf{Approach.} \cite{Uchigaki2012} propose to
build an ensemble model of single metric models. Concretely, they propose to
create the ensemble as
\begin{equation}
h(s) = \frac{\sum_{i=1}^p w_i \cdot h_i(m_i(s))}{\sum_{i=1}^p w_i}
\end{equation}
where $h_i$ is a Logistic Regression models trained only with metric $m_i$.
The weights $w_i$ are the contribution ratio of the sub-models to the classifcation as
determined by the goodness of fit. 

Moreover, the authors propose to preprocess the metric data with log
transformation (see Section~\ref{sec:turhan2009}). Furthermore, the authors
propose to standardize the data using z-score standardization, i.e., to transform
all data as follows:
\begin{equation}
\hat{m}_i(s) = \frac{m_i(s)-mean(m_i(S))}{std(m_i(S))}.
\end{equation}

\noindent\textbf{Approach type.} Other (classification model and instance
standardization).

\noindent\textbf{Classifier.} Single-metric ensemble model (see above) and
Logistic Regression.

\noindent\textbf{Data.} All twelve products from the NASA data.

\noindent\textbf{Case study setup.} The authors use three experiment
configurations: (1) a conventional multivariate Logistic Regression model; (2)
the proposed ensemble model without preprocessing the metrics; and (3) the proposed
ensemble model with preprocessing the metrics. For all three configurations, the
authors create pair-wise \ac{CPDP} models. 

\noindent\textbf{Performance measures.} $AUC_{Alberg}$.

\noindent\textbf{Results.} The authors report the performance of all
pair-wise predictions as well as the overall means. The conventional
multivariate Logistic Regression (1) achieves a mean performance of 0.703
$AUC_{Alberg}$, the ensemble model without preprocessing (2) of 0.740
$AUC_{Alberg}$, and the ensemble model with preprocessing (3) of 0.746
$AUC_{Alberg}$. The authors conclude that their model outperforms conventional
logistic regression. 

\subsection{Canfora\etal, 2013, 2015}
\label{sec:canfora2013}
\noindent\textbf{Approach.} In an article by \cite{Canfora2013}
later extended in \cite{Canfora2015}, the authors propose a multi-objective
approach for \ac{CPDP} based on the findings by \cite{Rahman2012}
regarding cost effectiveness of models (see
Section~\ref{sec:rahman2012}). Canfora\etal~consider two objectives: (1)
maximization of the effectiveness of the defect prediction, i.e., to find as
many defects as possible and (2) minimization of the cost of the
reviewing effort due to the prediction. Hence, the objective functions are
\begin{equation}
\begin{split}
\max_h &\sum_{s^* \in S^*} h(s^*)\cdot c^*(s^*) \\
\min_h &\sum_{s^* \in S^*} h(s^*)\cdot LOC(s^*).
\end{split}
\end{equation}
Since standard learning procedures only optimize the effectiveness of the defect
prediction, but ignore the costs, the authors used a genetic program for the
training of a classifier. Concretely, the authors interpret the hypothesis $h$
representing the classifier as the fitness function of the genetic program. The
structure of the classifier, e.g., the coefficients that define a regression
model, are estimated using the genetic algorithm. The authors
apply the vNSGA-II algorithm~\citep{Deb2002} to obtain a set of optimal
solutions, i.e., a set of Pareto optimal~\citep{Coello2006} solutions that can
be used for defect predition with different desired values for effectiveness and
costs. The authors call this approach \ac{MODEP}. Moreover, the authors propose
to standardize all data with z-score standardization (see
Section~\ref{sec:uchigaki2012}) before applying their approach.

\noindent\textbf{Approach type.} Cost curves and other (classification model).

\noindent\textbf{Classifier.} Logistic Regression and Decision Trees to
instantiate \ac{MODEP}, \ac{WPDP}, and \ac{CPDP}; local model with Association
Rules.

\noindent\textbf{Data.} 10 products of the JURECZKO data. The authors only
use the six Chidamber and Kemerer metrics and \ac{LOC} and discard the other
metrics contained in the data.

\noindent\textbf{Case study setup.} The authors used ten experiment
configurations: (1) Logistic Regression \ac{MODEP}; (2) Decision Tree
\ac{MODEP}; (3) a trivial model that ranks entities by their decreasing
\ac{LOC}; (4) a trivial model that ranks entities by their increasing \ac{LOC};
(5) a normal Logistic Regression \ac{CPDP} model; (6) a normal Decision Tree
\ac{CPDP} model; (7) a local model obtained with $k$-Means as
clustering algorithm, with $k=10$ determined by the Silhouette
coefficient~\citep{Rousseeuw1987} and Logistic Regression as classifier; (8) a
local model with MDS~\citep{Borg2005} as clustering algorithm and association
rules as classifier; (9) \ac{WPDP} with 10x10 cross-validation and Logistic Regression
as classifier; (10) \ac{WPDP} with 10x10 cross-validation and a Decision Tree as
classifier. The configurations (1)--(8) are using the data from all products,
except the target product as training data.

\noindent\textbf{Performance measures.} \emph{AUC} for trivial model
comparison (3)--(4), \emph{recall}, \emph{precision}, and review \emph{cost}
measured in KLOC for the other configurations. Please note that the authors use two
different definitions for \emph{recall} and \emph{AUC} in the journal
extension~\citep{Canfora2015}: they once consider the commonly used
\emph{recall} of classes, i.e., how many classes that contain defects are found, and once the
\emph{recall} of the overall defects that were found. The difference is that the
second accounts for classes that contain multiple defects.
 
\noindent\textbf{Results.} Since \ac{MODEP} does not compute a single
classifier, but a family of classifiers that is Pareto optimal in
terms of costs and effectiveness, the authors decided to pair the results of
\ac{MODEP} with results that achieve the same effectiveness or costs achieved
with other models. In the conference paper~\citep{Canfora2013}, the authors pair
based on effectiveness, i.e., for a given model that receives a certain \emph{recall},
the authors pick the \ac{MODEP} model that achieves the same \emph{recall}. In the
journal extension~\citep{Canfora2015}, the authors pair based on the costs. Due
to the large overall number of pairings considered, we do not report all
mean values here\footnote{6 pairings by \emph{recall} in the conference paper
with three performance metrics each, 28 pairings by costs in the journal
extension with four performance metrics each, and the comparison with trivial
models using one performance metric. This means, a total number of 
131 mean values would be required for full reporting here.}. Since the Logistic
Regression yields better results than the Decision Tree, we only report on the pairings in
which Logistic Regression is used. Moreover, the results for both \emph{recall}
and \emph{AUC} are very similar independent of whether defective classes or
number of defects are considered. Therefore, we report on the \emph{recall} and
\emph{AUC} related to classes, because it is commonly used in other
publications. Furthermore, the authors evaluated all configurations with and
without z-score standardization. Since the general conclusions hold for
\ac{MODEP} independent of the standardization, we leave out all non-standardized
results, since the performance with standardization is generally higher. 

First, we consider the comparison with the trivial model. The Logistic
Regression \ac{MODEP} (1) achieves a mean \emph{AUC} of 0.836, the trivial model
with decreasing sorting (3) a mean \emph{AUC} of 0.388, and the trivial model
with increasing sorting (4) a mean \emph{AUC} of 0.759. From this, the
authors conclude that \ac{MODEP} outperforms trivial approaches.

Second, we consider the pairings based on \emph{recall}. For the comparison
of the Logistic Regression \ac{MODEP} (1) with the normal \ac{CPDP} model
(5), the authors report a mean \emph{recall} of 0.615 for both. Logistic
Regression \ac{MODEP} achieves a mean \emph{precision} of 0.522 and a mean
\emph{cost} of 113.7 KLOC. The normal \ac{CPDP} model achieves a mean
\emph{precision} of 0.587 and a mean \emph{cost} of 127.5 KLOC. For the
comparison of the Logistic Regression \ac{MODEP} (1) with the local model (7),
the authors report a mean \emph{recall} of 0.455 for both. The Logistic
Regression \ac{MODEP} achieves a mean \emph{precision} of 0.537 and a mean
\emph{cost} of 93.8 KLOC. The local model achieves a mean \emph{precision} of
0.431 and a mean \emph{cost} of 115.7 KLOC. For the comparison of the
Logistic Regression \ac{MODEP} (1) with the \ac{WPDP} model (9), the authors
report a mean \emph{recall} of 0.446 for both. The Logistic Regression \ac{MODEP}
achieves a mean \emph{precision} of 0.547 and a mean \emph{cost} of 85.5
KLOC. The \ac{WPDP} model achieves a mean \emph{precision} of 0.678 and a mean
\emph{cost} of 95.2 KLOC. The authors conclude that their cost optimization by
using \ac{MODEP} performs better in terms of costs compared against
all three other scenarios, if the same values for \emph{recall}, i.e., the same
effectiveness is assumed.

Finally, we consider the pairings based on \emph{cost}. For the comparison
of the Logistic Regression \ac{MODEP} (1) with the normal \ac{CPDP} model
(5), the authors report a mean \emph{cost} of 124.5 KLOC for both. Logistic
Regression \ac{MODEP} achieves a mean
\emph{recall} of 0.937 and a mean \emph{precision} of 0.459. The normal
\ac{CPDP} model achieves a mean
\emph{recall} of 0.615 and a mean \emph{precision} of 0.587. For the comparison
of the Logistic Regression \ac{MODEP} (1) with the local model (8), the authors
report a mean \emph{cost} of 104.2 KLOC for both. The Logistic Regression
\ac{MODEP} achieves a mean \emph{recall} of
0.846 and a mean \emph{precision} of 0.378. The local model achieves a mean
\emph{recall} of 0.578 and a mean \emph{precision} of 0.520. For the
comparison of the Logistic Regression \ac{MODEP} (1) with the \ac{WPDP} model
(9), the authors report a mean \emph{cost} of 101.9 KLOC for both. The Logistic
Regression \ac{MODEP} achieves a mean
\emph{recall} of 0.741 and a mean \emph{precision} of 0.474. The \ac{WPDP}
model achieves a mean \emph{recall} of
0.577 and a mean \emph{precision} of 0.717. The authors conclude that
their cost optimization by using \ac{MODEP} performs better in terms of costs
compared against all three other scenarios, if the same costs are
assumed.

\subsection{Peters\etal, 2013a}
\label{sec:peters2013privacy}
\noindent\textbf{Approach.} \cite{Peters2013a} developed
the approach CLIFF+MORPH to facilitate data sharing while ensuring privacy as an
extension of the previously proposed MORPH~\citep{Peters2012} (see
Section~\ref{sec:peters2012}).
The CLIFF is a relevancy filter that uses the concept of \emph{power} of
subranges in the data.
First, equal frequency binning is applied to create $n=10$ subranges
$ranges(m_i)$ for each metric $m_i$ in the data\footnote{The authors note that
other values for $n$ are possible, but currently not explored.}. The power is
then calculated using the likelihood of a class based on the data in the bin,
i.e.,
\begin{equation}
power(c|r) = \frac{(P(r|c)\cdot P(c))^2}{P(r|c)\cdot P(c) + P(r|not(c)\cdot
P(not(c))))}
\end{equation}
with the class label $c \in \{0,1\}$ and  the range $r \in ranges(m_i)$. Then,
the overall power of an entity $s \in S$ is calculated as the product of the
power of the subranges in which the metrics $m_i$ fall, i.e.,
\begin{equation}
power(s) = \prod_{i=1,\ldots,d} power(c_S(s),r)
\end{equation}
with $r$ the range into which $m_i(s)$ falls within $ranges(m_i)$. CLIFF then
selects the $p$ percent of data with the highest power. Once instances are
selected with CLIFF, they are privatized with MORPH. 

\noindent\textbf{Approach type.} Relevancy filtering and other (data privacy).

\noindent\textbf{Classifier.} $k$-NN, Na\"{i}ve Bayes, and \ac{SVM}.

\noindent\textbf{Data.} Ten products from the JURECZKO data.

\noindent\textbf{Case study setup.}  
The authors use five experiment configurations:
(1) \ac{CPDP} without privitization; (2) \ac{CPDP} with MORPH; (3) \ac{CPDP}
with CLIFF+MORPH with $p \in {10, 20, 40}$
percent of the data selected; (4) \ac{CPDP} with data swapping \citep{Reiss1982}
with $p \in \{10, 20, 40\}$ of the data swapped; and
(5) \ac{CPDP} with $k$-anonymity~\citep{Sweeney2002} with $k \in \{2, 4\}$
indistinguishable members. For all of the above, the authors use each product
once as target product and the other nine products for training.
 
\noindent\textbf{Performance measures.} \emph{recall}, \emph{pf}, and
\emph{G-measure} for prediction performance. The privacy is evaluated by
measuring how many queries to the privatized data, e.g., for certain subranges
of attributes, return the same instances as to the original data.

\noindent\textbf{Results.} The values for \emph{recall}, \emph{pd}, and
\emph{G-measure} are reported for all ten products with all privitization
approaches. We only report the results for the Na\"{i}ve Bayes, as it performs
best overall. However, the performance of $k$-NN is not statistically
significantly different from Na\"{i}ve Bayes. \ac{CPDP} without privatization
(1) achieves a mean performance of 0.236 \emph{recall}, 0.091 \emph{pf}, and
0.335 \emph{G-measure}. \ac{CPDP} with MORPH (2) achieves a mean performance of 0.229
\emph{recall}, 0.089 \emph{pf}, and 0.327 \emph{G-measure}. CLIFF+MORPH
(3) achieves the best performance with $p=40$ percent of the data selected. The
mean performance is 0.659 \emph{recall}, 0.4 \emph{pf}, and 0.598
\emph{G-measure}. Data swapping (4) achieves the best performance with $p=10$
percent of the data swapped. The mean performance is 0.225 \emph{recall}, 0.084
\emph{pf}, and 0.32 \emph{G-measure}. $k$-anonymity achieves the best
performance with $k=2$. The mean performance is 0.195 \emph{recall}, 0.08
\emph{pf}, and 0.294 \emph{G-measure}. 
In terms of privacy, CLIFF+MORPH (3) and $k$-anonymity (5) are almost perfect
for all values of $p$ and $k$ with over 90\% of queries failing. MORPH (2) and
data swapping (4) are below 80\%. 

\subsection{Peters\etal, 2013b}
\label{sec:peters2013filter}
This publication by \cite{Peters2013b} was actually retracted due
to problems with the case study
setup\footnote{http://de.slideshare.net/timmenzies/msr13-mistake}. However, we
still list this work here for three reasons: (1) completeness of the review;  
(2) accurate statistics about classifiers, data usage, etc.; and (3) because
the proposed approach is being used for comparisons regardless of the
retraction (e.g., by \cite{Kawata2015}, discussed in
Section~\ref{sec:kawata2015}). However, we only describe the approach suggested
by Peters\etal, the data used and the performance metrics. We do not 
elaborate on the case study setup or the results.

\noindent\textbf{Approach.} \cite{Peters2013b} propose a
relevancy filter where the candidate data drives the filtering. For each
entity in the candidate data $s \in S^{cand}$ the closest entity in
the target data is selected. Through this, a subset of the target data is
defined as
\begin{equation}
\begin{split}
S_{closest}^* = \{&s^* \in S^*:\exists~s \in S^{cand}~|
\\&s^* = \arg\min_{\hat{s}^* \in S^*} dist(m(s), m(\hat{s}^*)) \}.
\end{split}
\end{equation}
Using $S_{closest}^*$, the same approach is used to determine the training data
as subset of the candidate data as training data:
\begin{equation}
\begin{split}
S^{train} = \{&s \in S^{cand}: \exists~s^* \in S_{closest}^*~|
\\&s =
\arg\min_{\hat{s} \in S^{cand}} dist(m(s^*), m(\hat{s}) \}.
\end{split}
\end{equation}

\noindent\textbf{Approach type.} Relevancy filtering.

\noindent\textbf{Classifier.} Logistic Regression, Na\"{i}ve Bayes, and Random
Forest.

\noindent\textbf{Data.} 56 products from the JURECZKO data.
 
\noindent\textbf{Case study setup.} Not applicable.

\noindent\textbf{Performance measures.} \emph{accuracy}, \emph{recall}, \emph{pf},
\emph{precision}, \emph{F-measure}, and \emph{G-measure}.
 
\noindent\textbf{Results.} Not applicable.

\subsection{Kocaguneli\etal, 2013}
\label{sec:kocoguneli2013a}

\cite{Kocaguneli2013} discuss the problem of experiments in a
cross-project setting in general, and do not propose a concrete approach or
conduct a case study, which is why we break with our reporting pattern. From a
brief consideration of previously considered approaches for the cross-project
context, not only for defect prediction, but also for effort prediction, the
authors formulate three synergies between those approaches which they believe
can advance the state of the art: (1) the synergy between supervised learning
and transfer learning, which they state is already being explored with some
success; (2) the synergy between semi-supervised learning and transfer learning,
i.e., only labeling a small subset of the target product context manually and
using transfer learning to augment the data; and (3) the combination of active
learning, e.g., to reduce the available data to the most fitting the context
before applying transfer learning. 

\subsection{Turhan\etal, 2013}
\label{sec:turhan2013}
\noindent\textbf{Approach.} \cite{Turhan2013} explore \ac{MPDP}
as an extension of their earlier
work~\citep{Turhan2011} (see Section~\ref{sec:turhan2011}).
The basic approach is identical to their previous work. However, they
additionally explore how using less within-project data affects the performance
of predictors. 

\noindent\textbf{Approach type.} Relevancy filtering.

\noindent\textbf{Classifier.} Na\"{i}ve Bayes.

\noindent\textbf{Data.} Seven products from the NASA data, all three
products from the SOFTLAB data, and 63 products from the JURECZKO data.
 
\noindent\textbf{Case study setup.} The authors compare three configurations
against each other: (1) \ac{WPDP} with 10x10 cross validation and (2) \ac{WPDP} with
10x10 cross validation, where the training data is extended with cross-project
data; and (3) \ac{WPDP} with only 10\%  of the within-project data for
training augmented with cross-project data. The third configuration is only
tested on a subset of 31 products of the JURECZKO data. The NASA and SOFTLAB
data are used together as a single data set. The authors use a Mann-Whitney-U
test to determine the statistical significance of their results.

\noindent\textbf{Performance measures.} \emph{recall}, \emph{pf}, and
\emph{balance}.

\noindent\textbf{Results.} The authors report all three performance
metrics for all products for the configurations (1) and (2). For configuration
three, the results are only reported for a subset of the JURECZKO data. For the
\ac{WPDP} (1), the authors report a mean performance of 0.743 \emph{recall}, 0.276
\emph{pf}, and 0.710 \emph{balance} on the NASA/SOFTLAB data  and of 0.620
\emph{recall}, 0.303 \emph{pf}, and 0.612 \emph{balance} on the JURECZKO data.
For \ac{WPDP} with 10x10 cross validation augmented with cross-project data (2),
the authors report a mean performance of 0.784 \emph{recall}, 0.275
\emph{precision}, and 0.620 \emph{balance} on the NASA/SOFTLAB data and of 0.675
\emph{recall}, 0.305 \emph{precision}, and 0.649 \emph{balance} on the JURECZKO
data. For \ac{WPDP} with only 10\% of within-project data augmented with
cross-project data, the authors report a mean performance of 0.706
\emph{recall}, 0.152 \emph{pf}, and 0.671 \emph{balance} on 31 products from the
JURECZKO data. Using the Mann-Whitney-U test, the authors determine that
augmenting the \ac{WPDP} with cross-project data, improves \emph{balance} in
comparison to normal \ac{WPDP} significantly in about half of the cases and
does not decrease the performance in the other cases. With the reduced data from
within the product, this finding still holds. 

\subsection{Singh\etal, 2013}
\label{sec:singh2013}
\noindent\textbf{Approach.} \cite{Singh2013} do not propose any
special approach, but evaluate pair-wise prediction without additional data treatment.

\noindent\textbf{Approach type.} None.

\noindent\textbf{Classifier.} C4.5 Decision Tree, Decision Table, $k$-NN,
Na\"{i}ve Bayes, Random Forest and \ac{SVM}.

\noindent\textbf{Data.} One Java product from the NASA data and one Java
product mined by the authors, for which the data is not publicly available. Only
Chidamber and Kemerer metrics were used from the available data. 

\noindent\textbf{Case study setup.} The authors use two experiment
configurations: (1) \ac{WPDP} with 10x10 cross validation; and (2) pair-wise
\ac{CPDP} with each product once as training and once as test data. 

\noindent\textbf{Performance measures.} \emph{recall}, \emph{precision},
\emph{pf}, and \emph{F-measure}.

\noindent\textbf{Results.} The best overall performance is achieved by
the decision tree. For \ac{WPDP} (1), the authors report a mean performance of
0.690 \emph{recall}, 0.650 \emph{precision}, 0.302 \emph{pf},  and 0.669
\emph{F-measure}. For \ac{CPDP} (2), the authors report a mean performance of
0.585 \emph{recall}, 0.693 \emph{precision}, 0.214 \emph{pf}, and 0.624
\emph{F-measure}. The authors conclude that \ac{CPDP} using Chidamber and
Kemerer metrics is possible. 

\subsection{Herbold, 2013}
\label{sec:herbold2013}
\noindent\textbf{Approach.} \cite{Herbold2013} propose a relevancy
filter based on the clustering of products using distributional characteristics. For
each product $S$, the \emph{characteristic vector} defined as
\begin{equation}
\begin{split}
char(S) = (&char_1(m_1(S)), \ldots,char_1(m_p(S)), \ldots,
\\&char_q(m_1(S)),\ldots char_q(m_p(S)))
\end{split} 
\end{equation}
is calculated and the
euclidean distances between these vectors are considered. While the authors
state that any distributional charactistic can be used, they only use
the mean value (i.e., $char_1=mean$) and the standard deviation (i.e.,
$char_2=sd$.
The authors propose two approaches for \ac{CPDP} based on this general idea:
the first approach is to use the $k$-NN algorithm to select $k$ products
that have the most similar distributional characteristics. 
The second approach is to apply the EM clustering algorithm~\citep{Dempster1977}
to create clusters of products based on the characteristic vectors of the candidate data
joined with the target product. The number of clusters is determined automatically by the EM
algorithm. The approach then selects those products as training data, that are within the same cluster as the target
product. In case the target product is the only product in
a cluster (i.e., in a cluster of size one), the EM algorithm is
configured such that one less cluster than automatically determined is
allowed and the clustering is repeated. 

Additionally, Herbold proposed a technique for the treatment of bias in the
training data in case the distribution of the dependent variable, i.e., the defect-proneness is not
roughly equal for defect-prone and non-defect-prone. With equal weighting, the
entities of the training data are weighted such that the total weight of the defect-prone
entities equals the total weight of the non-defect-prone entities, i.e., 
\begin{equation}
\label{eq:weighting}
\begin{split}
w_{dp} &= 0.5 \frac{|\{s \in S^{train}: c_{S^{train}}(s)=1\}|}{|S^{train}|} \\
w_{ndp} &= 0.5 \frac{|\{s \in S^{train}: c_{S^{train}}(s)=0\}|}{|S^{train}|}
\end{split}
\end{equation}
with $w_{dp}$ as weight for the defect-prone entities, $w_{npd}$ as weight
for the non-defect-prone entities, and $S^{train}, c_{S^{train}}$ are the
selected entities for training and the respective concept. The equal weighting
is applied after the proposed training relevancy filtering techniques, because
otherwise the property that both defect-prone and non-defect-prone entities have the same overall weight may be violated. Prior to the relevancy
filtering, the metric data is standardized to the  [0,1] using min-max
normalization, i.e.,
\begin{equation}
\hat{m}_i(s) = \frac{m_i(s)-\min_{s' \in S} {m_i(s')}} {\max_{s' \in
S}{m_i(s')}-\min_{s' \in S}{m_i(s')}}
\end{equation}
for all $s \in S$. 

\noindent\textbf{Approach type.} Relevancy filtering, instance weighting, and
other (instance standardization).

\noindent\textbf{Classifier.} Bayesian Network, C4.5 Decision Tree, Multilayer
Perceptron, Na\"{i}ve Bayes, Random Forest, and \ac{SVM} with an \ac{RBF}
kernel.

\noindent\textbf{Data.} 44 products from the JURECZKO data. 

\noindent\textbf{Case study setup.} The authors used four experiment
configurations: (1) \ac{WPDP} with 10x10 cross validation; (2) \ac{CPDP} with
all data from other products; (3) \ac{CPDP} with the $k$-NN product relevancy
filter; and (4) \ac{CPDP} with EM clustering for relevancy filtering. All of the
above are applied with and without the equal weighting for bias treatment and for all seven classifiers listed
above. For the \ac{CPDP} all products are used as training data that are not
other versions of the target product.

\noindent\textbf{Performance measures.} \emph{recall}, \emph{precision}, and
$succ_{0.7,0.5}$.

\noindent\textbf{Results.} The equal weighting produces significantly
better results consistently. Of the used classification schemes, the \ac{SVM}
with the \ac{RBF} kernel performs best. Therefore, we only report those results
in the following. The \ac{WPDP} (1) achieves a mean performance of 0.57
\emph{recall}, 0.63 \emph{precision}, and 0.37 $succ_{0.7,0.5}$. \ac{CPDP}
with all data (2) achieves a mean performance of 0.64 \emph{recall},
0.42 \emph{precision}, and 0.09 $succ_{0.7,0.5}$. The $k$-NN product relevancy filter
(3) performs best for $k=20$ with a mean performance of 0.74 \emph{recall},
0.40 \emph{precision}, and 0.18 $succ_{0.7,0.5}$. The EM clustering (4) achieves
a mean performance of 0.71 \emph{recall}, 0.40 \emph{precision}, and
0.13 $succ_{0.7,0.5}$. The author concludes that the proposed approaches
outperform using all data, but are still worse than \ac{WPDP}.

\subsection{Z. He\etal, 2013}
\label{sec:he2013}
\noindent\textbf{Approach.} \cite{He2013}
discuss if data obtained from open source software can be used to
predict defects within proprietary software. Additionally, they suggest an
approach for \ac{CPDP} that consists of three parts: relevancy filtering of
products, feature selection, and ensemble learning for classification. For
relevancy filtering, their idea is to select products based on the
separability measure between data sets proposed by \cite{Hido2008}.
To measure the separability of two products $S, S'$, 500 random samples $S_{sample}, S'_{sample}$ are drawn
from both products\footnote{Note, that depending on the size of the data set
this can lead to both oversampling and undersampling.}. Then, a logistic
regression classifier is trained with the metric data $M(S_{sample}),
M(S'_{sample})$ as input and the data set membership as classification. The
\emph{accuracy} of this classifier is evaluated with 5x5 cross validation.
Based on the \emph{accuracy}, the distance between products is defined as
\begin{equation}
dist_{sep}(M(S_{sample}), M(S'_{sample})) = 2\cdot(acc-0.5).
\end{equation}
The $k$ closest
neighbors according to this distance metric are used for training.

The second part of their approach is feature selection. Following the
suggestion by \cite{Hido2008}, the authors propose to filter
the available metrics and only use those for training, which are similar in the
training data and the target data.
To measure the similarity they suggest to analyze the classification model
trained to separate the data. The assumption is that the metrics with a high
information gain~\citep{Mitchell1997} for the classifier are responsible for the
differences between the training data and the target data. Therefore, the features with the highest
information gain are removed to improve the homogeneity between the products.

The third part of their approach is to use an
\emph{ensemble classifier}~\citep{Rokach2010} and train a different classifer
for the data of each product that was selected for training using the relevancy
filter. The authors propose to use \emph{bagging}~\citep{Breiman1996} for the
ensemble learning, i.e., training a classifier for each product in the selected
training data and using the majority vote of those classifiers for the overall
classification. 

Moreover, the authors propose to apply undersampling \citep{Drummond2003} to the
data to treat a bias towards non-defect-prone predictions due to the fact that
more instances are non-defect-prone than defect-prone. In case there are more non-defect-prone
instances than defect-prone instances, undersampling reduces the
non-defect-prone instances through sampling from them until they have the same
number as the defect-prone instances. Undersampling was also used
successfully in context of \ac{WPDP}, e.g., by \cite{Menzies2008}.
Additionally, the data is standardized to the interval  [0,1] using min-max
normalization (see Section~\ref{sec:herbold2013}).

\noindent\textbf{Approach type.} Relevancy filtering, stratification, and other
(classification model, instance standardization, feature selection, and data
source type influence).

\noindent\textbf{Classifier.} Logistic Regression, Na\"{i}ve Bayes, and Random
Forest. 

\noindent\textbf{Data.} 34 products from the JURECZKO data. 

\noindent\textbf{Case study setup.} The authors use five experiment configurations: (1)
\ac{CPDP} with the best possible open source product as training data for 
each proprietary product; (2) \ac{WPDP} with 5x5 cross validation; (3) 
\ac{CPDP} with all data from the open source products as training data for each
proprietary product; (4) \ac{CPDP} with the $k$-NN relevancy filter
proposed by \cite{Turhan2009} (see Section \ref{sec:turhan2009}); and
(5) \ac{CPDP} with their proposed approach. 

\noindent\textbf{Performance measures.} \emph{recall}, \emph{pf}, and
\emph{G-measure}.

\noindent\textbf{Results.} No numerical results are reported, instead
box-plots of the performance measures for the five models are shown. The
boxplots show that their approach (5) slightly outperforms
Turhan\etal's $k$-NN approach (4). They also performed a Mann-Whitney-U test 
between their suggested \ac{CPDP} procedure and the $k$-NN approach. Here,
they determined that in most cases their approach performs significantly better
or equal to the $k$-NN approach, and only is worse for four, seven,
respectively five proprietary products, for the Na\"{i}ve Bayes, Logistic
Regression, respectively Random Forest classifier. Regarding the number of
metrics that should be removed by the feature selection, the authors determined
that removing 60\%-80\% of the metrics yields the biggest improvement for \ac{CPDP}. In terms
of the overall performance, their suggested approach is slightly below the
\ac{WPDP} 5x5 cross-validation model (2). Both the NN approach and the approach
suggest in this article clearly outperform using all data (4).

\subsection{Nam\etal, 2013}
\label{sec:nam2013}
\noindent\textbf{Approach.} 
\cite{Nam2013} propose to use \ac{TCA} \citep{Pan2011}, a technique with
the goal to learn a transformation $\phi$ that minimizes the distances between
domains of the training data and target product, while maximizing the
variance between the data. In their work, they use a linear transformation for $\phi$. This
linear transformation is then applied to each entry in the training data before
training, thereby mapping the training data to the domain of the target data. 

Moreover, the authors propose to standardize the data before applying \ac{TCA}.
They consider min-max normalization (see Section~\ref{sec:herbold2013}), z-score
standardization (see Section~\ref{sec:uchigaki2012}), z-score-training
standardization defined as
\begin{equation}
\hat{m}_i(s) = \frac{m_i(s)-mean(m_i(S^{candidate}))}{std(m_i(S^{candidate}))},
\end{equation}
and z-score-target standardization defined as 
\begin{equation}
\hat{m}_i(s) = \frac{m_i(s)-mean(m_i(S^*))}{std(m_i(S^*))}.
\end{equation}

Additionally, the authors propose \emph{TCA+}. The aim of TCA+ is the automated
selection of the best strategy for standardization before applying TCA,
depending on how different data sets are. To determine the differences between data sets,
the authors define the \ac{DCV}. The \ac{DCV} consists of the mean, median, min,
max, and standard deviation of the distances between the entities within the
data sets. Based on the \ac{DCV}, the authors define conditions, which state that
if the distributional characteristics in the \ac{DCV} are much more, more,
slightly more, same, slightly less, less, or much less. TCA+ defines
five rules that state standardization should be used: one for each of
the four standardization approaches above and a fifth one that defines when no
normalization shall be applied.

\noindent\textbf{Approach type.} Other (instance standardization).

\noindent\textbf{Classifier.} Logistic Regression. 

\noindent\textbf{Data.} All five products of the AEEEM data without context
factors and all three products of the RELINK data.

\noindent\textbf{Case study setup.} The authors used seven experiment
configurations: (1) \ac{CPDP} with no data transformations; (2) \ac{CPDP} with
TCA without standardization; (3) \ac{CPDP} with TCA and min-max normalization;
(4) \ac{CPDP} with TCA and z-score standardization; (5) \ac{CPDP} with  TCA and
z-score-training standardization; (6) \ac{CPDP} with TCA and z-score-target
standardization; (7) \ac{CPDP} with TCA+; and (8) \ac{WPDP} with 100
random 50:50 splits of the data.
The \ac{CPDP} was performed  pair-wise between products of the same data
set.   

\noindent\textbf{Performance measures.} \emph{F-measure}.

\noindent\textbf{Results.} The mean performances in terms of the
\emph{F-measure} over all pair-wise \acp{CPDP} on
the AEEEM are 0.31 without any transformation (1), 0.14 for TCA
without standardization (2), 0.24 for TCA with min-max normalization (3), 0.41 for
TCA with z-score standardization (4), 0.38 for TCA with z-score-training
standardization (5), 0.38 for TCA with z-score-target standardization (6), and 0.41
with TCA+ (7). The \ac{WPDP} (8) has a mean performance of
0.42. On the RELINK data, the mean performances are 0.49 without
any transformation (1), 0.45 for TCA without standardization (2), 0.44 for TCA
with min-max normalization (3), 0.57 for TCA with z-score standardization (4),
0.49 for TCA with z-score-training standardization (5), 0.59 for TCA with
z-score-target standardization (6), and 0.61 with TCA+ (7). The
\ac{WPDP} (8) has a mean performance of 0.53.
The authors conclude that the z-score standardization provides a significant
improvement in comparison to no standardization. They note that overall,
TCA+ performs a bit better than always using the same standardization and that
the performance of TCA+ is comparable to the performance of the within-project
predictions.

\subsection{Panichella\etal, 2014}
\label{sec:panichella2014}
\noindent\textbf{Approach.} \cite{Panichella2014}
propose to create an ensemble classifier from different base classifiers.
Through a \ac{PCA}, the authors determined that different classifiers
can complement each other, because they are well-suited to detect defects in
different parts of the data. Concretely, they propose to train a set of
classifiers $h_1, \ldots, h_n$ for all $S^i \in S^{cand}$. Using the prediction
results $h_i(s)~\forall~s \in S^i, i=1,\ldots, n$ as input, an \ac{CODEP}
classifier is trained, e.g., a Logistic Regression over the results of the classifiers $h_1,
\ldots, h_n$. Hence, the \ac{CODEP} classifier is not trained directly on the
available metric data, but only indirectly since it is based on results by
classifiers trained on the metric data. 

\noindent\textbf{Approach type.} Other (classification model).

\noindent\textbf{Classifier.} \ac{CODEP} with Alternating Decision Trees,
Bayesian Networks, Decision Tables, Logistic Regresion, Multilayer Perceptron,
and \ac{RBF} Networks as classifiers $h_i$ and Logistic Regression and
Bayesian Networks as CODEP classifiers.

\noindent\textbf{Data.} 10 products of the JURECZKO data.

\noindent\textbf{Case study setup.} Within their case study, the authors compare
the performance of using the six classifiers above by themselves to the
performance of the two CODEP classifiers. The authors use
all data except the data from the target product as training data. 

\noindent\textbf{Performance measures.} \emph{AUC} and \emph{AUCEC} after
\cite{Rahman2012} (see Section~\ref{sec:rahman2012}).

\noindent\textbf{Results.} The authors report the \emph{AUC} and
\emph{AUCEC} values for all products and all classifiers, as well as the mean
performance achieved. For \emph{AUC}, the Logistic Regression model achieves a
performance of 0.76, the Bayesian Network of 0.45, the \ac{RBF}
Network of 0.72, the Multilayer Perceptron of 0.76, the
Alternating Decision Tree of 0.72, the Decision Table of 0.70, the Logistic
Regression \ac{CODEP} of 0.82, and the Bayesian Network \ac{CODEP} of 0.86. 
For \emph{AUCEC}, the Logistic Regression model achieves a
performance of 0.45, the Bayesian Network of 0.48, the \ac{RBF}
Network of 0.42, the Multilayer Perceptron of 0.45, the
Alternating Decision Tree of 0.47, the Decision Table of 0.49, the Logistic
Regression \ac{CODEP} of 0.54, and the Bayesian Network \ac{CODEP} of 0.55. The
authors conclude that both CODEP models outperform normal classifiers and yield
an advantage both in terms of \emph{AUC} and \emph{AUCEC}.

\subsection{F. Zhang\etal, 2014, 2015a}
\label{sec:zhang2014}
\noindent\textbf{Approach.} In an article by \cite{Zhang2014},
later extended in \cite{Zhang2015}, the authors propose clustering based on the context factors of a product, e.g., the
programming language, whether an \ac{ITS} was used, and the number of developers
in a product to perform rank transformation of the data. First, they create
groups of products by binning the products according to their product factors. Then, they use Cliff's $\delta$
\citep{Cliff1993}, a measure for the effect size in case of stastical
significant differences. If they determine a large effect size in the difference between two
product groups, they are assigned to different clusters. If the difference is
only small, they are merged into a single product group. This clustering is
performed for every metric $m_i$ on its own and, consequently, the
clusters are different for each metric. Once the
clusters are created, the authors propose a rank transformation  of the metric
values.
The transformation is based on the deciles of the metric values in a cluster as follows:
\begin{equation}
\hat{m}(i) =
\begin{cases}
 1 & \text{if}~m(i) \leq dec_1(cl(S,i)), \\
 k & \text{if}~m(i) \in (dec_{k-1}(cl(S,i)),dec_k(cl(S,i))], \\
10 & \text{if}~m(i) > dec_9(cl(S,i)),
\end{cases}
\end{equation}
where $dec_j$ is the $j$-th decile of $cl(S,i)$ for the product $S$ and the
metric $m_i$. To train a defect prediction model they use the transformed
metric values from all products. 

\noindent\textbf{Approach type.} Other (instance standardization and metric
type influence).

\noindent\textbf{Classifier.} Na\"{i}ve Bayes.   

\noindent\textbf{Data.} All 1385 products of the MOCKUS data and all
five products of the AEEEM data extended with context factors. 

\noindent\textbf{Case study setup.} The authors used nine experiment
configurations: (1) \ac{WPDP} with their rank transformation; (2) \ac{WPDP} with
log transformation (logarithm of all data is used, see
Section~\ref{sec:turhan2009}); (3) \ac{CPDP} with their rank transformation and
only product metrics; (4) \ac{CPDP} with their rank transformation and
only product and process metrics; (5) \ac{CPDP} with their rank transformation and
only product metrics and context factors; (6) \ac{CPDP} with their rank
transformation and only process metrics and context factors; (7) \ac{CPDP} with
their rank transformation with all metrics and context factors; (8) \ac{CPDP}
with their rank transformation on foreign data; and (9)
\ac{WPDP} on the foreign data. For configurations (1)--(2), 10x10 cross valididation is
used. For configurations (3)--(7) the authors do 10x10 cross validation with
10\% of the products as hold-out data. The configurations (1)--(7) use only the MOCKUS data.
The AEEEM data is used as foreign data for configurations (8) and (9).
For configuration (8), the authors train the prediction model on the MOCKUS data
and predict defects on the AEEEM data, configuration (9) is 10x10 cross
validation on the AEEEM data. 

\noindent\textbf{Performance measures.} \emph{recall}, \emph{precision},
\emph{pf}, \emph{F-measure}, \emph{G-measure}, \emph{MCC}, and \emph{AUC}.

\noindent\textbf{Results.} The authors report the mean values achieved
by the configurations. The \ac{WPDP} with rank transformation (1) has a mean
performance of  0.580 \emph{recall}, 0.525
\emph{precision}, 0.359 \emph{pf}, 0.534 \emph{F-measure}, 0.521
\emph{G-measure}, 0.214 \emph{MCC}, and 0.615 \emph{AUC}. The \ac{WPDP} with log transformation (2) has a mean
performance of 0.576 \emph{recall}, 0.519 \emph{precision}, 0.369 \emph{pf},
0.527 \emph{F-measure}, 0.511 \emph{G-measure}, 0.202 \emph{MCC}, and 0.609
\emph{AUC}. Due to the small and statistically not significant differences, the
authors determine that rank transformation and log tranformation perform similar
for \ac{WPDP}. 
For the \ac{CPDP} (3)--(7), the configuration with all metrics (7) yields the best results with a mean
performance of 0.591 \emph{recall}, 0.455 \emph{precision},
0.396 \emph{pf}, 0.510 \emph{F-measure}, 0.594 \emph{G-measure}, 0.186
\emph{MCC}, and 0.641 \emph{AUC}. Overall, they conclude that they achieve a
similar predictive performance as the \ac{WPDP} model. 
For the evaluation with data from another data set (8), the authors report a
mean performance of 0.888 \emph{recall}, 0.256 \emph{precision}, 0.653
\emph{pf}, 0.370 \emph{F-measure}, 0.473 \emph{G-measure}, 0.203 \emph{MCC}, and
0.741 \emph{AUC}. The \ac{WPDP} achieves a mean performance of 0.659
\emph{recall}, 0.323 \emph{precision}, 0.319, \emph{pf}, 0.420 \emph{F-measure},
0.651 \emph{G-measure}, 0.273 \emph{MCC}, and 0.734 \emph{AUC}. They observe a
lower \emph{precision}, \emph{F-measure}, \emph{G-measure}, and \emph{MCC}, as
well as a larger \emph{pf} for the \ac{CPDP} model. The only metrics where the
\ac{CPDP} model outperforms the \ac{WPDP} model are the \emph{recall}
(significantly) and \emph{AUC} (very small difference).

\subsection{Fukushima\etal, 2014, Kamei\etal 2015}
\label{sec:fukushima2014}
\noindent\textbf{Approach.} In an
article by \cite{Fukushima2014} and extended by
Kamei\etal~\cite{Kamei2015}, the authors study how well \ac{CPDP} is suited to
the prediction of defective changes, also known as \ac{JIT} defect prediction. For the selection of similar products from a set of
candidates for relevancy filtering, the authors propose a procedure based on the
Spearmean correlation~\citep{Kendall1970}. First, the three metrics $m_{i_1},
m_{i_2}, m_{i_3}$ with $i_1, i_2, i_3 \in \{1, \ldots, d\}$ with the highest correlation
between a candidate product $S$ and the target product are selected for each candidate product $S
\in S^{cand}$.
Then, the pair-wise correlations within a product $S$ are computed and used to
define a three-dimensional internal correlation vector, i.e.,
\begin{equation}
intcor_S = (cor(m_{i_1}, m_{i_2}), cor(m_{i_1}, m_{i_3}),
cor(m_{i_2}, m_{i_3})).
\end{equation}
The euclidean distance between the internal correlation vectors for a candidate
product and the target product is used to define the distance between products. The product with
the smallest distance is used as training data, i.e.,
\begin{equation}
S^{train} = \arg\min_{S \in S^{cand}} dist(intcor_S,
intcor_{S^*}).
\end{equation}
Furthermore, the authors explore the use of context factors for relevancy
filtering, i.e., of the organization that develops the product, the intended
audiance, the user interface, whether the product uses a database, and the
programming languages. Using the context factors, the authors define a vector of binary variables, that define whether a context factor is fulfilled by a product or
not, e.g., if a product is implemented by Eclipse, or if a product is written in
C++. They then use the euclidean distance between these context vectors to
define the similarity between products and propose to use the most similar for
training. 

In addition to using just the most similar product, the authors also propose to
use the $k$ most similar products together. Moreover, the authors propose a
sampling-based approach that favors data from more similar products. They
propose to sample $\frac{10-(r-1)}{10}$ percent of the data of a
product\footnote{The number 10 results from the number of products available
for training and should be changed when adopting this approach.}, whereas $r$ is
the rank of a product according to their distance measure. Hence, all of the
data of the most similar product is used, 90\% of the data of the second most similar,
and so on.

Additionally, the authors consider bagging for the prediction, i.e., training a
separate model for each product in the data set, same as \cite{He2013}.
In addition to normal bagging, where the vote of the classifier trained for each
product is worth the same, the authors propose a weighted voting scheme, which
is analogous to the sampling described above, i.e., the vote of the most similar
products is counted fully, the vote of the second most similar product only to
90\%, and so on.

For all of the above, the authors propose to use undersampling to deal with a
potential bias in the data. 

\noindent\textbf{Approach type.} Relevancy filtering, stratification and other
(classification model).

\noindent\textbf{Classifier.} Random Forest.

\noindent\textbf{Data.} The authors use data about the changes of eleven open
source large-scale open source products. The data contains nine product and
process metrics. The authors do not take the data as is, but perform an analysis
for collinearity using Spearman's correlation coefficient. Due to this, the
authors dropped one metric which was highly correlated and adopted
the normalization approach from \cite{Nagappan2006} to merge three
correlated metrics into two uncorrelated metrics. This means that in total six metrics are
used.
The data is not publicly available online. 

\noindent\textbf{Case study setup.} 
The authors use 13 different setups experiment configurations: (1) \ac{WPDP}
with 10x10 cross-validation; (2) pair-wise \ac{CPDP} between with all possible combinations for defect prediction; (3)
\ac{CPDP} with the most similar product using their Spearman correlation
procedure (4) \ac{CPDP} using the context factors as similarity
measure; (5) \ac{CPDP} using all data from other products for
training; (6) \ac{CPDP} using the three most similar products for
training; (7) \ac{CPDP} using the five most similar products for
training; (8) \ac{CPDP} using the similarity-based sampling approach proposed;
(9) \ac{CPDP} with rank transformations as proposed by Zhang\etal, 2014
(see Section~\ref{sec:zhang2014}); (10) \ac{CPDP} with bagging; 
(11) \ac{CPDP} with bagging and the similarity-based voting scheme; (12)
\ac{WPDP} same as (1), but without undersampling the data; and (13)
pair-wise \ac{CPDP} same as (2), but without undersampling the data. 

\noindent\textbf{Performance measures.} \emph{recall}, \emph{precision},
\emph{F-measure}, and \emph{AUC}.

\noindent\textbf{Results.} The authors report all values for all
performance metrics for the \ac{WPDP} with 10x10 cross validation (1) and the
pair-wise \ac{CPDP} (2).
The \ac{WPDP} (1) achieves a mean performance of 0.669 \emph{recall},
0.423 \emph{precision}, 0.502 \emph{F-measure}, and 0.778 \emph{AUC}.
The pair-wise \ac{CPDP} (2), the achieves a mean performance of 0.585
\emph{recall}, 0.353 \emph{precision}, 0.404 \emph{F-measure}, and
0.67 \emph{AUC}.
For the approaches (3)--(11) the exact results are not reported.
Instead, bean plots of the percentage of \emph{AUC} achieved in
comparison to the \ac{WPDP} are given. From their results, the authors conclude
that similarity measured using context factors (4) yields better results than
similarity measured using the Spearman correlation procedure (3), but both
perform statistically significantly worse than within-project prediction. In
contrast, the results when using multiple products together for training
(5)--(8) are not significantly different from the \ac{WPDP} results and,
thereby, provide a strong advantage over simple \ac{CPDP}.
Moreover, the authors observe that simply using all products (5) is only
rarely outperformed by the more sophisticated relevancy filters
(6)--(8).
Furthermore, they determined that rank transformations (9) do not perform well
in the \ac{JIT} setting. Additionally, they determined that the bagging-based
approaches (10)--(11) do not perform statistically significantly different than
\ac{WPDP} and, thereby, provide an advantage over simple \ac{CPDP}. 
For the evaluation of \ac{WPDP} and pair-wise \ac{CPDP} without
undersampling (12)--(13), the authors report all values for all performance
metrics. The \ac{WPDP} without undersampling (12) achieves a mean
performance of 0.303 \emph{recall}, 0.645 \emph{precision},
0.385 \emph{F-measure}, and 0.780 \emph{AUC}. The pair-wise \ac{CPDP} without
undersampling (13) achieves a mean performance of 
0.204 \emph{recall}, 0.523 \emph{precision}, 0.234 \emph{F-measure}, and 0.7
\emph{AUC}. From these results, the authors conclude that the undersampling does
not impact the \emph{AUC}, but makes a difference for \emph{recall} and
\emph{F-measure}, which are both significantly better with undersampling.

\subsection{Mizuno and Hirata, 2014}
\label{sec:mizuno2014}
\noindent\textbf{Approach.} \cite{Mizuno2014} proposed a text-based
approach for \ac{CPDP}, inspired by spam filtering. Their first step is to separate the code
lines from the comment lines. Moreover, the authors distinguish between
end-of-line comments (e.g., //), block comments (e.g., /* */), and documentation
comments (e.g., /** */). Then, they reduce the source code to its basic
structure by removing all comments and whitespaces and replacing all identifiers
and numbers with I, strings with S and characters with C. Each line and comment
is then considered to be a token for the classification. 

\noindent\textbf{Approach type.} Other (classification model and metric type
influence).

\noindent\textbf{Classifier.} Text classification.

\noindent\textbf{Data.} The authors use 25 products from the JURECZKO data
and three products from the ECLIPSE data. Only the defect information of those
data sets is used, the metrics were discarded and instead the source code was mined
and tokenized by the authors themselves. The tokenized data is not publicly
available.

\noindent\textbf{Case study setup.} The authors use three experiment
configurations: (1) pair-wise \ac{CPDP} between all products from different
products; (2) \ac{CPDP} with all products from different products as training
data; (3) \ac{WPDP} with all data from older versions of the product as
training data. 

\noindent\textbf{Performance measures.} \emph{recall},
\emph{precision}, \emph{F-measure}, \emph{accuracy}. 

\noindent\textbf{Results.} The authors report the mean performance over
all products per product for five of the eight products and, thereby for 21 of
the 28 products. For the other seven products no
results are reported. The mean performance of pair-wise \ac{CPDP} (1) over these
products is 0.298 \emph{recall}, 0.581 \emph{precision}, 0.310 \emph{F-measure},
and 0.362 \emph{accuracy}. For \ac{CPDP} (2), the mean performance is 0.580
\emph{recall}, 0.465 \emph{precision}, 0.427 \emph{F-measure}, and 0.656
\emph{accuracy}. For \ac{WPDP} (3), the mean performance is 0.431 \emph{recall},
0.527 \emph{precision}, 0.349 \emph{F-measure}, and 0.676 \emph{accuracy}. 
The authors conclude that the \ac{CPDP} is better than \ac{WPDP} in terms of
\emph{recall}, but worse in terms of \emph{precision}. Moreover, they determine
that the source code tokens and end-of-line comment tokens are best suited for
defect prediction.

\subsection{Ryu\etal, 2014}
\label{sec:ryu2014}

\noindent\textbf{Approach.} \cite{Ryu2014}\footnote{Online first
publication December 2014.} propose to use data weighting in combination with
boosting. For the data weighting, they propose to use the data gravitation
approach by \cite{Ma2012} (see Section~\ref{sec:ma2012}). Within
this work, the gravity weights are the similarity weights. They do not affect
the classification or training process and are, hence, not related to the weights generated for boosting.
Then, 50\% of the training data are drawn randomly and 20\% of the
instances with the highest similarity weights are used as internal hold-out data
for model validation. Once the weights are determined and the validation data is
picked, the training loop starts. The first step of the loop is to resample
the training data in order to treat a potential bias. The resampling is done
through a combination of oversampling for data with high similarity weigths and
undersampling for data with low similarity weights. Then, a \ac{SVM} is trained
using the remaining training data. Using AdaBoost~\citep{Freund1997}, the
weights of all entities $s \in S^{train}$ that violate the classification are increased in order to get a
classifer that does not repeat the mistakes. The overall classification is the
weighted mean of the \acp{SVM} trained in each loop iteration. Prior to all
of the above, the authors propose to use z-score standardization (see
Section~\ref{sec:uchigaki2012}).

\noindent\textbf{Approach type.} Instance weighting and other (classification
model and instance standardization).

\noindent\textbf{Classifier.} \ac{SVM} with a linear kernel which is internally
used for the boosting; C4.5 Decision Tree,
$k$-NN, Logistic Regression, Multilayer Perceptron, Na\"{i}ve Bayes, PART
Decision List, and Random Forest for comparison. 

\noindent\textbf{Data.} Seven products from the NASA data and all three
products from the SOFTLAB data. 

\noindent\textbf{Case study setup.} The authors use four experiment
configurations: (1) their approach; (2) a \ac{SVM} with normal boosting; (3)
Na\"{i}ve Bayes with data gravitation as proposed by \cite{Ma2012}; and
(4) the above classifiers as comparison to non-boosting techniques. As
training data, the authors used all of the NASA products as training data for
predictions of the SOFTLAB data. Moreover, the authors use each product
in the NASA data once as target product with all other products from the
NASA data for training. The SOFTLAB data is not used for the training of predictors. 

\noindent\textbf{Performance measures.} \emph{recall}, \emph{pf}, \emph{AUC}, and
\emph{H-measure}. 

\noindent\textbf{Results.} The authors report the values for
configurations (1), (2), and (4) for all performance metrics. For configuration
(3) only \emph{AUC} is fully reported, for \emph{recall} and \emph{pf} only the
results on the SOFTLAB data are reported and no values for \emph{H-measure}.
Hence, we only summarize the result for \emph{AUC} for configuration (3). For
their approach (1), the authors report a mean performance of 0.581
\emph{recall},
0.306 \emph{pf}, 0.753 \emph{AUC}, and 0.335 \emph{H-measure}. For the
\ac{SVM} with normal boosting (2), the authors report a mean performance of 
0.337 \emph{recall}, 0.095 \emph{pf}, 0.744 \emph{AUC}, and 0.332
\emph{H-measure}. For Na\"{i}ve Bayes with data gravition (3), the authors
report a mean performance of 0.691 \emph{AUC}. Of the classifiers used for
comparison (4), Logistic Regression yields the best performance. The authors
report a mean performance of 0.364 \emph{recall}, 0.128 \emph{pf},
0.784 \emph{AUC}, and 0.361 \emph{H-measure}. Using the
A-statistic~\citep{Vargha2000, Arcuri2011} the authors compared their approach
to the others. On the one hand, the authors determined that their approach is outperformed in terms of \emph{pf} by the competitors. On the other hand, their
approach outperforms all competitors in terms of \emph{recall}. In terms of
\emph{AUC} and \emph{H-measure}, their approach outperforms all others except
Logistic Regression, which is similar. 

\subsection{P. He\etal, 2015}
\label{sec:he2015}
\noindent\textbf{Approach.} \cite{He2015} investigate strategies for the
selection of metrics. To this aim, they define the $TOP_k$ metrics as
the $k$ most-often used metrics by trained defect predictors. For example, if a
separate classifier is trained for each product in a data set, the authors count
how often a metric $m_i \in M$ is used. This means that
their approach can only be applied, after defect predictors were already trained
and that their approach utilizes knowledge about the prediction results. 
In order to determine how many metrics should be selected, i.e., a good value
for $k$, the authors introduce the notion of coverage
\begin{equation}
coverage(TOP_k) = \frac{1}{|S^{cand}|} \sum_{S \in S^{cand}}
\frac{|CFS_S \cap TOP_k|}{|CFS_S \cup TOP_k|},
\end{equation}
where $CFS_S$ are the metrics selected by the \ac{CFS}~\citep{Hall1998}
approach.

Additionally, the authors propose to optimize the $TOP_k$ by taking correlations
between the metrics into account. Concretely, they determine the $coverage$ for
all subsets of $TOP_k$ and pick the one with the best coverage, such that it
does not contain strongly correlated metrics, i.e.,
\begin{equation}
\begin{split}
OPTTOP_k =& \arg\max_{M' \subset TOP_k} \{coverage(M'): \\&\not\exists~m, m'
\in M', m\neq m' | cor(m,m')>\phi\}
\end{split} 
\end{equation}
with $\phi$ a threshold that defines when two metrics are strongly correlated.
Furthermore, the authors propose to use log transformation for all metrics
before creating prediction models.  

\noindent\textbf{Approach type.} Other (feature selection).

\noindent\textbf{Classifier.} Bayesian Network, C4.5 Decision Tree, Decision
Table, Logistic Regression, Na\"{i}ve Bayes, and \ac{SVM} with a polynomial
kernel.

\noindent\textbf{Data.} 34 products from the JURECZKO data.

\noindent\textbf{Case study setup.} The authors use three
experiment configurations: (1) \ac{WPDP} with the latest
version before the target product as training data; (2) \ac{WPDP} with all
prior versions of the target product as training data; (3) selection of the best
subset of three products for \ac{CPDP} using the brute-force strategy and
knowledge about the prediction performance on the target product as proposed by
He\etal~\citep{He2012} (see Section~\ref{sec:he2012}). All three
configurations were used with all classifiers, as well as all metrics, $TOP_5$,
$OPTTOP_5$, and $CFS_S$, max-relevance selction~\citep{Peng2005} and
minimal-redundancy-maximal-relevance~\citep{Peng2005}. To determine $OPTTOP_5$,
the authors use $\phi=0.6$ as threshold for strong correlation.

\noindent\textbf{Performance measures.} \emph{recall}, \emph{precision}, 
\emph{F-measure}, $succ_{0.7, 0.5}$ and \emph{consistency}.

\noindent\textbf{Results.} The results are evaluated mainly visually
using box-plots supported by the median of some results. 
As $TOP_5$ metrics of the JURECZKO data, the
authors determine \ac{CBO}, \ac{RFC}, \ac{LCOM}, \ac{Ce}, and \ac{LOC} (see
\cite{Jureczko2010}). The results of the case study show that the
$TOP_5$ metric set is comparable to the full metric set, in the sense that it
yields results that are within 90\% of the original performance. This finding
holds true for all three configurations and all classifiers. In comparison to
the existing approaches, i.e., $CFS_S$, max-relevance, and
minimal-redundancy-maximal-relevance, the authors determine that $TOP_5$ yields
a better prediction performance. Regarding the classifiers, Na\"{i}ve Bayes is
the best classifier for \ac{WPDP}, i.e., configurations (1) and (2), Decision
Table for \ac{CPDP}, i.e., configuration (3). This finding holds true with and
without the metric selection. Through their optimization, the authors determine
CBO, LCOM, and LOC as $OPTTOP_5$, because they are not strongly correlated and
form the subset of $TOP_5$ with the best $coverage$. In terms of prediction
performance $OPTTOP_5$ is determined to be close to $TOP_5$ in terms of
\emph{precision}, as well as predictions that are successfull according to
$succ_{0.7, 0.5}$. For \ac{WPDP} (1)--(2), $TOP_5$ outperforms $OPTTOP_5$ in
terms of \emph{recall} and \emph{F-measure}. For \ac{CPDP} (3), the differences
between $TOP_5$ and $OPTTOP_5$ are not stastically significant. Using
\emph{consistency} and an \ac{ANOVA} analysis, the authors determine
that the $TOP_5$ metrics are consistent for different classifiers and not classifier specific. 

\subsection{Peters\etal, 2015}
\label{sec:menzies2015}
\noindent\textbf{Approach.} \cite{Peters2015} suggested LACE2 is an
extension of CLIFF+MORPH~\citep{Peters2013a} (see
Section~\ref{sec:peters2013privacy}) that accounts for multi-party data sharing instead of single party data sharing,
i.e., multiple data owners take their turns to add data, then hand-off the data
to the next data owner, so that they can add data. To facilitate sharing in a
multi-owner environment, the authors propose to only add data, that is not
yet represented in the data set with a strategy that adopts the Leader-Follower
algorithm \citep{Duda2012} called \emph{LeaF}. 
To determine if an entity $s \in S$ is already represented in the privatized
data $S^{privat}$, 100 entities $S^{100} \subset S^{privat}$ are sampled. Then, the median of the distances
to their nearest unlike neighbors (see Section~\ref{sec:peters2013privacy}) is
calculated. An entity $s$ is added to the data, if the distance to its
nearest unlike neighbor is greater than the median of the 100 samples. The
complete LACE2 approach first applies CLIFF for filtering, then LeaF to
determine which new entities add information to the data, and finally MORPH to
privatize the data. This procedure is repeated by each data owner in turn, to
create the overall body of knowledge. 

\noindent\textbf{Approach type.} Relevancy filtering and other (data privacy).

\noindent\textbf{Classifier.} $k$-NN.

\noindent\textbf{Data.} Not applicable, since no experiment configurations
addresses \ac{CPDP}.

\noindent\textbf{Case study setup.} The authors do not perform any experiments on
the performance of \ac{CPDP}, all executed scenarios in the case study consider
only the \ac{WPDP} scenario with cross-validation.

\noindent\textbf{Performance measures.} Not applicable, since no experiment
configuration addresses \ac{CPDP}.

\noindent\textbf{Results.} Not applicable for prediction performance.
Regarding privacy, LACE2 outperforms CLIFF+MORPH, but both are on a very high
level.

\subsection{Chen\etal, 2015}
\label{sec:chen2015}

\noindent\textbf{Approach.} \cite{Chen2015} propose an approach
for \ac{MPDP} called \ac{DTB}.
They propose to apply $k$-NN filtering to select similar
instances~\cite{Turhan2009} (see Section~\ref{sec:turhan2009}), and
then to use oversampling with SMOTE~\citep{Chawla2002} to treat a potential bias
in the data. Following that, they propose to determine weights for all instances
using the data gravitation approach proposed by \cite{Ma2012} (see
Section~\ref{sec:ma2012}). Finally, they propose to use transfer boosting based
on TrAdaBoost~\citep{Dai2007}. TrAdaBoost works with two data sets: a small data
set from the same domain as the test data and a larger data set that might not
be perfectly suited for the target domain. The boosting favors data from the
small set by assigning higher weights, and assigns very low weights to data from
the larger set that contradicts the current hypothesis, i.e., noisy data. In the
considered scenario, the small data set are 10\% of the data from the target
product $S^*$ and the large data set are the candidate products $S^{cand}$.

\noindent\textbf{Approach type.} Instance weighting and other (classification
model).

\noindent\textbf{Classifier.} Logistic Regression, Na\"{i}ve Bayes, and Random
Forest.

\noindent\textbf{Data.} 15 products from the JURECZKO data.

\noindent\textbf{Case study setup.} The authors use seven experiment
configurations: (1) \ac{MPDP} with their proposed approach; (2) \ac{CPDP} with
all cross-project data without further treatment; (3) \ac{CPDP} with $k$-NN
relevancy filter after \cite{Turhan2009} (see Section~\ref{sec:turhan2009}; (4) \ac{MPDP}
as proposed with $k$-NN relevancy filtering after \cite{Turhan2013}
(see \ref{sec:turhan2013}); (5) \ac{CPDP} with data gravitation after
\cite{Ma2012} (see Section~\ref{sec:ma2012}); (6) \ac{WPDP} with 10\% of
the data for training, the rest as test data; and (7) \ac{WPDP} with 60\% of the
data for training, the rest as test data. Na\"{i}ve Bayes is used as base
classifier for configurations (1)--(5). For these configurations, the authors randomly draw
10\% of the entities from the target product as within-project data for training
and use all data from the other products as cross-project data. The remaining
90\% of the target product are used as test data. This is repeated 20 times and
mean values are considered. The same strategy for creating training and test
data is also followed for (6) and (7), although 60\% instead of 10\% are drawn
as training data for (7) and the authors use Logistic
Regression, Na\"{i}ve Bayes and Random Forest for these two configurations. 

\noindent\textbf{Performance measures.} \emph{recall}, \emph{pf},
\emph{G-measure}, and \emph{MCC}.

\noindent\textbf{Results.} The authors report the results for all
products and performance measures, as well as the mean performances. For
their proposed approach (1), the authors report a mean performance of 
0.702 \emph{recall}, 0.330 \emph{pf}, 0.664 \emph{G-measure}, and
0.282 \emph{MCC}.
For \ac{CPDP} without data treatment (2), the authors report a mean
performance of 0.856 \emph{recall}, 0.704 \emph{pf},
0.425 \emph{G-measure}, and 0.126 \emph{MCC}. For \ac{CPDP} with the $k$-NN
relevancy filter (3), the authors report a mean performance of
0.811 \emph{recall}, 0.604 \emph{pf}, 0.498 \emph{G-measure}, and
0.167 \emph{MCC}.
For \ac{MPDP} with the $k$-NN relevancy filter (4), the authors report a mean
performance of 0.815 \emph{recall}, 0.601 \emph{pf},
0.501 \emph{G-measure}, and 0.169 \emph{MCC}. For \ac{CPDP} with data
gravitation (4), the authors report a mean performance of
0.644 \emph{recall}, of 0.350 \emph{pf}, of 0.615 \emph{G-measure}, and
0.226 \emph{MCC}. Using the Mann-Whithney-U test, the authors determine that
their approach is significantly better than the competitors in terms of the
\emph{G-measure}. In terms of \emph{MCC}, their model outperforms all models
except data gravitation, which is not statistically significantly different.
For the comparison with normal \ac{WPDP}, i.e., of their approach (1) with
configurations (6) and (7), the authors report that their approach outperforms
all classifiers when only 10\% of the data is used for training (6).
Moreover, the proposed approach outperforms Logistic Regression and Random
Forest when 60\% of the data is used (7) and is similar to Na\"{i}ve Bayes.

\subsection{Kawata\etal, 2015}
\label{sec:kawata2015}
\noindent\textbf{Approach.} 
\cite{Kawata2015} proposed a relevancy filter based on
\ac{DBSCAN}~\citep{Ester1996}. \ac{DBSCAN} 
determines regions of high density in the data. Kawata\etal~propose to combine
the metric data of the candidate products and the target product, i.e.,
$M(S^{cand}\bigcap S^*)$ and then apply the DBSCAN algorithm to this data. As training data, all entities are selected that
fall into the same cluster as any instance in the target data. Or vice versa:
all data, that does not fall into the same cluster as at least one instance of
the target product is discarded. 

\noindent\textbf{Approach type.} Relevancy filtering.

\noindent\textbf{Classifier.} 1-NN, Logistic Regression Logistic
Regression, Na\"{i}ve Bayes, and Random Forests.

\noindent\textbf{Data.} 56 products from the JURECZKO data. 

\noindent\textbf{Case study setup.}  The authors use four experiment configurations:
(1) \ac{CPDP} with the DBSCAN relevancy filter as described above; (2)
\ac{CPDP} with the $k$-NN relevancy filter proposed by \cite{Turhan2009} (see
Section~\ref{sec:turhan2009}); (3) \ac{CPDP} with the filtering approach
proposed by \cite{Peters2013b} (see
Section~\ref{sec:peters2013filter}); and (4) \ac{CPDP} with no relevancy
filtering. The authors used products with less than 100 entities as target
products, and all products with more than 100 products together as candidate data.

\noindent\textbf{Performance measures.} \emph{recall}, \emph{precision},
\emph{F-measure}, \emph{G-measure}, and \emph{AUC}.

\noindent\textbf{Results.} The authors report the median performance 
over all products for all results. Of the used classifiers, Na\"{i}ve Bayes
performs best. Therefore, we only repeat the results for this classifier in the following. The DBSCAN relevancy filter (1) achieves
a median performance of 0.500 \emph{recall}, 0.353 \emph{precision}, 0.444
\emph{F-measure}, 0.572 \emph{G-measure}, and 0.624 \emph{AUC}. The $k$-NN
relevancy filter (2) achieves a median performance of 0.462 \emph{recall}, 0.600
\emph{precision}, 0.462 \emph{F-measure}, 0.558 \emph{G-measure}, and 0.624
\emph{AUC}. The relevancy filter by Peters\etal~(3) achieves a median
performance of 0.600 \emph{recall}, 0.474 \emph{precision}, 0.471
\emph{F-measure}, 0.544 \emph{G-measure}, and 0.612 \emph{AUC}. \ac{CPDP}
without relevancy filtering (4) achieves a median performance of
0.077 \emph{recall}, 0.600 \emph{precision}, 0.143 \emph{F-measure}, 0.143
\emph{G-measure}, and 0.533 \emph{AUC}. The authors conclude that their
approach is comparable to the competitors. 

\subsection{Y. Zhang\etal, 2015b}
\label{sec:zhang2015comb}
\noindent\textbf{Approach.} \cite{Zhang2015a}
compare different ensemble predictors for \ac{CPDP}. Using the success of
\ac{CODEP}~\citep{Panichella2014} (see Section~\ref{sec:panichella2014}) as
motivation, the authors set out to compare different ensemble predictors with \ac{CODEP}. To
this aim, the authors propose the usage of Average Voting, Maximum Voting,
Bagging, Boosting, and Random Forests.

\noindent\textbf{Approach type.} Other (classification model).

\noindent\textbf{Classifier.} Average Voting, Maximum Voting, and Logistic
Regression \ac{CODEP} based on Alternating Decision Trees, Bayesian Networks,
Decision Tables, Logistic Regresion, Multilayer Perceptron, and \ac{RBF}
Networks; Bagging based on C4.5 Decision Trees; Bagging based on Na\"{i}ve
Bayes; Boosting based on C4.5 Decision Trees; Boosting based on Na\"{i}ve
Bayes; and Random Forest.

\noindent\textbf{Data.} Ten products of the JURECZKO data.

\noindent\textbf{Case study setup.} The authors use one experiement configuration
in which they use all products for training except that target product. Using
this configuration, the authors compare the performance of eight ensemble
classifiers listed above. 

\noindent\textbf{Performance measures.} \emph{F-measure} and $NofB_{20\%}$.

\noindent\textbf{Results.} The authors report the \emph{F-measure} and
$NofB_{20\%}$ values for all products and all classifiers, as well as
the mean performance achieved. For \emph{F-measure}, the Average Voting achieves
a mean performance of 0.299, the Maximum Voting of 0.412, the Logistic
Regression \ac{CODEP} of 0.301, the C4.5 Decision Tree Bagging of 0.245, the Na\"{i}ve
Bayes Bagging of 0.298, the C4.5 Decision Tree Boosting of 0.302, the Na\"{i}ve
Bayes Boosting of 0.298, and the Random Forest of 0.308. 
For $NofB_{20\%}$, the Average Voting achieves
a mean performance of 38.1, the Maximum Voting of 37.1, the Logistic
Regression \ac{CODEP} of 35.2, the C4.5 Decision Tree Bagging of 40.6, the
Na\"{i}ve Bayes Bagging of 34.4, the C4.5 Decision Tree Boosting of 35.4, the
Na\"{i}ve Bayes Boosting of 22.8, and the Random Forest of 37.2. The authors
conclude that Maximum Voting, C4.5 Decision Tree Bagging, and the Random Forest
outperform \ac{CODEP} both in terms of prediction performance determined by
\emph{F-measure} as well as number of bugs found determined by $NofB_{20\%}$.

\subsection{Amasaki\etal, 2015}
\label{sec:amasaki2015}
\noindent\textbf{Approach.} \cite{Amasaki2015} propose to
filter the metrics using the synonym pruning proposed for effort
prediction~\citep{Kocaguneli2013}. The general idea is to compute the distances
between all pairs of entities for all metrics $m_i$, i.e.,
$dist(m_i(s),m_i(s'))~\forall~s, s' \in S$ with $s \neq s'$. Then, all metrics
are discarded, where no instance has the closest metric value to any other
instance, i.e., only the metrics
\begin{equation}
\begin{split}
M^{red} &= \{m_i \in M: \exists~s, s' \in S, s \neq s' |\\
&dist(m_i(s),m_i(s')) = min_{m_j \in M} dist(m_j(s),m_j(s')) \}
\end{split}
\end{equation}
are kept.

The authors propose to remove outliers using a similar procedure,
i.e., to keep only instances for which any metric value is closest to any
another instance, i.e., 
\begin{equation}
\begin{split}
S^{red} &= \{s \in S: \exists~m_i \in M^{red}, s' \in S |\\
&dist(m_i(s),m_i(s'))=min_{s'' \in S} dist(m_i(s''),m_i(s')\}.
\end{split}
\end{equation}
Moreover, the authors propose to use log transformation for all metrics before
applying either of the above filters. 

\noindent\textbf{Approach type.} Outlier detection and other (feature
selection).

\noindent\textbf{Classifier.} Logistic Regression, Na\"{i}ve Bayes, Random
Forest, and \ac{SVM} with an \ac{RBF} kernel.

\noindent\textbf{Data.} 44 products from the JURECZKO data.

\noindent\textbf{Case study setup.} The authors combine the suggested metric and
instance filtering with two existing \ac{CPDP} approaches from the state of the
art, i.e., the $k$-NN instance relevancy filter proposed by
\cite{Turhan2009} (see Section~\ref{sec:turhan2009}) and the $k$-NN
product relevancy filter proposed by \cite{Herbold2013} (see
Section~\ref{sec:herbold2013}).
They compare the performance of those techniques with and without their approach for data treatment. Hence, the
authors use four experiment configurations: (1) $k$-NN
instance relevancy filter with data treatment; (2)  $k$-NN instance
relevancy filter without data treatment; (3) $k$-NN project relevancy filter
with data treatment; and (4) $k$-NN project relevancy filter without data
treatment. 

\noindent\textbf{Performance measures.} \emph{recall}, \emph{precision},
\emph{F-measure}, and \emph{AUC}.

\noindent\textbf{Results.} The authors report the median values for all
classifiers and both \ac{CPDP} approaches with and without their data treatment.
We restrict the reporting here to the result achieved with Logistic Regression,
which performed best of the classifiers, although the \ac{SVM} was a close
second. For the $k$-NN instance relevancy filter with data treatement
(1), the authors report a median performance of 0.661 \emph{recall}, 0.374
\emph{precision}, 0.489 \emph{F-measure}, and 0.638 \emph{AUC}. Without
data treatment (2), the authors report median performance of 0.689 
\emph{recall}, 0.383 \emph{precision}, 0.510 \emph{F-measure}, and 0.662 \emph{AUC}. From
this, the authors conclude that their data treatment might actually lead to
worse results if combined with the $k$-NN instance relevancy filter. For
the $k$-NN product relevancy filter (3), the authors report median performance
of 0.620 \emph{recall}, 0.424 \emph{precision}, 0.503 
\emph{F-measure}, and 0.660 \emph{AUC}. Without data treatment (4), the
authors report median performance of 0.625 \emph{recall}, 0.376
\emph{precision}, 0.494 \emph{F-measure}, and 0.639 \emph{AUC}. The authors
note a statistically significant improvement of \emph{AUC} using a
Mann-Whitney-U test. From this, the authors conclude that there might be an improvement for
this relevancy filter.

\subsection{Ryu\etal, 2015a}
\label{sec:ryu2015boosting}
\noindent\textbf{Approach.} \cite{Ryu2015a} proposed an extension of
their earlier work on the application of boosting for \ac{CPDP}~\citep{Ryu2014}
(see Section~\cite{Ryu2015a}).
The general approach is the same. There are two differences to their prior work.
First, they do not assume that an \ac{SVM} is used as internal classifier which
is boosted, but allow any classifier. Second, the way the weights for boosting
are calculated is changed, such that it can account for the costs of
misclassifications. To this aim, they introduce a cost adjustment function
$\beta$ that penalizes misclassifications of data with different distributions. This is a variant of
AdaCost~\citep{Fan1999}, which penalizes all misclassifications the same. Prior
to all of the above, the authors propose to use z-score standardization (see
Section~\ref{sec:uchigaki2012}).

\noindent\textbf{Approach type.} Instance weighting and other (classification
model and instance standardization).

\noindent\textbf{Classifier.} Na\"{i}ve Bayes, AdaCost, and
TransferBoost~\citep{Eaton2011}.

\noindent\textbf{Data.} 15 products from the JURECZKO data.

\noindent\textbf{Case study setup.} The authors use six experiment
configurations:
(1) \ac{MPDP} with their approach and Na\"{i}ve Bayes as internally boosted
algorithm; (2) \ac{CPDP} with the $k$-NN relevancy filter proposed by
\cite{Turhan2009} (see Section \ref{sec:turhan2009}) and Na\"{i}ve
Bayes as classifier; (3) \ac{CPDP} with with SMOTE~\citep{Chawla2002} treatment
and Na\"{i}ve Bayes as classifier; (4) \ac{CPDP} with Na\"{i}ve Bayes without
further data treatment; (5) \ac{CPDP} with AdaCost; and (6) \ac{CPDP} with
TransferBoost.
For all of the above, the authors choose each data set once as target product.
As training data, the authors use the data of the remaining products. For
configuration (1) the authors additionally use between 5\%, 10\%, and 25\% of
the data of the target product to enable their mixed-project approach.

\noindent\textbf{Performance measures.} \emph{recall}, \emph{pf},
\emph{G-measure}, and \emph{balance}.

\noindent\textbf{Results.} The authors report the values for products,
configurations and performance metrics, as well as the median performance
achieved over all products. The performance between using different amounts of
within-project data is similar, therefore, we summarize only the results for 5\%
within-project data here, because it is closest to strict \ac{CPDP}. For their
approach (1), the authors report a mean performance of 0.627 \emph{recall},
0.627 \emph{pf}, 0.643 \emph{G-measure}, and 0.633 \emph{balance}. For
Na\"{i}ve Bayes with $k$-NN filtering (2), the authors report a mean performance
of 0.736 \emph{recall}, 0.528 \emph{pf}, 0.560 \emph{G-measure}, and
0.556 \emph{balance}. For Na\"{i}ve Bayes with SMOTE (3), the authors report a
mean performance of 0.401 \emph{recall}, 0.130 \emph{pf},
0.576 \emph{G-measure}, and of 0.559 \emph{balance}. For simple Na\"{i}ve Bayes
(4), the authors report a mean performance of 0.260 \emph{recall},
0.075 \emph{pf}, 0.453 \emph{G-measure}, and 0.470 \emph{balance}. For
AdaBoost (5), the authors report a mean performance of 0.527 \emph{recall},
0.199 \emph{pf}, 0.625 \emph{G-measure}, and 0.606 \emph{balance}. For
TransferBoost (6), the authors report a mean performance of 0.829 \emph{recall},
0.631 \emph{pf}, 0.431 \emph{G-measure}, and 0.460 \emph{balance}. The
authors use Mann-Whitney-U tests in combination with the A-statistic to evaluate
differences between the configurations. The authors conclude that the
cost-sensitive boosting is outperformed by TransferBoost and Na\"{i}ve Bayes
with $k$-NN relevancy filter in terms \emph{recall} and by AdaCost, Na\"{i}ve
Bayes with and without SMOTE in terms of \emph{pf}. However, using
\emph{G-measure} and \emph{balance}, the proposed approach performs best among
the configurations.

\subsection{Ryu\etal, 2015b}
\label{sec:ryu2015selection}
\noindent\textbf{Approach.} \cite{Ryu2015b} proposed a relevancy filter
approach based on an adoption for string distances. The distance between two
entities is defined as the number of different metric values after \cite{Raman2003}, i.e., 
\begin{equation}
dist(s,s') = |\{m_i \in M: m_i(s) \neq m_i(s')\}|
\end{equation}

In the proposed approach, the authors first train a classifier on selected
training data. For the relevancy filter, first outliers are removed using
Mahalanobis distance \citep{Mahalanobis1936}. Then, all entities that are not
in the $\epsilon$ neighborhood of an entity of the target product are removed,
i.e., the training data is constructed as
\begin{equation}
S^{train} = \{s \in \bigcup S^{cand}: \exists~s^* \in S^*: dist(s,
s^*)\leq \epsilon\}.
\end{equation}
Using this data, the authors train a classifier. However, the classifier is not
simply applied to all instances $s^* \in S^*$. Instead, the authors
differentiate between three cases, following the LASER
approach~\citep{Raman2003}: (1) exactly one training instance has minimal
distance; (2) multiple instances with the same classification have the minimal
distance; and (3) multiple instances with different classifications have the
minimal distance. In case (1), a neighborhood with distance $\epsilon$ is
searched instead of only the nearest neighbors. In case the neighbor now has
multiple neighbors, the behavior is the same as for cases (2) and (3). If there
is only one instance in the neighborhood, the classifier is applied. In case
(2), the classification of the neighborhood is also used for the instance. In case
(3), the classifier is applied.

\noindent\textbf{Approach type.} Relevancy filtering and other (classification
model)

\noindent\textbf{Classifier.} Na\"{i}ve Bayes.

\noindent\textbf{Data.} 7 products from the NASA data. 

\noindent\textbf{Case study setup.} The authors use five
experiment configurations:
(1) \ac{CPDP} with their proposed approach; (2) \ac{CPDP} with LASER and without
their relevancy filter; (3) \ac{CPDP}  with $k$-NN relevancy filter proposed by \cite{Turhan2009}; (4) \ac{CPDP}
without any data treatment; and (5) \ac{WPDP} with 2x2 cross validation 100
times repeated. For \ac{CPDP} in configurations (1)--(3), all products except the
target product are used as training data.

\noindent\textbf{Performance measures.} \emph{recall}, \emph{pf}, and
\emph{balance}. 

\noindent\textbf{Results.} The authors report the values for
all products as well as the median
performance achieved over all products for all performance metrics. The proposed
approach (1) achieves a mean performance of 0.664 \emph{recall}, 0.335 \emph{pf}
and 0.657 \emph{balance}. LASER (2) achieves a mean performance of 0.494
\emph{recall}, of 0.249 \emph{pf} and 0.553 \emph{balance}.
The $k$-NN filter by Turhan\etal~(3) achieves a mean performance of
0.788 \emph{recall}, 0.496 \emph{pf} and 0.612 \emph{balance}. \ac{CPDP}
without any data treatment (4) achieves a mean
performance of 0.830 \emph{recall}, 0.516 \emph{pf} and 0.605 \emph{balance}.
\ac{WPDP} (5) achieves a mean performance of
0.771 \emph{recall}, 0.437 \emph{pf} and 0.640 \emph{balance}. The
authors use Mann-Whitney-U tests in combination with the A-statistic to evaluate
differences between the configurations. The results indicate that their proposed approach performs best overall in terms of
\emph{balance}. 

\subsection{Nam and Kim, 2015a}
\label{sec:nam2015hetero}

\noindent\textbf{Approach.} \cite{Nam2015a} address the
problem of combining different data sets, that may contain different sets of
metrics. They consider a single candidate product
$S$ with metrics $m_1, \ldots, m_p$  and a target product $S^*$ with a different
set of metrics $m^*_1, \ldots, m^*_{p^*}$. First, they propose to perform feature
selection as proposed by \cite{Gao2011} to reduce the number of
metrics. Then, they suggest to calculate how well each metric in the candidate
product is correlated to each metric in the target product, i.e., the pair-wise
correlations $cor(m_i, m^*_j)$ for $i=1,\ldots, p$ and $j=1, \ldots, p^*$. All
pairs with a correlation higher than a certain cutoff threshold are kept as
candidate metric pairs. Using maximum weigthed bipartite
matching~\citep{Matouek2006}, they select the candidate metric pairs that
achieve the highest correlations without any duplicates in the metrics, i.e., no metric
from the candidate product or target product occurs twice. The authors consider
three different correlation measures: Spearman rank correlation;
the $p-value$ of a Kolmogorow-Smirnov test~\citep{Massey1951}; and
percentile-based matching defined as $\frac{\sum_{k=1}^9 percentileratio(10 \cdot k)}{9}$ where
$percentileratio(n) = \frac{sp}{bp}$ is the ratio of the $n$-th
smaller percentile and the bigger percentile of the candidate and target
product. 

\noindent\textbf{Approach type.} Other (data set combination).

\noindent\textbf{Classifier.} Bayesian Networks, C4.5 Decision Trees, Logistic
Model Trees, Logistic Regression, Random Forests, Simple Logistic model based on
LogitBoost, and \ac{SVM}.

\noindent\textbf{Data.} Five products from the NASA
data, five products from the SOFTLAB data, ten products from the JURECZKO data,
all three products from the RELINK data, and all five products from the AEEEM
data. In total, 28 products from five different data sets were used. 

\noindent\textbf{Case study setup.} The authors use four experiment
configurations. (1) pair-wise \ac{CPDP} across data sets,
i.e., all products from one data set were predicted with all products from the
other four data sets; (2) \ac{WPDP} using 2x2 cross-validation 500 times repeated; (3) pair-wise\footnote{This is
not explicitly stated in the paper and our assumption.} \ac{CPDP} with the
common metrics between data sets; and (4) pair-wise \ac{CPDP} across data sets
based on an approach proposed by \cite{He2014}\footnote{Not included in this overview,
because this is not a peer-reviewed publication.} for combining data from
different data sets. The authors used the Mann-Whitney-U test to evaluate the
statistical significance of results.
This is done by using a Mann-Whitney-U test to check if the
performance on the 1000 data splits was significantly different. Depending on
the mean value, significant difference then means a win or a loss. In case of no
significant difference of the results, the approaches tie. Moreover, they
evaluated the percentage of pair-wise combinations, to which their matching
based on correlations could be applied, since too low correlations between data
sets mean that no matches will be found and, thereby, the approach cannot be
applied. 

\noindent\textbf{Performance measures.} \emph{AUC} and win/tie/loss record, i.e.,
how often their approach outperforms others.

\noindent\textbf{Results.} The authors note that the Kolmogorov-Smirnov
test with a cutoff of 0.05  and Logistic Regression as classifier yields the
best results.
Therefore, the authors, and consequently this overview, only reports on the
\emph{AUC} performance achieved with that correlation measure. The authors
report the median \emph{AUC} for each product, each data set, and for all
products together. We only report the aggregated results for all data sets,
the performances on the each data set are similar to each other. The proposed
Kolmogorov-Smirnov based matching (1) achieves a median \emph{AUC} of
0.724, the \ac{WPDP} (2) of 0.657, the \ac{CPDP} with common metrics (3) of
0.636 and the \ac{CPDP} based on the matching by \cite{He2014} (4) of
0.555. The overall performance improvement was found to be statistically
significant. Regarding the win/tie/loss record,
the authors report 66.2\% of wins for the Kolmogorov-Smirnov based matching (1)
against \ac{WPDP} (2), 66.2\% of wins against \ac{CPDP} using common metrics
(3), and 82.0\% of wins against \ac{CPDP} with matching according to
\cite{He2014}. However, this comparison is only done on 378 of the 600
pair-wise combinations, because no matching could be determined on the other 222
pairs. For AEEEM, 48\% of the combinations were matches, for RELINK 88\%, for
JURECZKO 100\%, for NASA 52\%, and for SOFTLAB 100\%. 

\subsection{Jing\etal, 2015}
\label{sec:jing2015}
\noindent\textbf{Approach.} \cite{Jing2015} proposed an approach to
create a \ac{UMR} which allows the usage of data accross
data sets with different metrics. 
Assuming that $M^{train}$ are the metrics available for
the training products and $M^*$ the metrics for the target product. Then $M^{both} =
M^{train} \cap M^*$, $M^{trainonly} = M^{train} \setminus M^*$ and
$M^{*only} = M^* \setminus M^{train}$. Since the metric values of $M^{*only}$
are missing for the training products, the authors define $m(s) = 0~\forall s
\in S^{train}, m \in M^{*only}$ and vice versa for the metrics of the training
products not available for the target product $m(s) = 0~\forall s \in
S^*, m \in M^{trainonly}$. This way, a \ac{UMR} is defined with values for all
metrics in $M = M^{train} \cup M^*$. The authors then apply
\ac{CCA}~\citep{Hardoon2004} to find linear transformations $w^{train}, w^*$ for
the training and test data that maximize the correlation between the two data sets. This is done
by solving the eigenvalue problem
\begin{equation}
\begin{bmatrix} & C^{both} \\ C_{both} & \end{bmatrix}\begin{bmatrix} w^{train}
\\ w^* \end{bmatrix} = \lambda \begin{bmatrix}C^{train} & \\ & C^*
\end{bmatrix}\begin{bmatrix} w^{train} \\ w^* \end{bmatrix}, 
\end{equation}
where $C^{train}, C^*$, and $C^{both}$ are the co-variance matrices of the
\ac{UMR} of training data, target data, and both together. 
The defect prediction is then performed using the transformed data. 
Prior to all of the above, the authors propose to use z-score standardization
(see Section~\ref{sec:uchigaki2012}).

\noindent\textbf{Approach type.} Other (data set combination).

\noindent\textbf{Classifier.} $k$-NN, Logistic Regression, Na\"{i}ve Bayes,
Random Forest, and \ac{SVM}.

\noindent\textbf{Data.} Three products from the NASA data, all three
products from the SOFTLAB data, all three products from the RELINK data, and all
five products from the AEEEM data. 

\noindent\textbf{Case study setup.} The authors use three experiment
configurations: (1) \ac{CPDP} with their approach; (2) \ac{CPDP} with the $k$-NN
relevancy filter proposed by \cite{Turhan2009} (see
Section~\ref{sec:turhan2009}); (3) \ac{CPDP} with data gravitation propoposed by
\cite{Ma2012} (see Section~\ref{sec:ma2012}); (4) \ac{CPDP} with TCA+
proposed by \cite{Nam2013} (discussed in Section~\ref{sec:nam2013});
and (5) \ac{WPDP} with 50\% randomly sampled training data and the rest as test
data 20 times repeated. The authors apply configuration (1) to all data.
Configurations (2)--(4) are only used for data sets with
overlapping metrics, i.e., if $M^{both} \neq \emptyset$, which is the case for
all pairings except the ones with the AEEEM
data. Configuration (5) is only used if there are no overlapping metrics, i.e.,
$M^{both} = \emptyset$, which is the case for all pairings with the AEEEM data.
For all \ac{CPDP} approaches, the authors consider both pair-wise prediction
with products from different data sets, as well as prediction with all data
from one data set for products of other data sets. TCA+ is only used for
pair-wise predictions. As classifier, the authors use $k$-NN for their approach
(1) and Na\"{i}ve Bayes for the others (2)--(5). The authors also report the
performance of the other classifiers listed above for their approach for five
predictions.

\noindent\textbf{Performance measures.} \emph{recall}, \emph{pf},
\emph{F-measure}, and \emph{MCC}

\noindent\textbf{Results.} The authors report nearly all values for
products and configurations for \emph{recall}, \emph{pf}, and \emph{F-measure},
as well as the mean values achieved for all predictions with the same number of common
metrics\footnote{Some combinations are missing in the tables, e.g., all
predictions for the product AR5 of the SOFTLAB data and the prediction
of the product Safe from RELINK data using the SOFTLAB data. Moreover,
different combinations of three products from the AEEEM data were used
to predict products from the other data. No explanation about the selection of
these subsets is given.}.
The results achieved with all data are generally better
than the results achieved with pair-wise predictions, which is why we restrict
our summary to the results achieved with all data from a data set for training.
This means we do not report the results of TCA+ (4), which are only available
for pair-wise predictions. However, in general, TCA+ was somewhere between the
$k$-NN (2) and data gravition (3) performance. 

For the data with 28 overlapping metrics (NASA
with SOFTLAB), the authors report for their approach (1) a mean
performance of 0.81 \emph{recall}, 0.02 \emph{pf}, and
0.84 \emph{F-measure}.
For the $k$-NN relevancy filter (2), the authors report a mean
performance of 0.65 \emph{recall}, 0.09 \emph{pf}, and
0.59 \emph{F-measure}.
For data gravitation (3), the authors report a mean performance of
0.74 \emph{recall}, 0.37 \emph{pf}, and of 0.44 \emph{F-measure}.
For the data with 3 overlapping metrics (NASA and SOFTLAB with
RELINK), the authors report for their approach (1) a mean
performance of 0.67 \emph{recall}, 0.22 \emph{pf}, and
0.55 \emph{F-measure}. For the $k$-NN filter (2), the authors report a mean
performance of 0.36 \emph{recall}, 0.21 \emph{pf}, and
0.34 \emph{F-measure}.
For data gravitation (3), the authors report a mean performance of
0.38 \emph{recall}, 0.31 \emph{pf}, and 0.30 \emph{F-measure}. For the
data with no overlapping metrics (NASA, SOFTLAB, and RELINK with AEEEM), the
authors report for their approach (1) a mean performance of
0.62 \emph{recall}, 0.17 \emph{pf}, and 0.52 \emph{F-measure}. For the
\ac{WPDP} (5), the authors report a mean performance of 0.70 \emph{recall},
0.27 \emph{pf}, and 0.48 \emph{F-measure}. The authors conclude that their
approach outperforms all others, even \ac{WPDP} in the settings where there are
no common metrics between the data sets.

Moreover, the authors list the results for five prediction combinations achieved
with the other classifers. $k$-NN performs best in the reported examples. For
the same five examples, the authors report the \emph{MCC} of their approach
in comparison to others. Their approach always performs best, with one exception
where it is beaten by the $k$-NN approach. 

\subsection{Cao\etal, 2015}
\label{sec:cao2015}
\noindent\textbf{Approach.} \cite{Cao2015} proposed a transfer learning
approach using neural networks. The first step of their approach is to remove
outliers. All entities from both the training and target product are removed,
where 80\% of the metrics are more than the interquartile range outside of the
upper or lower quartile, i.e., $m_i(s) < Q_i^1-IQ_i \vee m_i(s) > Q_i^3+IQ_i$
with $Q_i^1, Q_i^3$ the upper and lower quartile and $IQ_i = Q_i^3-Q_i^1$ the
interquartile range. Then, TCA~\citep{Pan2011} is applied, similar to the
proposal by \cite{Nam2013} (see Section~\ref{sec:nam2013}).
However, in comparison to Nam\etal, Cao\etal~propose to use a kernelized TCA
instead of a linear mapping. Then, a neural network is trained based on the
mapped data. The neural network is trained in a way, that it takes a potential
bias in the data to non-defect-proneness into account. 

\noindent\textbf{Approach type.} Outlier detection and other
(instance standardization and classification model).

\noindent\textbf{Classifier.} Logistic Regression and Neural Network (see above). 

\noindent\textbf{Data.} All three products from the RELINK data and all five
products from the AEEEM data.

\noindent\textbf{Case study setup.} The authors use four
configurations\footnote{The authors use one more configuration in which they
compare themselves to an approach called VAB-SVM. However, we could not find
the cited paper in the list of accepted publications of the referenced
conference.}: (1) \ac{CPDP} with their
approach; (2) \ac{CPDP} with TCA and Logistic Regression as classifier; (3)
\ac{CPDP} with TCA+ after \cite{Nam2013}; and (4) Logistic
Regression without any data treatment. All configurations perform pair-wise
\ac{CPDP} between products from the same data set. The authors use
\ac{WPDP} to tune the parameters of their neural network for the \ac{CPDP}. We do not report
the \ac{WPDP} results, because no information was given if and how the data was
split.

\noindent\textbf{Performance measures.} \emph{F-measure} and \emph{AUC}. 

\noindent\textbf{Results.} The authors report the \emph{F-measure} for
products and configurations, as well as the mean \emph{F-measure} achieved over
all products of the same data set. On the RELINK data, the authors report a mean
\emph{F-measure} of 0.66 for their approach (1), 0.59 for TCA with Logistic
Regression (2), 0.61 for TCA+ (3), and 0.49 for Logistic Regression without data
treatment. On the RELINK data, the authors report a mean
\emph{F-measure} of 0.41 for their approach (1), 0.41 for TCA with Logistic
Regression (2), 0.41 for TCA+ (3), and 0.32 for Logistic Regression without data
treatment. Values for \emph{AUC} are not reported, only the ROC curves for some
predictions are shown. The authors conclude that their approach outperforms the
competition on the RELINK data, but not on the AEEEM data. They attribute this
to the higher class imbalance of the AEEEM data.  

\subsection{Jureczko and Madeyski, 2015}
\label{sec:jureczko2015}
\noindent\textbf{Approach.} \cite{Jureczko2015} analyzed
the impact of metrics on defect prediction results depending on different
sources of the data, i.e., proprietary industrial products, open source products, and
academic products. In order to analyze the importance of each metric for a
linear regression defect prediction model, the authors define the importance
factor
\begin{equation}
IF(m_i) = \frac{a_i \cdot mean(m_i(S))}{\sum_{m_j \in M} |a_j \cdot
mean(m_j(S))|}
\end{equation}
with $a_i$ the coefficient in the linear regression model for the metric $m_i$. 

\noindent\textbf{Approach type.} Other (metric type influence and data source
type influence).

\noindent\textbf{Classifier.} Linear regression. 

\noindent\textbf{Data.} 83 products from the JURECZKO data.

\noindent\textbf{Case study setup.} The authors use twelve experiment
configurations: (1) \ac{CPDP} of industrial products based on
industrial products; (2) \ac{CPDP} of industrial products based on open source
products; (3) \ac{CPDP} of industrial products based on academic
products; (4) \ac{CPDP} of industrial products based on open source and
academic products; (5) \ac{CPDP} of open source products based on open source
products; (6) \ac{CPDP} of open source products based on industrial
products; (7) \ac{CPDP} of open source products based on academic products; (8)
\ac{CPDP} of open source products based on industrial and academic products; (9)
\ac{CPDP} of academic products based on academic products; (10) \ac{CPDP} of
academic products based on industrial products; (11) \ac{CPDP} of academic
products based on open source products; (12) \ac{CPDP} of academic products
based on industrial and open source products. The authors use
each of the models to predict the defects for each type of product. The results
are compared using t-tests with Bonferroni correction~\citep{Dunn1961}. 

\noindent\textbf{Performance measures.} $NofC_{80\%}$.

\noindent\textbf{Results.} 
The prediction of the industrial products (1)--(4)
performs best with a mean $NofC_{80\%}$ of 50.82 using the industrial model
(1), followed by using the open source and academic model (4) with $NofC_{80\%}$ of 52.96, the open source model (2) with $NofC_{80\%}$ of 55.38 and the academic
model (3) with $NofC_{80\%}$ of 73.59. The only statistically significant
difference is for the academic model, which is worse than the others.
 
The prediction of the open source products (5)--(8) performs best with a mean
$NofC_{80\%}$ of 54.00 using the open source model (5), followed by using the
industrial and academic model (8) with $NofC_{80\%}$ of 57.26, the industrial
model (6) with $NofC_{80\%}$ of 57.67 and the academic model (7) with
$NofC_{80\%}$ of 65.17. The only statistically significant difference is for the
academic model, which is worse than the others.

The prediction of the academic products (9)--(12) performs best with a mean
$NofC_{80\%}$ of 50.60 using the open source model (11), followed by using the
industrial and open source model (12) with $NofC_{80\%}$ of 53.19, the academic
model (9) with $NofC_{80\%}$ of 55.02 and the industrial model (10) with
$NofC_{80\%}$ of 56.34. None of the differences in prediction performance are
statistically significantly different.

Using the importance factor, the authors determine that the metric \ac{RFC} is
important in all the models based on all kinds of data. In case of academic and
open source products, \ac{Ca} and \ac{Ce} are also very important. For
industrial products \ac{LCOM} and \ac{CBO} are important. 

\subsection{Herbold, 2015}
\label{sec:herbold2015}

\cite{Herbold2015} proposed a tool for the benchmarking of \ac{CPDP}
techniques. Since this is a pure tool paper which does not
propose any approach or conduct a case study, we break with our reporting
pattern. The proposed CrossPare tool provides a framework that models the
general workflow of defect prediction experiments. CrossPare allows the
definition of relevancy filters, data transformations, and weighting schemes. It
is built around WEKA~\citep{WEKA} and allows the internal usage of all features
from WEKA. To define \ac{CPDP} experiments, scripts are defined using an XML dialect. 

\subsection{Nam and Kim, 2015b}
\label{sec:nam2015clami}
\noindent\textbf{Approach.} \cite{Nam2015b} proposed with
CLAMI a technique for unsupervised defect prediction. While this is not a
defect prediction technique, it fits our inclusion criteria since it is a
fully automated technique that does not require labeled data from within
the project. CLAMI actually consists of two parts, \ac{CLA} and \ac{MI}.

\ac{CLA} clusters the entities based on how many metric values are greater than the median,
i.e.,
\begin{equation}
clust(s) = |\{m_i: m_i(s)>median(m_i(S))\}|
\end{equation}
for $s \in S$. $clust(s)$ can also be interpreted as an integer. The
median of these cluster integers is used to define the labeling of entities. All
entities whose cluster integer is higher
than the median cluster integer are labeled as defective, i.e., 
\begin{equation}
c^{CLA}(s) =
\begin{cases}
1 & \mbox{if}~clust(s)>median(clust(S)) \\
0 & \mbox{otherwise.}
\end{cases}
\end{equation}

\ac{MI} selects metrics and entities for the training, that do not violate the
above defined classification and, thereby, removes noise from the data.
Only the metrics with the fewest violations of the classification, i.e., that
minimize
\begin{equation}
M^{MI} = \arg\min_{m_i \in M}|\{s \in S: m_i(s)>median(m_i(S))\}|
\end{equation}
are selected. Then, all entities that still violate the classification are
removed, i.e.,
\begin{equation}
\begin{split}
S^{MI} = \{&s \in S: c^{CLA}(s)=1~\wedge \\
& m_i(s)>median(m_i(S))~\forall~m_i \in M^{MI}\} \\
\cup~\{&s \in S: c^{CLA}(s)=0~\wedge \\
& m_i(s)\leq median(m_i(S))~\forall~m_i \in M^{MI}\}.
\end{split}
\end{equation}

\noindent\textbf{Approach type.} Outlier detection and other (feature selection
and unsupervised learning).

\noindent\textbf{Classifier.} Bayesian Networks, C4.5 Decision Trees, Logistic Model Trees, Logistic
Regression, Random Forests, Simple Logistic model based on LogitBoost, and
\ac{SVM}.

\noindent\textbf{Data.} All three products from the RELINK data and all
four products from the NETGENE data.

\noindent\textbf{Case study setup.} 
The foundation of the case study setup are 500 2x2 cross validations data
splits, through which the authors define 1000 data splits with 50\% of the data.
Based on these splits, the authors use five experiment configurations: (1)
\ac{CLA} only applied to a split; (2) CLAMI applied to a split; (3)
\ac{WPDP} with the counterpart of the split used for training and the split itself for
validation, i.e., the normal 2x2 cross validation; (4) comparison to a
threshold-based technique adopted from
\cite{Marinescu2004} and \cite{Catal2009}; and (5) an
expert-based approach proposed by \cite{Zhong2004}. All of the above
are performed with a Logistic Regression classifier. Only CLAMI (configuration
2) is also performed with the other classifiers. To evaluate the differences between the classifiers, the authors use the Friedman test with the Nemeyi test as post-hoc
test~\citep{Demsar2006} when they compare multiple models and the Mann-Whitney-U
test when performing pairwise comparisons.

\noindent\textbf{Performance measures.} \emph{recall}, \emph{precision},
\emph{F-measure}, and \emph{AUC}. Moreover, the authors consider the mean
rank of the five configurations achieved with a metric.

\noindent\textbf{Results.} The authors report the mean results of the
performance metrics over the 1000 data splits for all products. \emph{AUC} is
not reported for configurations (1), (4), and (5). For \ac{CLA}
(1), the authors report a mean performance of 0.692 \emph{recall},
0.594 \emph{precision}, and 0.630 \emph{F-measure}. 
For CLAMI (2), the authors report a mean performance of 0.709 \emph{recall},
0.595 \emph{precision}, 0.636 \emph{F-measure}, and 0.724 \emph{AUC}.
For \ac{WPDP} (3), the authors report a mean performance of 0.569 
\emph{recall}, 0.616 \emph{precision}, 0.584 \emph{F-measure}, and
0.694 \emph{AUC}.
For the threshold-based approach (4), the authors report a mean performance
of 0.260 \emph{recall}, 0.749 \emph{precision}, and
0.251 \emph{F-measure}. For the expert-based
approach (5), the authors report a mean performance of 0.606 \emph{recall},
0.773 \emph{precision}, and 0.647 \emph{F-measure}. Regarding the rankings of
the technique, \ac{CLA} and CLAMI are quite similar. They are the best ranked
techniques in terms of \emph{recall} with mean ranks of 2.143 for \ac{CLA} and
2.000 for CLAMI. However, they are the worst ranked techniques in terms of
\emph{precision} with mean ranks of 3.71 for \ac{CLA} and 3.857 for CLAMI. The
good rankings in terms of \emph{recall} are sufficient to reach the second and
third ranks in terms of \emph{F-measure} with 2.357 for \ac{CLA} and 2.429 for
\emph{CLAMI}, only beaten by the expert-based approach with a mean rank of
1.929. The authors conclude that \ac{CLA} and CLAMI can compete even with
\ac{WPDP} performance and the expert-based approach and only have drawbacks in
terms of precision.

The comparison of different classifiers for CLAMI determined that Na\"{i}ve
Bayes and Logistic Model Trees perform statistically significantly
better than \acp{SVM} and C4.5 Decision Trees, both in terms of \emph{AUC} and
\emph{F-measure}. The Random Forest, Bayesian Network and Logistic Regression
are not signficantly different from the other classifiers and, thereby, form a
mid-field. 

\subsection{Altinger\etal, 2015}
\label{sec:altinger2015}
\noindent\textbf{Approach.} \cite{Altinger2015} evalute how well
\ac{CPDP} works in the automotive domain. In the concrete setting, very
restrictive development processes with automatically generated code from models,
as well as strict coding guidelines were considered. The assumption of the
authors was that \ac{CPDP} should work well in this setting, since the
restrictive setting should lead to homogeneous source code. The authors propose
to use undersampling and test different data normalization techniques, as well
as a relevancy filter from the literature. 

\noindent\textbf{Approach type.} Relevancy filtering and other (instance
standardization).

\noindent\textbf{Classifier.} \ac{SVM} with \ac{RBF} kernel.

\noindent\textbf{Data.} Two products from the AUDI data. 

\noindent\textbf{Case study setup.} The authors use six experiment
configurations: (1) \ac{WPDP} where the data from the first 50\% of versions is
used for training and the rest for evaluation; (2) \ac{CPDP} with $k$-NN
relevancy filter proposed by \cite{Turhan2009} (see
Section~\ref{sec:turhan2009}); (3) \ac{CPDP} with min-max normalized data (see Section~\ref{sec:herbold2013}); (4) \ac{CPDP} with z-score standardized
data (see Section~\ref{sec:uchigaki2012}); (5) \ac{CPDP} with z-score
standardized data based on the target product after \cite{Nam2013} (see Section \ref{sec:nam2013}); (6)
\ac{CPDP} with data standardization after \cite{Watanabe2008}
(see Section~\ref{sec:watanabe2008}). Since only two products are used,
all \acp{CPDP} are pair-wise between the products. 

\noindent\textbf{Performance measures.} \emph{recall}, \emph{precision}, and
\emph{F-measure}.

\noindent\textbf{Results.} The authors report the values for all
products, configurations, and performance metrics. With \ac{WPDP} (1), the
authors report a mean performance of 0.740 \emph{recall},
0.185 \emph{precision}, and 0.295 \emph{F-measure}. For \ac{CPDP} with the
$k$-NN relevancy filter (2), the authors report a mean performance of 0.615 
\emph{recall}, 0.195 \emph{precision}, and 0.300 \emph{F-measure}. For
\ac{CPDP} with min-max normalization, and both z-score standardizations (3)--(5)
the authors report a mean performance of 0.615 \emph{recall},
0.180 \emph{precision}, and 0.275 \emph{F-measure}. For standardization after
Watanabe\etal~(6), the authors report a mean performance of
0.585 \emph{recall}, 0.210 \emph{precision} and 0.315 \emph{F-measure}.
The authors conclude that the overall performance of all models is lacking. Due
to this, the authors further investigated the cause for this and determined that the correlation
between the metrics and the defect information is weak and, moreover, the
information gain due to the metrics is rather low. Hence, the metrics do not
seem to contain the required information about the defects. Moreover, the
authors performed \ac{PCA} to visually analyze if the defective regions in the
data are overlapping, which is a requirement for \ac{CPDP}. Here, the authors
determined that the defective regions were partially disjunctive, which also
explains the weak performance.

\section{Discussion}
\label{sec:metastudy}

Using the data collected from the literature review, we now answer our research
questions and comment on the complexity of case study reporting. 

\subsection{Research Questions}

\subsubsection*{RQ1: Which approaches were already considered for \ac{CPDP}?}

\begin{figure}
\includegraphics[width=\linewidth]{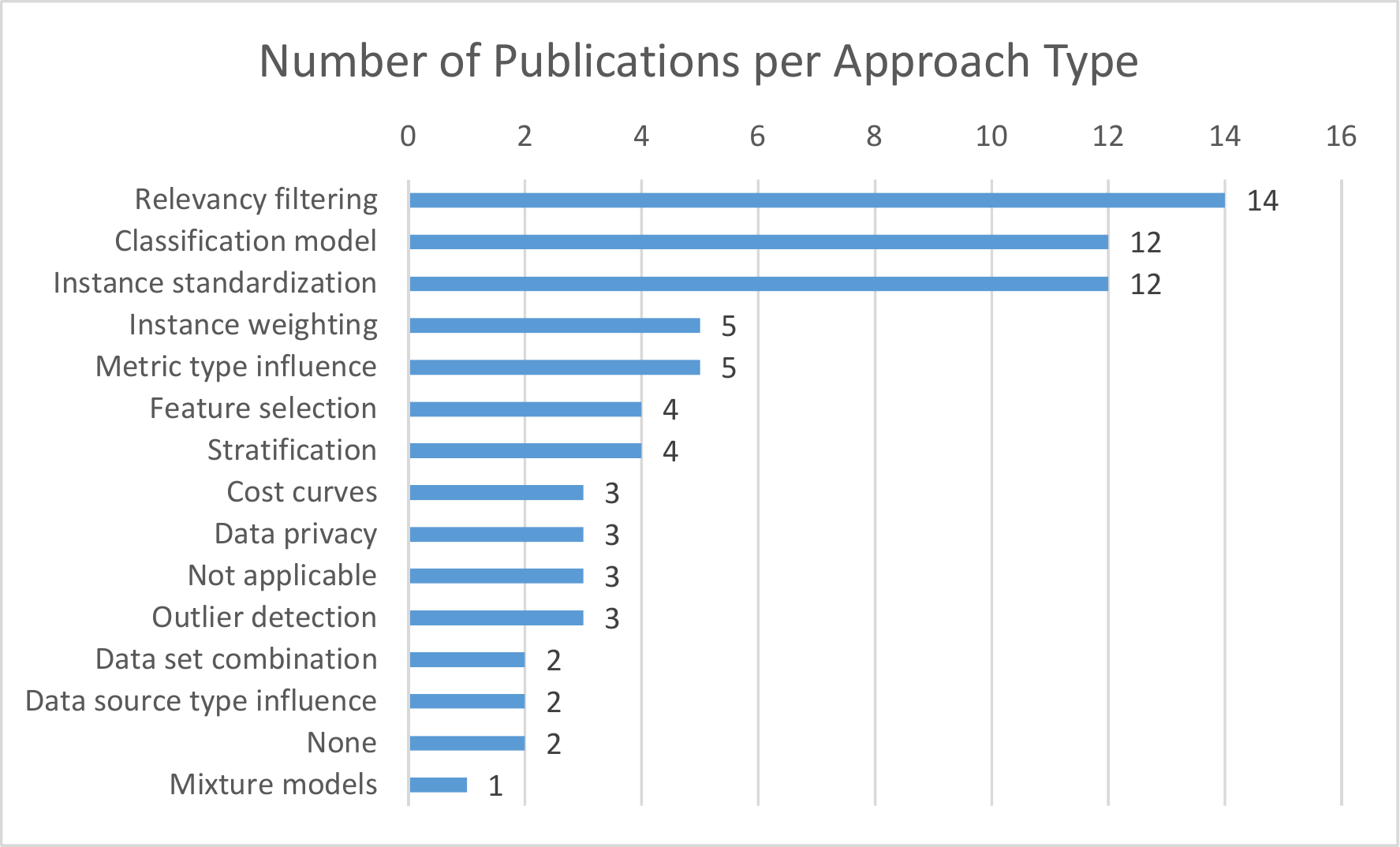}
\caption{Overview of the approach types}
\label{fig:number-approachtypes}
\end{figure}

Figure~\ref{fig:number-approachtypes} gives an overview of the approaches
considered in the state of the art. The taxonomy proposed by
\cite{Turhan2011} does not cover all considered approaches. We
identified seven types of approaches that extend the existing taxonomy:
\begin{itemize}
  \item \emph{instance standardization} of
  the training and/or test data according to specific rules with the intention to
  reduce differences between projects;
  \item \emph{classification models} which are tailored for \ac{CPDP};
  \item \emph{feature selection} to determine a subset of metrics best suited
  for \ac{CPDP};
  \item \emph{data privacy} without inhibiting the prediction performance;
  \item \emph{metric type influence} on \ac{CPDP} performance;
  \item \emph{data source type influence}, e.g., the how open source products
  influence the prediction of proprietary product; and
  \item \emph{data set combination} with none or partially overlapping features.
\end{itemize}

Among the approach types, relevancy filtering, classification models, and
instance standardization are the topics that gained most attention with 12--14
publications each. The topic instance weighting, 
metric type influence, feature selection, stratification,
cost curves, data privacy, and outlier detection received mild attention with
3--5 publications each. Influence of data source types and data set combination
are only seldomly studied with two publications each topic. However, we note
that the first publications on data set combination was in August 2015 indicating
that this is a new topic. 
Mixture models were nearly ignored as a potential solution, with only one work focusing on them.

\subsubsection*{RQ2: Which base classifiers were the most popular for \ac{CPDP}
studies?}

\begin{figure}
\includegraphics[width=\linewidth]{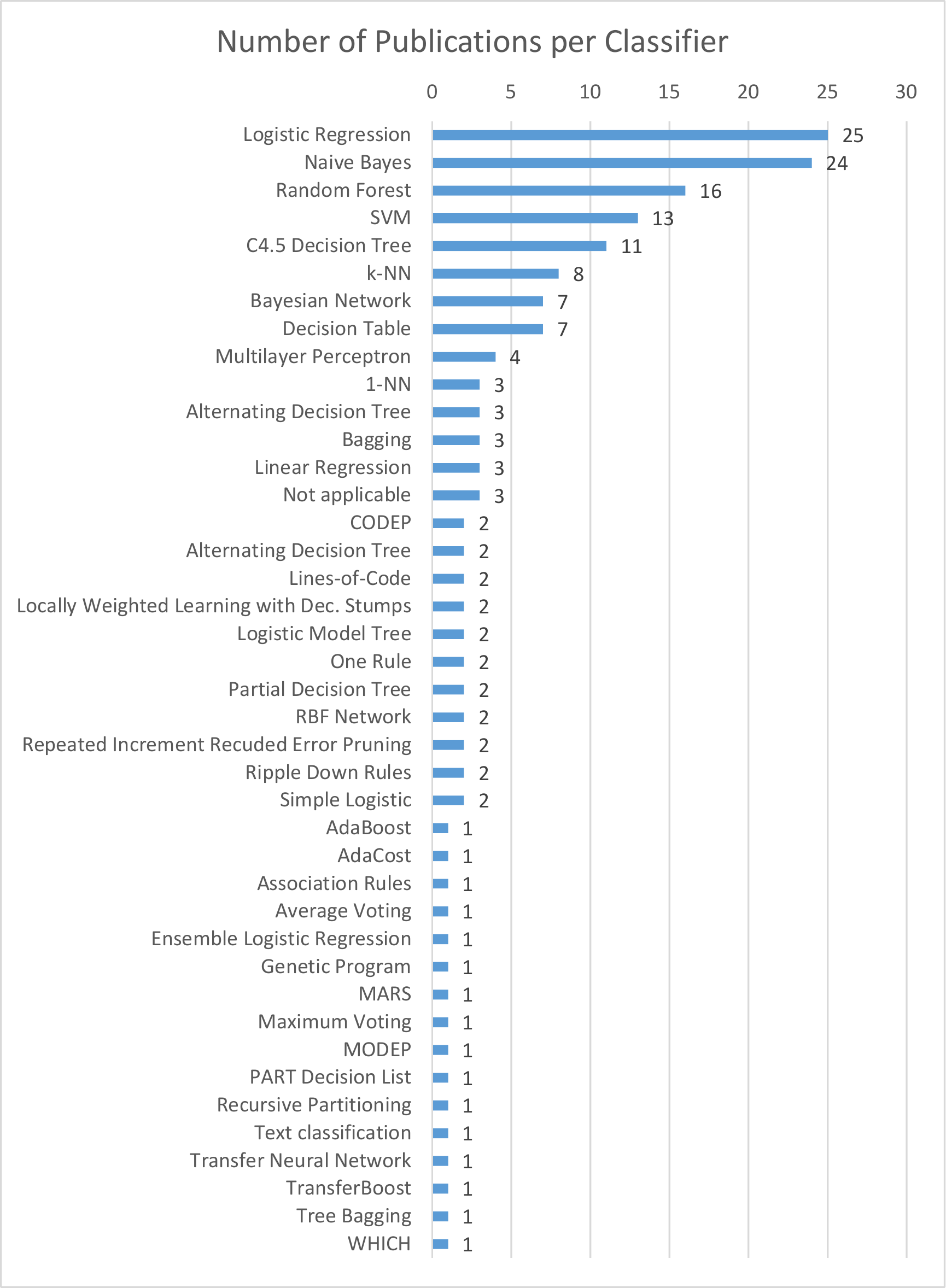}
\caption{Overview of the base classifiers}
\label{fig:number-classifiers}
\end{figure}

Figure~\ref{fig:number-classifiers} gives an overview on the classifiers used in
the publications. 38 different base classifiers were considered in total.
The two most popular are Logistic Regression and Na\"{i}ve Bayes, which were
both used in 24--25 publications, followed by Random Forest, \ac{SVM}, and C4.5
Decision Tree with 11--16 publications. $k$-NN, Bayesian Network, Decision
Table, Linear Regression, Multilayer Perceptron, 1-NN, Alternating Decision
Tree, and Bagging received mild attention with 3--7 publications considering them. All other
models, including those proposed within the literature specifically for
\ac{CPDP} were only used at most two times.
Regarding which classifiers are best, the literature is inconsistent. As candidates,
Na\"{i}ve Bayes, Logistic Regression, Random Forest, \ac{SVM}, $k$-NN, as well
as all algorithms specifically proposed for \ac{CPDP} should be considered based
on the results reported in the publications. 

\subsubsection*{RQ3: Which data sets were used within \ac{CPDP} studies?}

\begin{figure}
\includegraphics[width=\linewidth]{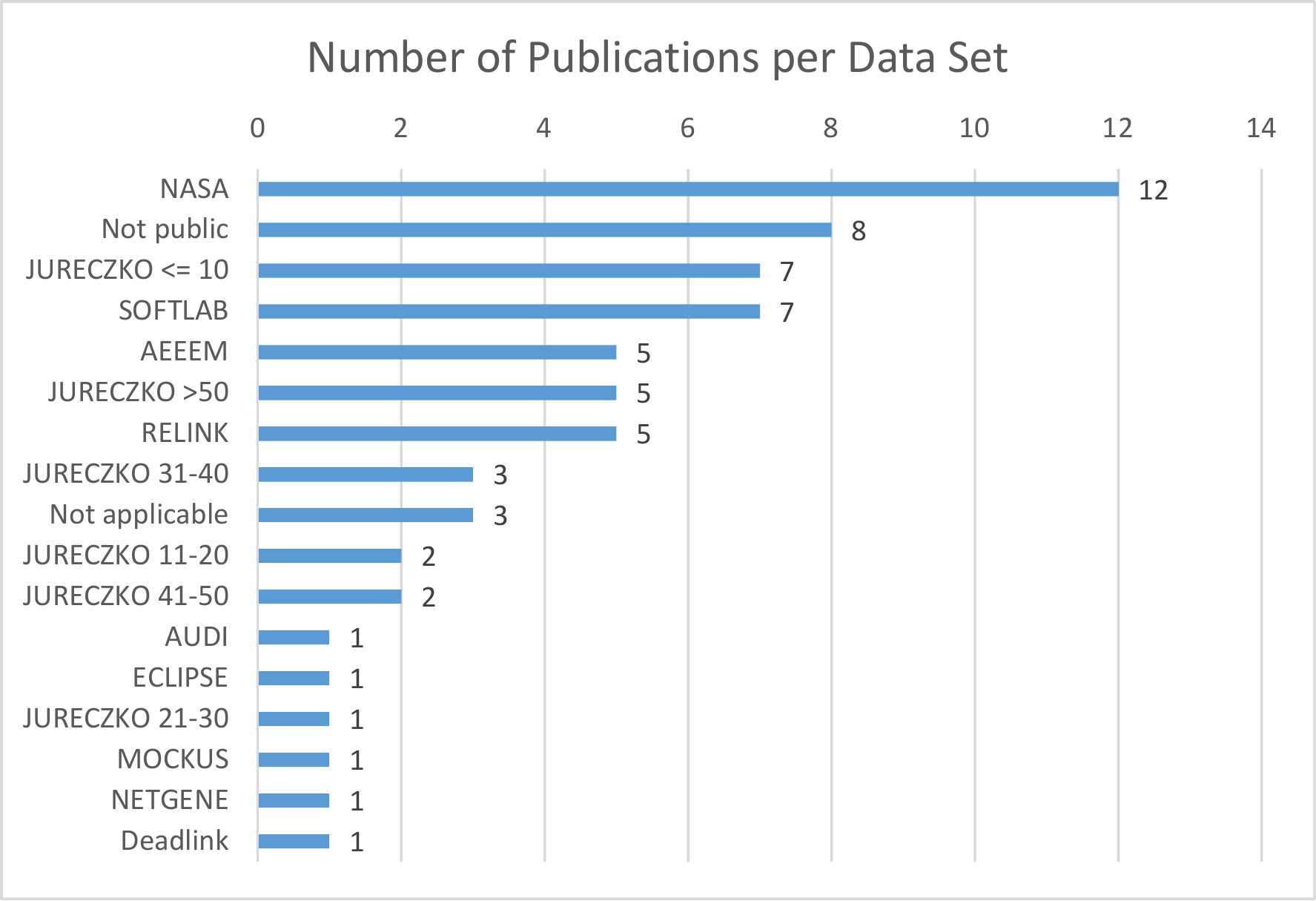}
\caption{Overview of the data set usage within case studies}
\label{fig:number-data}
\end{figure}

Figure~\ref{fig:number-data} gives an overview on the usage of the data sets.
For the JURECZKO data, the figure does not show the
overall usage of the data, but a more fine-grained look where less than or equal to 10 products, 11-20 products, 21-30 products,
31-40 products, 40-50 products and more than 50 products are used. This is to
account for our observation that different subsets of the data were used within the literature
and we think that the size can be used as relevant indicator to differentiate
between the subsets.

The JURECZKO data was the most popular in the literature and was used in 20
publications. However, most publications used subsets of different size.
Seven of these publications used less than or equal to 10 products,
only five used more than 50 products. The second most popular data set is the NASA data which was
used twelve times, followed by SOFTLAB, which was used seven times. The newer
datasets RELINK and AEEEM were used five times  each and gained traction in the
later publications, but were, overall, used less than the JURECZKO, NASA, and SOFTLAB data.
The other newer data sets MOCKUS, NETGENE, and AUDI were all used in one
publication each. The ECLIPSE data was only used once, even though it has been
available longer than all other data sets except NASA.
Additionally, one data set should have been public, but the link was dead. Data
that is not public was used by eight publications. 

\subsubsection*{RQ4: Which performance metrics were used to assess \ac{CPDP}?}

\begin{figure}
\subfigure[Metrics that measure the overall performance]
{\includegraphics[width=\linewidth]{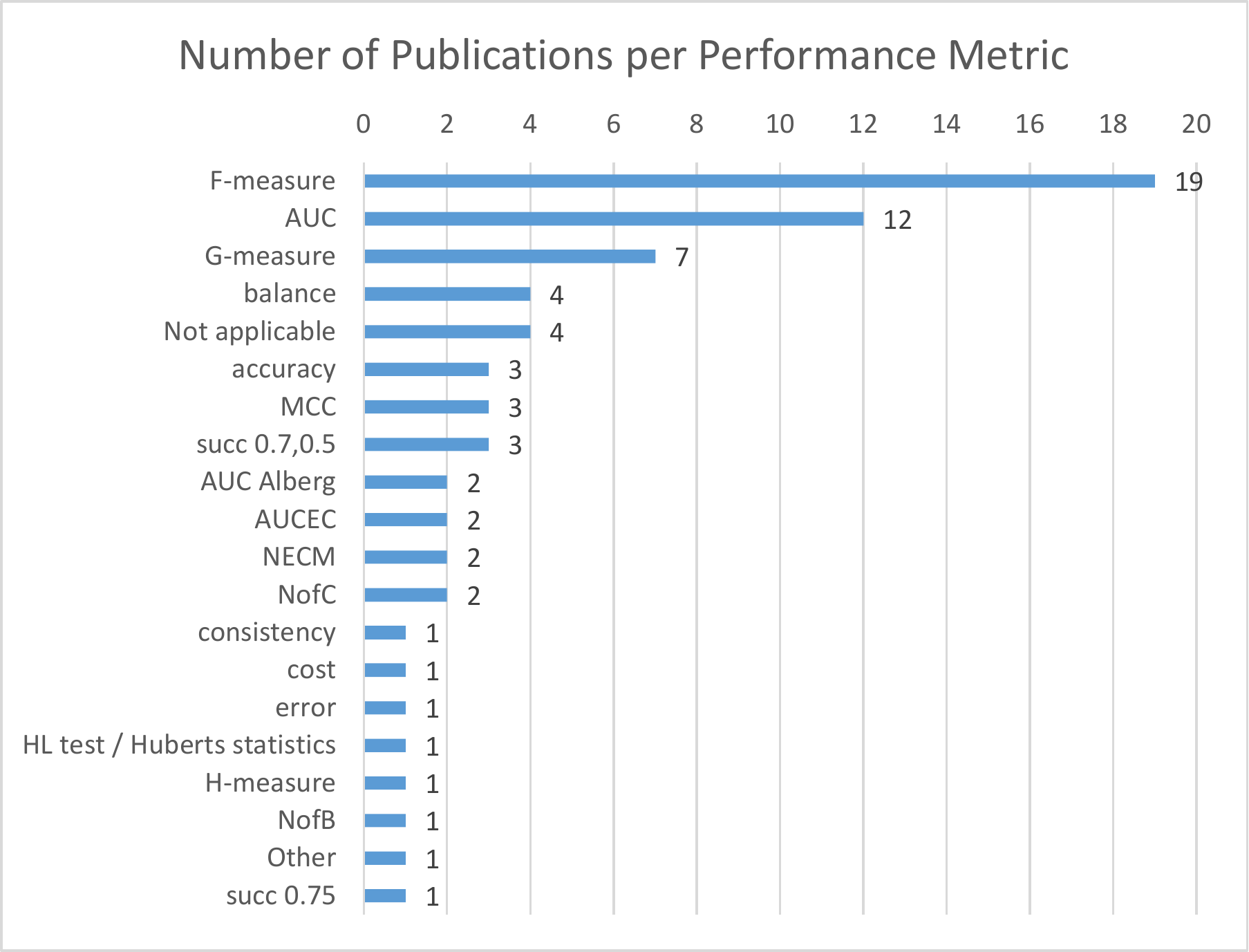}
\label{fig:number-metrics-overall}}
\subfigure[Metrics that measure specific aspects of the performance]
{\includegraphics[width=\linewidth]{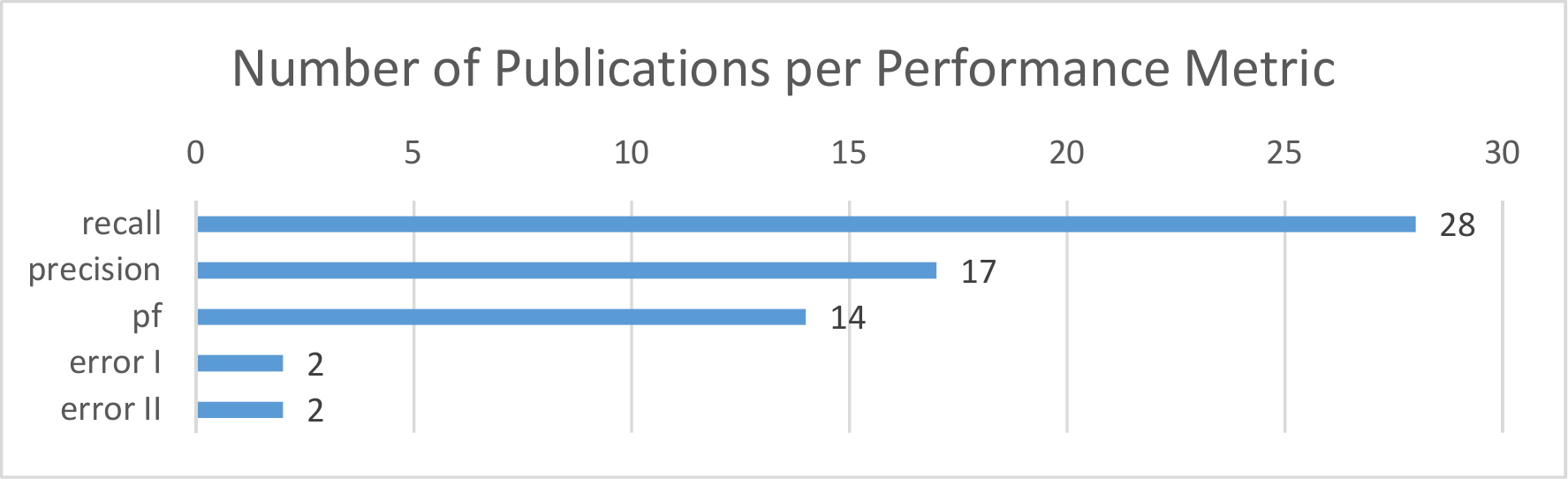}
\label{fig:number-metrics-specific}}
\caption{Overview of the performance metrics used in case studies}
\label{fig:number-metrics}
\end{figure}

Figure~\ref{fig:number-metrics} gives an overview on the performance metrics
that were used within case studies. We distinguish between two types of
metrics: overall performance metrics (Figure \ref{fig:number-metrics-overall})
and metrics that consider specific aspects (Figure
\ref{fig:number-metrics-specific}).
Of the metrics that consider the overall performance, \emph{F-measure} is the most popular and used in 19
publications, followed by \emph{AUC}, which is used in twelve publications, and
\emph{G-measure}, which is used in seven publications. The other measures
received only mild or minor attention and were used in at most four publications.
However, we note that costs were considered in eight publications, although
with different metrics. 

Of the aspect-specific metrics, \emph{recall} was by far the most important
metric and used in 28 of the publications, followed by \emph{precision} and
\emph{pf} which were in 17, resp. 14 publications and are almost equally
popular.
The other metrics were each only used two times. 

Interestingly, \emph{F-measure} which is the harmonic mean of \emph{recall} and
\emph{precision} was used more often than \emph{G-measure}, which is the
harmonic mean of \emph{recall} and $1-pf$, while \emph{precision} and \emph{pf}
were used almost equally often. 

\subsubsection*{RQ5: To which baselines were proposed approaches compared?}

\begin{figure}
\subfigure[General baselines]
{\includegraphics[width=\linewidth]{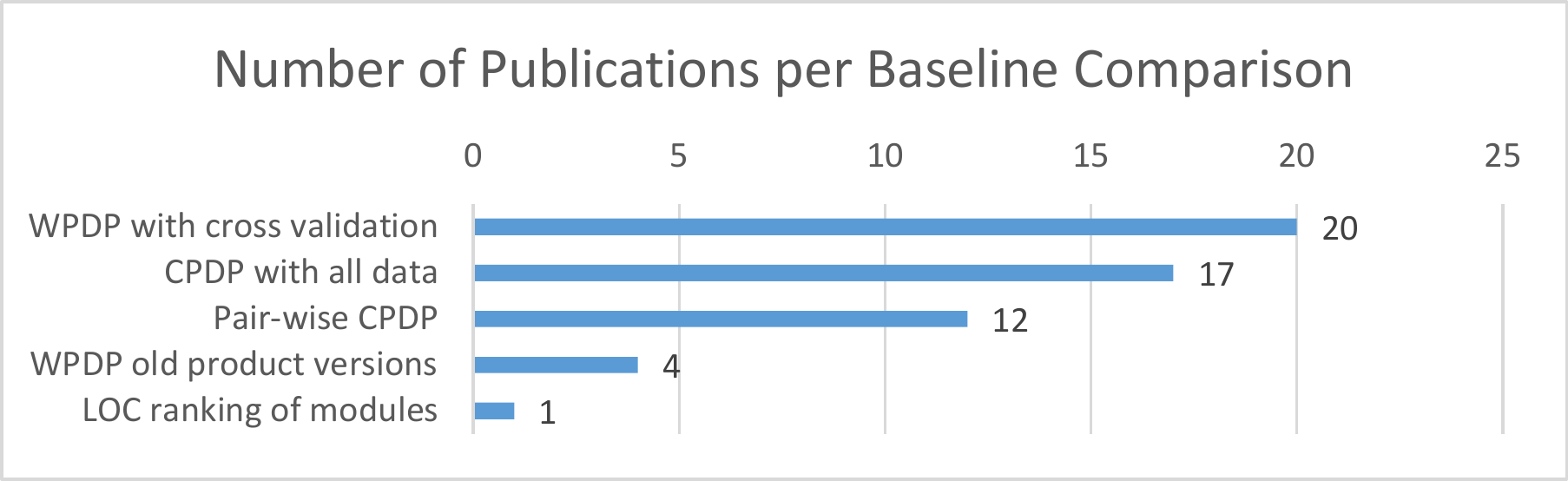}
\label{fig:number-baselines-general}}
\subfigure[Previously proposed approaches as baseline]
{\includegraphics[width=\linewidth]{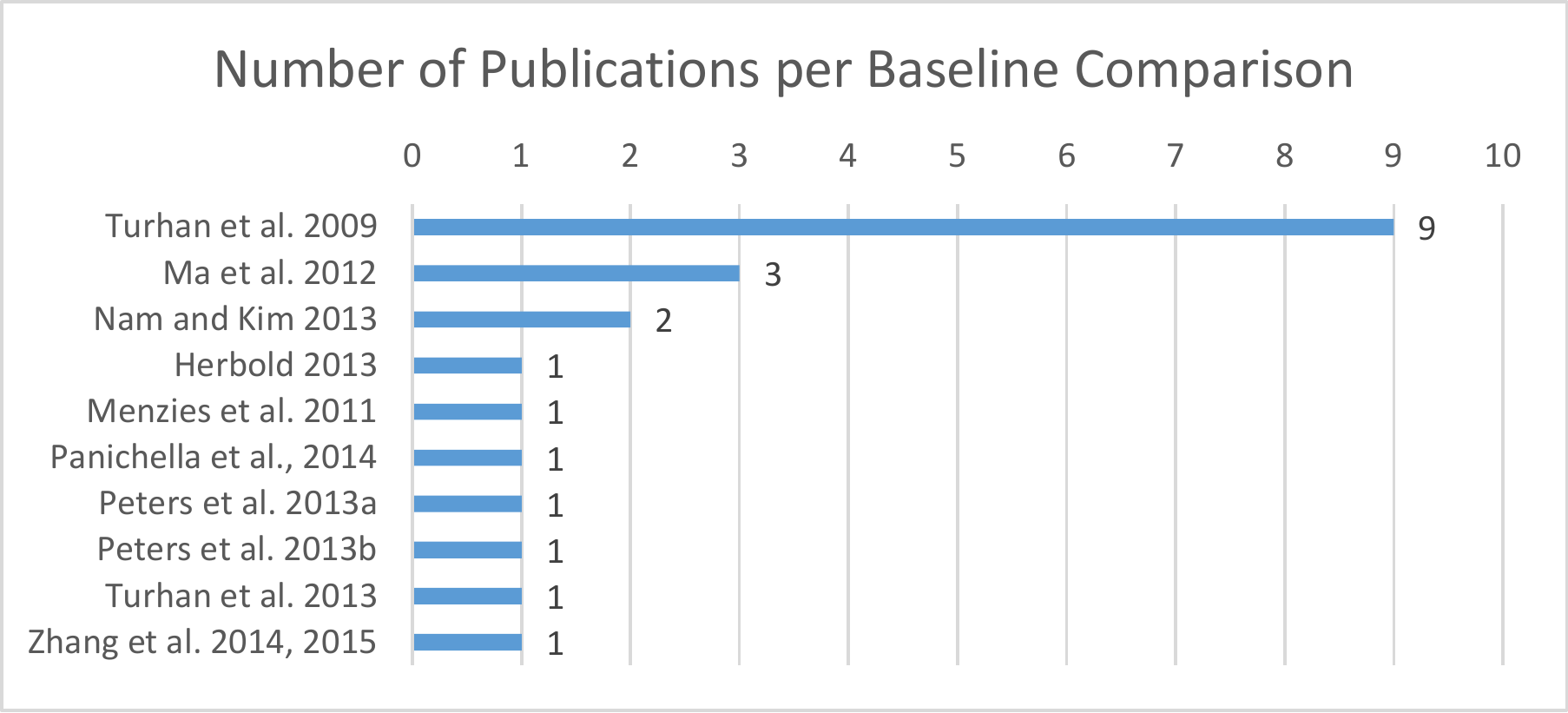}
\label{fig:number-baseline-approaches}}
\caption{Baseline configurations used for comparisons in case studies.}
\label{fig:number-baselines}
\end{figure}

Figure~\ref{fig:number-baselines} gives an overview on the baselines against
which proposed methods were compared. We distinguish between general baselines,
i.e., \ac{WPDP} are \ac{CPDP} without any specific approach and approaches that
were previously proposed in a publication on \ac{CPDP}. As general
baseline, \ac{WPDP} with cross-validation is the most popular and
used in 20 publications, followed by \ac{CPDP} with all non-target product data
as training data used in 17 publications and pair-wise \ac{CPDP} between the
products used in twelve publications and \ac{WPDP} with old versions of the same
project used in four publications. LOC ranking of modules was used only
once.

Of the approaches previously proposed for \ac{CPDP}, the $k$-NN relevancy
filter proposed by \cite{Turhan2009} (discussed in Section~\ref{sec:turhan2009}),
is used 9 times as baseline and, thereby, somewhat regularly. Nine more
approaches were also used between 1--3 times as baseline.
 
\subsection{Lessons learned}

Our mapping study showed that the work carried out for \ac{CPDP} is very
diverse. While this is, in general, a positive points because many different
techniques, data sets, and performance measures are being considered, it makes
comparisons between approaches almost infeasible in form of a metastudy that
synthesizes the results. The following points are especially problematic in that
regard: 
\begin{itemize}
  \item \ac{CPDP} studies are only comparable, if the exact same data is used.
  Due to different data sets that are nowadays available, many publications are based on different data sets,
  the largest subset of publications that uses the exact same data we found were
  six. Papers that use multiple data sets often have a difference of more than 
  10\% in prediction performance between the data sets. Even if only a subset
  is used, one cannot compare the prediction results just on the subset, because
  the training data used would still be different. Hence, one cannot just
  assume that for publications A and B with studies based on $data(A) \neq
  data(B)$, that approach A outperforms approach B if $performance(A) >
  performance(B)$. Only conjectures are possible, which would impose a major
  threat to validity of the conclusion.
  \item The diverse usage of performance metrics makes comparisons hard. While
  \emph{recall} seems to be a generally accepted metric for \ac{CPDP} studies,
  it only covers one aspect of the evaluation, i.e., if bugs are found. For all
  other metrics, there seems to be a strong disagreement within the community
  with at least four competing parties: 1) those in favor of the
  \emph{F-measure} and \emph{precision} in addition to \emph{recall}; 2) those
  in favor of the \emph{G-measure} with \emph{pf} in addition to \emph{recall};
  3) those in favor of the threshold-free \emph{AUC}; and 4) those in favor of
  cost-sensitive metrics. Studies based on different metrics are hard to
  compare. To some degree, a later inference of the confusion matrices is
  possible~\citep{Bowes2012, Bowes2013} and, thereby, the calculation of metrics
  that are not reported. However, this is not possible with all metrics, e.g.,
  cost sensitive metrics and \emph{AUC} cannot be calculated that way. 
\end{itemize}

In addition to these problems with comparability, we also observed a general
lack of replication of previously proposed techniques. Only ten of the
approaches proposed in the state of the art were ever re-used in a case study
by another publication, the other approaches were, apparently, never replicated
as part of another public case study, which in itself is a major threat to the
external validity of the results.

\section{Conclusion}
\label{sec:conclusion}

In this article, we provided a systematic mapping study off the
state-of-the-art of \ac{CPDP}. Our review detected 49 publications that address
the topic of \ac{CPDP}. We systematically summarized the
contributions of all publications, including the approach proposed,
information about the case study setup, and the results achieved. Through our
findings, we were able to extend the taxonomy by \cite{Turhan2012} on how
cross-project problems can be addressed. Moreover, we detected problems
regarding the comparability of \ac{CPDP} results across publications due
to a diverse usage of data sets, performance metrics, and base classifiers
combined with a lack of replication of previously proposed approaches for
comparison. 

\bibliographystyle{spbasic}      
\bibliography{literature}

\appendix
\begin{acronym}[DBSCAN]
\section{Table of Acronyms}
\acro{ANOVA}{ANalysis Of VAriance}
\acro{AST}{Abstract Syntax Tree}
\acro{AUC}{Area Under the ROC Curve}
\acro{Ca}{Afferent Coupling}
\acro{CBO}{Coupling Between Objects}
\acro{CCA}{Canonical Correlation Analysis}
\acro{Ce}{Efferent Coupling}
\acro{CFS}{Correlation-based Feature Subset}
\acro{CLA}{Clustering and LAbeling}
\acro{CODEP}{COmbined DEfect Predictor}
\acro{CPDP}{Cross-Project Defect Prediction}
\acro{DBSCAN}{Density-Based Spatial Clustering}
\acro{DCV}{Dataset Characteristic Vector}
\acro{DTB}{Double Transfer Boosting}
\acro{fn}{false negative}
\acro{fp}{false positive}
\acro{HL}{Hosmer-Lemeshow}
\acro{ITS}{Issue Tracking System}
\acro{JIT}{Just In Time}
\acro{LCOM}{Lack of COhession between Methods}
\acro{LOC}{Lines Of Code}
\acro{MDP}{Metrics Data Program}
\acro{MI}{Metric and Instances selection}
\acro{MODEP}{MultiObjective DEfect Predictor}
\acro{MPDP}{Mixed-Project Defect Prediction}
\acro{NN}{Nearest Neighbor}
\acro{PCA}{Principle Component Analysis}
\acro{RFC}{Response For a Class}
\acro{SCM}{SourceCode Management system}
\acro{SVM}{Support Vector Machine}
\acro{TCA}{Transfer Component Analysis}
\acro{tn}{true negative}
\acro{tp}{true positive}
\acro{RBF}{Radial Basis Function}
\acro{ROC}{Receiver Operating Characteristic}
\acro{UMR}{Unified Metric Representation}
\acro{VCB}{Value-Cognitive Boosting}
\acro{WPDP}{Within-Project Defect Prediction}
\end{acronym}

\end{document}